\documentclass[a4paper,11pt]{article}

\usepackage{graphicx}
\usepackage{float}
\usepackage[T1]{fontenc}
\usepackage{tikz}
\usepackage{epsfig}
\usepackage{xcolor}
\usepackage{amsmath,jheppub,mathtools}
\usepackage{cancel}
\usepackage{subfloat}
\usepackage{amsfonts}
\usepackage{braket}
\usepackage{verbatim}
\usepackage{scalerel}
\usepackage{booktabs}
\usepackage{epstopdf}
\usepackage{caption}
\usepackage{graphicx}
\usepackage{subcaption}
\usepackage{enumitem}
\usepackage{microtype} 
\usepackage[numbers]{natbib}
\usepackage[titletoc,toc,title]{appendix}
\usepackage{hyperref}
\usepackage{cleveref} 
\usepackage[dvipsnames]{xcolor}

\usepackage[svgnames]{xcolor}
\makeatletter
\gdef\@fpheader{}
\makeatother

\hypersetup{
    colorlinks=true,
    linkcolor=BrickRed,     % Internal links (sections/equations)
    citecolor=Bittersweet,       % Citations (bibliography)
    urlcolor=Periwinkle,      % Web links
    filecolor=red
}
\allowdisplaybreaks
\title{\boldmath Holographic Duals of $SL^{(q)}(2, \mathbb{R})$ Deformed CFTs}

\author[a]{Somnath Porey,}
\author[b]{Baishali Roy,}
\author[c]{Rahul Roy,}
\author[c]{Bobby Ezhuthachan}
\affiliation[a]{Chennai Mathematical Institute, H1 SIPCOT IT Park, Siruseri 603103, India}
\affiliation[b]{Department of Physics, Indian Institute of Technology - Kanpur, Kanpur 208016, India}
\affiliation[c]{Ramakrishna Mission Vivekananda Educational and Research Institute, Belur Math, Howrah-711202, West Bengal, India}
\emailAdd{sporey@cmi.ac.in}
\emailAdd{baishalir@iitk.ac.in}
\emailAdd{rahul.roy.phy23@gm.rkmvu.ac.in}
\emailAdd{bobby.phy@gm.rkmvu.ac.in}

\date{}

\abstract{ We study the holographic bulk dual of 2D CFTs evolved by $sl^{(q)}(2,R)$-deformed Hamiltonians. Working in the Hamiltonian-adapted frame, in which the $sl^{(q)}(2,R)$-deformed effective Floquet Hamiltonian generates time translations, we construct the dual Bañados geometries for the three distinct classes of dynamics: the heating phase, the non-heating phase and the phase boundary. Specifically, we show that in the heating phase, the bulk patch dual to the region between two neighbouring fixed points has a time-dependent causal horizon and asymptotes to a static AdS Rindler geometry at late times, while the non-heating phase has no horizon. We discuss how the late-time behaviour of the entanglement entropy, particularly its linear growth in the heating phase, can be understood from the bulk in two different ways: in global AdS, from the portion of the geodesic behind the horizons, and in a quotient description, from an effective shockwave produced by a boosted conical defect. Finally, for ($q=1$), we complete the holographic analysis of Ref. \cite{Das:2022pez} by reproducing the exact CFT expression of the OTOC from the bulk and explaining why it exhibits an exponential growth {\it only} when the operators lie on the side of the two fixed points.}

\def\tr{\text{Tr}}

\DeclareMathOperator{\sech}{sech}

\newcommand{\Lie}{\mathcal L}
\begin{document}
\maketitle
\section{Introduction}

%\subsection{Prior works and new contributions}

In this paper, we study the bulk dual of $sl^{(q)}(2, \mathbb{R})$-deformed  CFTs on a circle, by which we mean CFTs whose Hamiltonian lies in the $sl^{(q)}(2, \mathbb{R})$ subalgebra of the Virasoro algebra. Such deformed Hamiltonians arise as the effective generators of ${\it stroboscopic\; time}$ evolution in certain periodically driven Floquet CFTs \cite{FO18, FO20_2, FO20_1, Das:2021gts}\footnote{Recently they have been generalized to higher dimensions \cite{Das:2025wjo}, non-relativistic CFTs\cite{Das:2026ifh} as well as for more general class of deformations\cite{FO20_1, Lapierre:2020ftq}} . These Hamiltonians belong to one of three conjugacy classes of $sl^{(q)}(2,\mathbb{R})$, which we refer to as the heating-phase ($hp$), non-heating phase ($nh$) and the phase-boundary($pb$), following the terminology in the literature on Floquet CFTs, where they describe novel {\it dynamical} {\it phases} distinguished by the late-time behavior of observables such as the total energy ($E$), entanglement entropy ($EE$) and the out-of-time-order correlation functions ($OTOCs$). The aim of the present work is to understand the emergence of these dynamical features from a holographic perspective.

Many of the novel  dynamical features of these non-equilibrium system has its origin in the fact that these deformed Hamiltonians break translational symmetry on the spatial circle, and its signature show up as interesting spatial inhomogeneities in the temporal growth of correlation functions. 
%For instance, in the $hp$, the $sl^{(q)}(2,R)$ deformed Hamiltonian has $2q$ fixed points under the Hamiltonian flow. 
For the case of $q=1$, while the vacuum $EE$ shows no temporal growth\footnote{ Since the vacuum remains invariant under a $sl(2,R)$ deformation},  signatures of the spatial inhomogeneity show up in the behaviour of higher point correlation functions such as the OTOC. In the $hp$, the $sl^{(q)}(2,R)$ deformed Hamiltonian has $2q$ fixed points under the Hamiltonian flow. As was shown in \cite{Das:2022jrr}, the normalized OTOC  $\left(\frac{\langle V(t)W(0)V(t)W(0)\rangle}{\langle V(t) V(t)\rangle\langle W (0) W (0)\rangle}\right)$ in the vacuum of a large-$c$ $sl(2,\mathbb{R})$ deformed CFT in the $hp$, shows an exponential growth at late time, characteristic of chaotic systems ${\it only}$ when the two operators $V$ and $W$ are located on the same side of the fixed points of the Hamiltonian flow, and not when they lie  on opposite sides of the fixed points (See Figure~ \ref{fig:heating_arcs}).  

In \cite{Das:2022pez}, some features of the OTOC computation such as the lyapunov exponent of the OTOC when the two operators are placed on the same side of the two fixed points was reproduced from a bulk computation, however a bulk derivation of the exact CFT expression of the OTOC  and a holographic explanation of the opposite-side behavior remained open.   From a holographic perspective, this late time exponential growth of the OTOC in the vacuum of the $sl(2,\mathbb{R})$ deformed CFTs may seem surprising, since in the gravity picture such a growth is understood to be caused due to scattering near a bulk horizon, yet the vacuum is invariant under $sl(2,\mathbb{R})$ transformations and so the bulk remains global $AdS$. As explained in \cite{Das:2024vqe} however, the relevant horizon becomes apparent, when the  vacuum correlation functions are studied in an observers frame whose time translations are generated by the deformed Hamiltonian, which we refer to in the rest of the paper as the $Hamiltonian$-$adapted$ frame. 
%For $q=1$, this non-trivial temporal growth of the vacuum $EE$ is absent, since the vacuum remains invariant under a $sl(2,R)$ deformation. However even in this case, interesting signatures of the spatial inhomogeneity can be seen in the behaviour of higher point correlation functions such as the OTOC. As was shown in\cite{Das:2022jrr}, the normalized OTOC 
%$\left(\frac{\langle V(t)W(0)V(t)W(0)\rangle}{\langle V(t) V(t)\rangle\langle W (0) W (0)\rangle}\right)$ in the vacuum of a large-$c$ $sl(2,R)$ deformed CFT in the $hp$, shows an exponential growth at late time under the $hp$ Hamiltonian, characteristic of chaotic systems ${\it only}$ when the two operators $V$ and $W$ were placed on the same side of the fixed point, and not when they lie  on different sides of the fixed point (See Figure~ \ref{fig:heating_arcs}). In \cite{Das:2022pez}, some features of the OTOC computation such as the lyapunov exponent of the OTOC when the two operators are placed on one side of the fixed point was reproduced from a bulk computation, however a bulk derivation of the exact CFT expression of the OTOC  and a holographic explanation of the opposite-side behavior remained open. 

\subsection*{A short summary of the paper}
In this paper, we extend the analysis of \cite{Das:2022pez, Das:2024vqe}, to ($q>1$), describing the dual geometries in the three dynamical phases in the Hamiltonian-adapted frame. For these cases, the vacuum entanglement entropy $EE$ is time dependent and show a non trivial dependence on the fixed point structure of the Hamiltonian. In particular in the $hp$, one finds that the vacuum $EE$ shows a late time linear growth when the subregion include fixed points, while it saturates to a constant value when the fixed points lie outside the interval.
%\footnote{The bulk dual of the $sl^{(q)}(2,R)$ states in the $hp$ have been studied previously, for eg in []. However, there these states are interpreted as a one parameter family of generalized $sl_q(2,R)$ coherent descendant states of an undeformed 2D CFT, and not as time evolved states under a deformed Hamiltonian, as we do here. So, in that work, the  dual description of these states are as a one parameter family of three dimensional metrics, with the parameter not indentified as time. }  
We present a holographic derivation of this vacuum $EE$ from our bulk description. Our derivation shows how the fixed point structure govern its late-time behaviour, from a bulk analysis. Moreover, for ($q=1$), we complete the holographic analysis of Ref. \cite{Das:2022pez} by deriving the exact CFT expression of the OTOC from the bulk and explaining why it exhibits no exponential growth when the operators lie on opposite sides of the two fixed points. 
%From a holographic perspective, this late time exponential growth of the OTOC in the vacuum of the $sl(2,R)$ deformed CFTs may seem surprising, since in the gravity picture such a growth is understood to be caused due to scattering near a bulk horizon, yet the vacuum is invariant under $sl(2,R)$ transformations and so the bulk remains global $AdS$. As explained in \cite{Das:2024vqe} however, the relevant horizon becomes apparent, when the  vacuum correlation functions are studied in an observers frame whose time translations are generated by the deformed Hamiltonian, which we refer to in this paper as the $Hamiltonian$-$adapted$ frame. 

Following the procedure of \cite{Das:2024vqe}, we first transform to the Hamiltonian-adapted coordinates in each phase, and then use the Roberts map to obtain the appropriate bulk Ba\~{n}ados metric in these coordinates. For the case of $q=1$ in the heating phase, the observer only sees an AdS Rindler patch of the full $AdS$ space time, with the bifurcation surface of the Rindler horizon corresponding to the RT surface anchored at the fixed points in the boundary. Thus, the horizon emerges as an artifact of the observer's restricted causal diamond in AdS. In this case, there exist two such patches, restricted to the two sides of the fixed points on the circle, each of which map to two AdS Rindler exteriors. 

This generalizes in the ($q>1$) case, where in the $hp$, the dual geometry describes the bulk causal wedge for the spatial slice between two nearest fixed points. In this case  there are $q$ identical  pair of patches, which we label as $P$ and $S$. The geometry in each patch is described by a time dependent metric with a time dependent causal horizon, which asymptotes to a static AdS Poincare at late times. The time dependence is due to the fact that unlike the $q=1$ case, the $sl^{(q)}(2,R)$ Hamiltonian does not keep the vacuum invariant. The fact that they both asymptote to a static AdS Rindler exterior accounts for the CFT result that the $EE$ of an interval restricted to lie between two nearest neighbour fixed points saturate to a constant value at late times. 

In the phase boundary, $q$ of the fixed points merge and we end up with $q$ patches. For $q=1$, thus there is only one patch and the bulk dual as seen in the corresponding observer frame is Poincare patch of AdS. For $q>1$,  these patches are described in the bulk as time dependent geometries which asymptote in each patch to AdS Rindler. The Euclidean time circles, shrink to zero size as we go from the heating phase to the phase boundary, thus the horizon temperature vanishes in the $pb$. In the case of the non heating phase, there are no fixed points of the Hamiltonian flow and the dual bulk description in the Hamiltonian-adapted frame are given by time dependent horizonless Ba\~{n}ados geometries. 

We then obtain the temporal growth behaviour of the vacuum $EE$, from this bulk geometry in two ways. In the first method, we discuss in detail, how the various patches are embedded in the global AdS spacetime. By comparing the maps between the global coordinates and the Hamiltonian-adapted frame, one can track the time dependence of the geodesic length between the end points of the boundary intervals, we can explicitly reproduce the nontrivial dependence of the entanglement entropy on the number of  fixed points straddled between the intervals. While locally, $q>1$ and $q=1$ algebras are isomorphic, globally, the $sl^{q}(2,\mathbb{R})$ action, introduces a further $\mathbb{Z}_q$ quotienting of the spatial circle. This introduces a branch cut structure in the  Euclidean map relating the usual cylinder coordinates to the Hamiltonian-adapted frame. In the second method, we exploit the branch cut structure to compute the $EE$. In more detail, for the $q>1$ case, this map has a branch point, continuing around which connects the different $P$ patches or different $S$ patches. These {\it boundary} {\it conditions} which relate the different patches, can be instead imposed via a twist operator insertion at the branch point. one can then consider an effective single pair of $P$ and $S$ patches. In this effective {\it single} {\it patch} picture, we show that the late time linear growth in $EE$ is a consequence of computing the geodesic length in an ${\it effective}$ shockwave produced by the bulk dual of this twist operator, similar to the analysis of the OTOC in \cite{Shenker:2013pqa}. In the $q=1$ case, these features are absent due to the absence of any branch points in the map, consistent with the fact in that case, the vacuum $EE$ is time independent.
%However, even in this case, the OTOC shows a non trivial dependence on the position of the operators relative to the fixed points of the $hp$ Hamiltonian. We show, how this is consistent with bulk interpretation of the two patches as two AdS Rindler exteriors, completing the bulk account of the position-dependent OTOC behaviour in this case. 

\subsection*{Prior works and new contributions}
The bulk dual of the $sl^{(q)}(2,\mathbb{R})$ states in the $hp$ have been studied previously, for eg in \cite{Caputa:2022zsr}. However, there these states are interpreted as a one parameter family of generalized $sl^{(q)}(2,\mathbb{R})$ coherent descendant states of an undeformed 2D CFT, and not as time evolved states under a deformed Hamiltonian, as we do here. So, in that work, the  dual description of these states are as a one parameter family of three dimensional metrics, with the parameter not identified as time. In \cite{Jiang:2024hgt}, the authors use a different method to describe the bulk geometry. In that paper, the time dependence of the bulk enters through a time-dependent cutoff. They also study bulk horizons and the motions of endpoints of the extremal surfaces. However, in this work we explicitly construct the subregion-adapted horizon in terms of the Hamiltonian-adapted coordinates and identify the parts of the geodesics contributing to the late-time growth. In \cite{Das:2024lra}, the initial state considered was the thermal state, and the evolution of the event horizon in the corresponding bulk dual under $sl^{(2)}(2,\mathbb{R})$ drive was studied. On the other hand, our analysis is a direct generalization of the methods introduced in \cite{Das:2022pez}, where the bulk duals are obtained by moving into the frame of the observer which evolves under the deformed Hamiltonian.   

\subsection*{Organization of the paper} 
This paper is organized as follows: In Section~\ref{Sec:Slq2r}, we begin with a brief discussion of the $sl^{(q)}(2,\mathbb{R})$-deformed CFTs in the Hamiltonian-adapted coordinates; in \ref{CFTEE} we show, as an example, how the well-known 
features of the Floquet CFTs in the literature, %from observables like  late-time behaviour of the Entanglement Entropy in $hp$ from this Hamiltonian-adapted frame.
such as the late-time behaviour of the entanglement entropy in the $hp$, can be understood from this Hamiltonian-adapted frame.
%how the well-known late-time behaviour of the entanglement entropy in the $hp$ can be understood in terms of these Hamiltonian-adapted coordinates.
%% In Section~\ref{holodualSec}, we discuss the holographic bulk dual description of the $sl^{(q)}(2,\mathbb{R})$-deformed CFTs: metrics corresponding to each phases and some of their important features like time evolution of horizons in $hp$, absence of horizon in $nh$ and the dual description of $pb$ as limiting case of that of the $hp$. We also discuss the embedding of individual observers' patches in $hp$ in global AdS$_3$ which is used in the  Section~\ref{twopicture}, to derive the entanglement entropy from the bulk geometry. Mostly based on the $hp$, in \ref{quotientdesciption} we show from an alternate and simplified quotient how can we get the late time linear growth EE in $hp$ using an effective shockwave description in \ref{shockwavedesc}. We end this section with some comments the EE in $pb$ and $nh$ in \ref{pbnhpcommentSec}. 
 In Section~\ref{holodualSec}, we discuss the detailed bulk dual geometries of the different phases of the $sl^{(q)}(2,\mathbb{R})$-deformed CFTs, including the explicit metrics corresponding to each phase and some of their important features, such as the presence or otherwise of horizons. In particular, we discuss the time evolution of the horizon in the $hp$ and the $pb$.
 %and show that the absence of a horizon in the $nh$, and 
 %and the dual description of the $pb$ as a limiting case of that of the $hp$. 
 %the $pb$ as a limit of the $hp$.
 We also discuss the embedding of individual observers’ patches in the $hp$ in global AdS$_3$, which is used in Section~\ref{twopicture} to derive the entanglement entropy from the bulk . 
 %In \ref{quotientdesciption} we show from an alternate quotient description how we can get the late-time linear growth of the EE in the $hp$, using an effective shockwave description in Section~\ref{shockwavedesc}. 
 In \ref{quotientdesciption} and \ref{shockwavedesc}, we use an alternate quotient description to obtain the late-time linear growth of the EE in the $hp$.
 We end this section with some comments on the EE in the $pb$ in \ref{pbnhpcommentSec}.
In Appendix~\ref{otoc}, we discuss the OTOC for the special case $q=1$ from the bulk description, and show that it exactly reproduces the CFT result obtained previously. %In Appendix~\ref{AppEE}, we discuss some comparisons between the Hamiltonian-adapted coordinates and the plane coordinates commonly used in the Floquet CFT literature. 
In Appendix~\ref{AppEE}, we compare the Hamiltonian-adapted coordinates with the coordinates commonly used in the Floquet CFT literature. In Appendix~\ref{appendixPIComparison}, we discuss the path-integral interpretation of the quotient description of \ref{quotientdesciption}. In Appendix~\ref{heating horizon derivation}, we find the Killing vectors corresponding to a particular boundary subregion and show that its horizon matches the null cone obtained from the Roberts map. We end this paper with an analysis of the KSW criteria in Appendix~\ref{kswa}, which justifies our choice of the physical slices in Section~\ref{Sec:Slq2r} and \ref{holodualSec}.

\section{Holographic dual of
\texorpdfstring{$sl^{(q)}(2,\mathbb{R})$}
{sl(q)(2,R)} deformed CFTs}\label{Sec:Slq2r}
%\subsection{CFT}
%labelled by coordinate $x_c$.
Our physical system is a two dimensional CFT on a circle  of length $2\pi$. In Euclidean signature, the two dimensional  space is topologically a cylinder, with complex coordinates $(\omega,\bar{\omega}),\; (\omega = \tau_c +ix_c)$, where $(x_c \sim x_c+2\pi)$ is the coordinate label of the spatial circle. Equivalently, one can also describe the geometry via a conformal mapping $(\zeta =e^{\omega})$ to the complex plane with coordinates $(\zeta, \bar{\zeta})$. In these coordinates, the deformed CFT Hamiltonian generates a one parameter ($\sigma_0$) family of conformal transformation on the plane. We now introduce the Hamiltonian-adapted coordinates $z = \sigma_0 +ix$, in which the time translation along the  Euclidean time $\sigma_0$ is generated by the system Hamiltonian, which in our case is:
\begin{equation}\label{q}
   H = \alpha (L_0 + \bar{L}_0)  +\beta (L_q +L_{-q} +\bar{L}_{q} +\bar{L}_{-q}),
\end{equation}
These two coordinates  $\sigma_0$ and $x$ are related to the coordinates of the plane through :
\begin{eqnarray}
   \alpha \left( \zeta\frac{\partial}{\partial \zeta} + \bar{\zeta}\frac{\partial}{\partial \bar{\zeta}}\right) +\beta \left(\zeta^{q+1}\frac{\partial}{\partial \zeta} + \zeta^{-q+1}\frac{\partial}{\partial \zeta}  + \bar{\zeta}^{q+1}\frac{\partial}{\partial \bar{\zeta}} + \bar{\zeta}^{-q+1}\frac{\partial}{\partial \bar{\zeta}}\right) &=&\frac{\partial}{\partial \sigma_0} \nonumber \\
 \alpha \left( \zeta\frac{\partial}{\partial \zeta} - \bar{\zeta}\frac{\partial}{\partial \bar{\zeta}}\right) +\beta \left(\zeta^{q+1}\frac{\partial}{\partial \zeta} + \zeta^{-q+1}\frac{\partial}{\partial \zeta}  - \bar{\zeta}^{q+1}\frac{\partial}{\partial \bar{\zeta}} - \bar{\zeta}^{-q+1}\frac{\partial}{\partial \bar{\zeta}}\right) &=&-i\frac{\partial}{\partial x} 
\end{eqnarray}
%Where $\zeta =e^{\omega}$ is the exponential map to  the plane. The solution to the above equation, gives $\omega(z)$. 
Depending on the conjugacy class, classified by the sign of $\mathcal{D} = 4\beta^2-\alpha^2$, there exist three class of solutions for $\omega(z)$. 
%The dynamics under the Hamiltonian \ref{q} can be studied either in the original cylinder coordinates or in the Hamiltonian-adapted coordinates. As we mentioned in the introduction, we use the 
%\footnote{We will also refer to this as the \textit{observer frame}.}. 
% In appendix [], we give a detailed comparison between the dynamics as seen in the two frames.  
We now discuss the maps in each of the three phases in more detail. 
%While the Hamiltonian acts a conformal transformation of any point on the cylinder (or plane) in the original coordinates, by construction it acts as the generator of time translations in the Hamiltonian-adapted coordinates. %See appendix [] for a detailed discussion on the dynamics as seen from both these frames and and how they are related.  
%We now discuss the maps in each of the three phases in more detail. 
%This which be useful in understanding the dynamics as seen by in these frames in the respective phases.   

\paragraph{Heating Phase}
In the heating phase $(\mathcal{D}>0)$,  after solving (\ref{q}) we get the following map:

\begin{equation}\label{theHPmap}
 \zeta^{\pm}\equiv F_{\pm}=F(x^{\pm})=\left(-\frac{\alpha}{2\beta}+\frac{\sqrt{\mathcal{D}}}{2\beta} \tan \frac{k(\sigma^0+i x^{\pm})}{2} \right)^{\frac{1}{q}}; 
\end{equation}
In the above, we have analytically continued the Euclidean time $\sigma^0\to \sigma^0+it$ and  defined $k=q\sqrt{\mathcal{D}}$, while $x^{\pm} = t\pm x$. We now define
%\footnote{We consider $\alpha>0$ and $\beta>0$. Therefore, for this parametrization $\frac{\pi}{2}<\phi<\pi$.} 
$\cos{\phi}\equiv -\frac{\alpha}{2\beta}$ and $\sin{\phi}\equiv \frac{\sqrt{\mathcal{D}}}{2\beta}$. In terms of $\phi$, the above map takes the following form:
\[F_\pm=\left(\frac{\cos (\frac{k}{2} (\sigma^0+ix^\pm)-\phi)}{\cos (\frac{k}{2} (\sigma^0+ix^\pm))}\right)^{\frac{1}{q}}.\]
The map becomes pure phase at $\sigma_1^*=\phi/k$ and $\sigma_2^*=(\phi+\pi)/k$. In the original cylinder coordinates, this corresponds to the analytic continuation $\tau_c \rightarrow i\tau$. As we show in the next section, on these two slices the Schwarzian derivatives and thus the bulk dual metric become real. For these reasons we refer to these Euclidean slices as \textit{physical slices}. In the rest of the paper, we will refer to these two slices as the $\mathrm{P}$ and $\mathrm{S}$ arcs.

%We further show in Appendix~ \ref{kswa}, that analytic continuation around any other Euclidean slice, would give rise to complex metrics which moreover do not satisfy the KSW criteria.

\begin{figure}[h]
    \centering
    \includegraphics[width=0.45\textwidth]{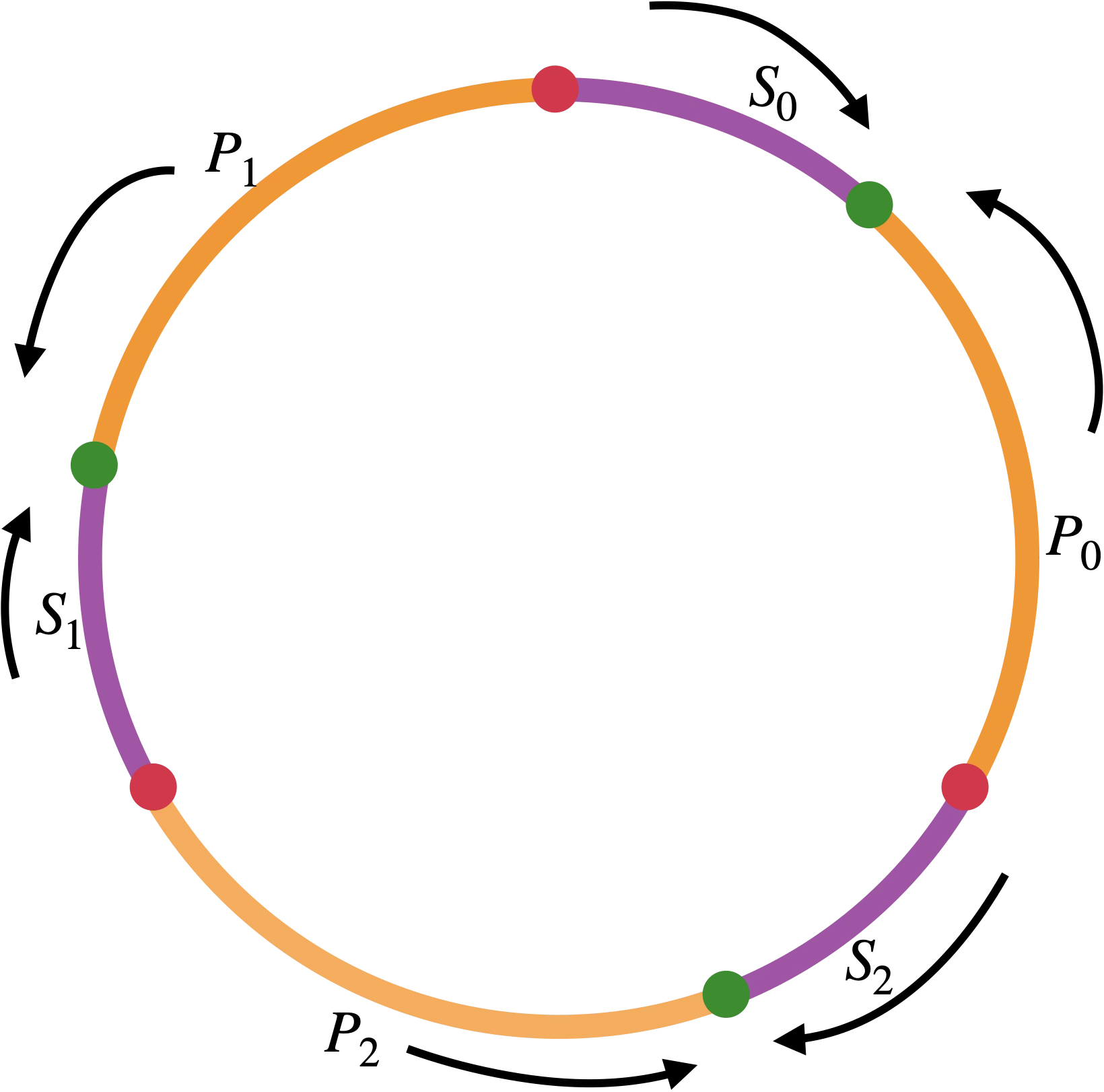}
    \caption{Representation of the dynamics in the heating phase along the spatial circle on the plane at Euclidean time slices $\sigma_1^*=\phi/k$ and $\sigma_2^*=(\phi+\pi)/k$ for $q=3$ . The red dots denote repulsive fixed points while green dots denote attractive ones. The arrows show flow direction of the points in the corresponding segments.}
    \label{fig:heating_arcs}
\end{figure}

On these physical slices, the map is
\begin{equation}
\begin{aligned}
\zeta^\pm &=F_{\pm}= e^{i\Theta_{\mathrm h}^\pm},
\qquad
\Theta_{\mathrm h}^\pm
= \frac{\psi_\eta(x^\pm)+2\pi n^\pm}{q},\\[3pt]
\psi_\eta(x)
&= 2\tan^{-1}\!\left[
\tan\frac{\phi}{2}
\left(\tanh\frac{kx}{2}\right)^\eta
\right]
+\pi(1-\eta)H(-x),\\[3pt]
\eta&=
\begin{cases}
+1, & \mathrm P,\\
-1, & \mathrm S.
\end{cases}
\end{aligned}
\label{eq:HTheta}
\end{equation}
with $n^{\pm}\in\mathbb{Z}$
labeling the branch of $(\cdot)^{1/q}$ and $H$  being the heaviside step function. From the zeroes of the flow equation along the spatial circle: $\frac{d\Theta_{\mathrm{h}}}{dx^\pm}=0$ we get attractive and repulsive fixed points at 
\begin{equation}
\Theta^{\rm att (\pm)}=\frac{\phi+2\pi n_{\pm}}{q},\qquad
\Theta^{\rm rep{\pm}}=\frac{-\phi+2\pi n_{\pm}}{q},\qquad n_{\pm}=0,\dots,q-1 .
\label{eq:fps}
\end{equation}
From (\ref{eq:HTheta}), we find that $\psi_\mathrm{P}(\pm\infty)=\pm\phi$. On the other hand $\psi_{\mathrm{S}}(-\infty)=2\pi-\phi$ and $\psi_{\mathrm{S}}(\infty)=\phi$. Thus, in both cases, the dynamics are bounded between an attractive and a repulsive fixed point. If one fixes the spatial coordinate $x$, these points are reached only asymptotically at $t\to \pm \infty$. 
% \textcolor{red}{This can be seen directly from the map. For $\alpha=0$, \eqref{theHPmap} reads $\zeta^q=\tan\frac{k(\sigma^0+ix)}{2}$, so $\sigma^0+ix=\frac{2}{k}\arctan\zeta^q$. Taking the imaginary part gives
% $$x=\frac{1}{2k}\log\frac{(1-i\zeta^q)(1+i\bar\zeta^{\,q})}{(1+i\zeta^q)(1-i\bar\zeta^{\,q})} .$$
% At the fixed points, $\zeta^q=\pm i$ (\eqref{eq:fps} at $\phi=\pi/2$), either the denominator or the numerator vanishes, so the fixed points sit at $x=\pm\infty$. On the $\tau=0$ slice, $\zeta=e^{ix_c}$, this gives
% $$e^{kx}=\frac{1+\tan\frac{qx_c}{2}}{1-\tan\frac{qx_c}{2}},$$
% which maps the arc between two consecutive fixed points, $|x_c|<\frac{\pi}{2q}$, onto $x\in\mathbb R$. For general $\alpha$, the argument is the same, with the fixed points $\zeta^q=e^{\pm i\phi}$.
% }

Since the attractive fixed point is common between consecutive P and S arcs, the points on these arcs will flow in opposite directions with respect to each other. To summarize, the $1/q$ power of the map divides the unit spatial circle into $ q$ segments, each of width $\frac{2\pi}{q}$, and due to the dynamics, each of these segments gets fragmented into two subarcs, each corresponding to P and S segments (see Fig. [\ref{fig:heating_arcs}]). The map has two branch points at $\zeta^q=0$ and $\zeta^q=\infty$, which lie outside the physical slice. In the Hamiltonian-adapted coordinates, these points are at $x^\pm=0$ and $\sigma^0=(2\phi+\pi)/k$ and $\sigma^0=\pi/k$ respectively. They play important roles while studying entanglement entropy of regions with endpoints on different segments. See Section~ \ref{CFTEE} where we study the entanglement entropy in terms of the Hamiltonian-adapted coordinates in detail.

Since the analytically continued $F$ and $\bar F$ on the two physical slices, i.e. the P- and S-lines, become pure phases ($F_\pm=\zeta^\pm=e^{i\Theta_{\rm h}^\pm}$) with real $\Theta_{\rm h}^\pm$. 
%they are no longer complex conjugates of each other. 
Such phases cannot be used directly in the Roberts map (see Section~\ref{holodualSec}), which by construction is formulated for real boundary light-cone coordinates \cite{Roberts:2012aq}. We will therefore find it convenient to map them to real coordinates by the Cayley transformation \eqref{eq:cayley}, $\chi^\pm=\tan\frac{\Theta_{\rm h}^\pm}{2}$ which we discuss in detail in Section~\ref{holodualSec}.

\paragraph{Non-heating phase}
In the non-heating phase ($\mathcal{D}<0$), the coordinate map  is given by:
\begin{equation}\label{thenHPmap}
 \zeta^{\pm}=F_{\pm}(x^{\pm})=\left(-\frac{\alpha}{2\beta}-\frac{\sqrt{\mathcal{\varepsilon}}}{2\beta} \coth \frac{q\sqrt{\varepsilon}(\sigma^0+i x^{\pm})}{2} \right)^{\frac{1}{q}}; 
\end{equation}
where $\varepsilon=-\mathcal{D}>0$. Again, if we define $\cosh\mu = \frac{\alpha}{2\beta}$ and  $\sinh \mu= \frac{\sqrt{\mathcal{\varepsilon}}}{2\beta}.$ 
the map becomes
\[F_{\pm}(x^\pm)=\left(-\frac{\sinh{\left(\frac{q\sqrt{\varepsilon}(\sigma^0+i x^{\pm})}{2}+\mu\right)}}{\sinh{\left(\frac{q\sqrt{\varepsilon}(\sigma^0+i x^{\pm})}{2}\right)}}\right)^{\frac{1}{q}}\]
We can immediately identify the similarities and differences between the map and the heating phase. In this case $F_\pm$ is a pure phase at the Euclidean slice $\sigma^*=-\frac{\mu}{q\sqrt{\varepsilon}}$. We can express the map on this Euclidean slice as
\begin{equation}
\zeta^{\pm}=e^{i\Theta^{\pm}_{\textrm{nh}}},\qquad
\Theta^{\pm}_{\textrm{nh}}=\frac{\psi_{\textrm{nh}}(x^{\pm})+2\pi n^{\pm}}{q},\qquad
\psi_{\textrm{nh}}(x)=2\arctan\Bigl(\coth\frac{\mu}{2}\tan\frac{q\sqrt\varepsilon\,x}{2}\Bigr),
\label{eq:Theta2}
\end{equation}

Therefore, the circle on the $\zeta$-plane is again divided into $q$-segments. However, now the equation\hskip 0.2cm $\frac{d\Theta_{\mathrm{nh}}^\pm}{dx^\pm}
=
\frac{\sqrt\varepsilon}{\sinh\mu}\Big[\cosh\mu+\cos(q\Theta_{nh}^{\pm})\Big]=0$ has no solutions for real $\Theta$, so the fixed points lie outside the physical slice. So any point can cross the whole circle in finite coordinate time, unlike the other two phases. The branch points of the map are at $x^\pm=0$ and at Euclidean times $\sigma^0=-\frac{2\mu}{q\sqrt{\varepsilon}}$ and $\sigma^0=0$ respectively. 
% \textcolor{green}{
% On the slice $\sigma^*=-\frac{\mu}{q\sqrt{\varepsilon}}$, the corresponding Banados factor turns out to be
% \begin{equation}
% \mathcal{L}_{\pm}=\frac{\varepsilon}{2}\Bigg[
% q^2-
% \frac{(q^2-1)\sinh^2\mu}
% {\left(
% \cosh\left(q\sqrt{\epsilon}(\sigma^*+ix^\pm)-\mu\right)
% -\cosh\mu
% \right)^2}
% \Bigg].
% \end{equation}
% In the analytically continued plane, it is also written as pure phase.}

\paragraph{Phase boundary}

On the phase boundary ($\mathcal{D}=0\ \textrm{or}\ \alpha=2\beta$), the coordinate transformation that takes us to the  Hamiltonian-adapted coordinate is
% \textcolor{green}{
% \begin{equation}
%     \zeta^{\pm}=F_{\pm}(x^{\pm})=\left(-\frac{\alpha}{2\beta}-\frac{1}{q \beta (\sigma^0+i x^{\pm})}\right)^{\frac{1}{q}}
% \end{equation}}
\begin{equation}
    \zeta^{\pm}=F_{\pm}(x^{\pm})=\left(-\frac{\frac{q \alpha}{2} (\sigma^0+i x^{\pm})+1}{\frac{q \alpha}{2} (\sigma^0+i x^{\pm})}\right)^{\frac{1}{q}}
\end{equation}
In this case, the map becomes a pure phase on the Euclidean slice $\sigma^0=-\frac{1}{q\alpha}$,  with
\begin{equation}
\zeta^{\pm}=e^{i\Theta^{\pm}_{\textrm{pb}}},\qquad
\Theta^{\pm}_{\textrm{pb}}=\frac{\psi_{\textrm{pb}}(x^{\pm})+2\pi n^{\pm}}{q},\qquad
\psi_{\textrm{pb}}(x)=2\arctan(q\alpha x),
\end{equation}
In the above, we used the fact that, on the phase boundary $\alpha=2\beta$, we take $\alpha>0$. 
\begin{figure}[h]
    \centering
    \includegraphics[width=0.45\textwidth]{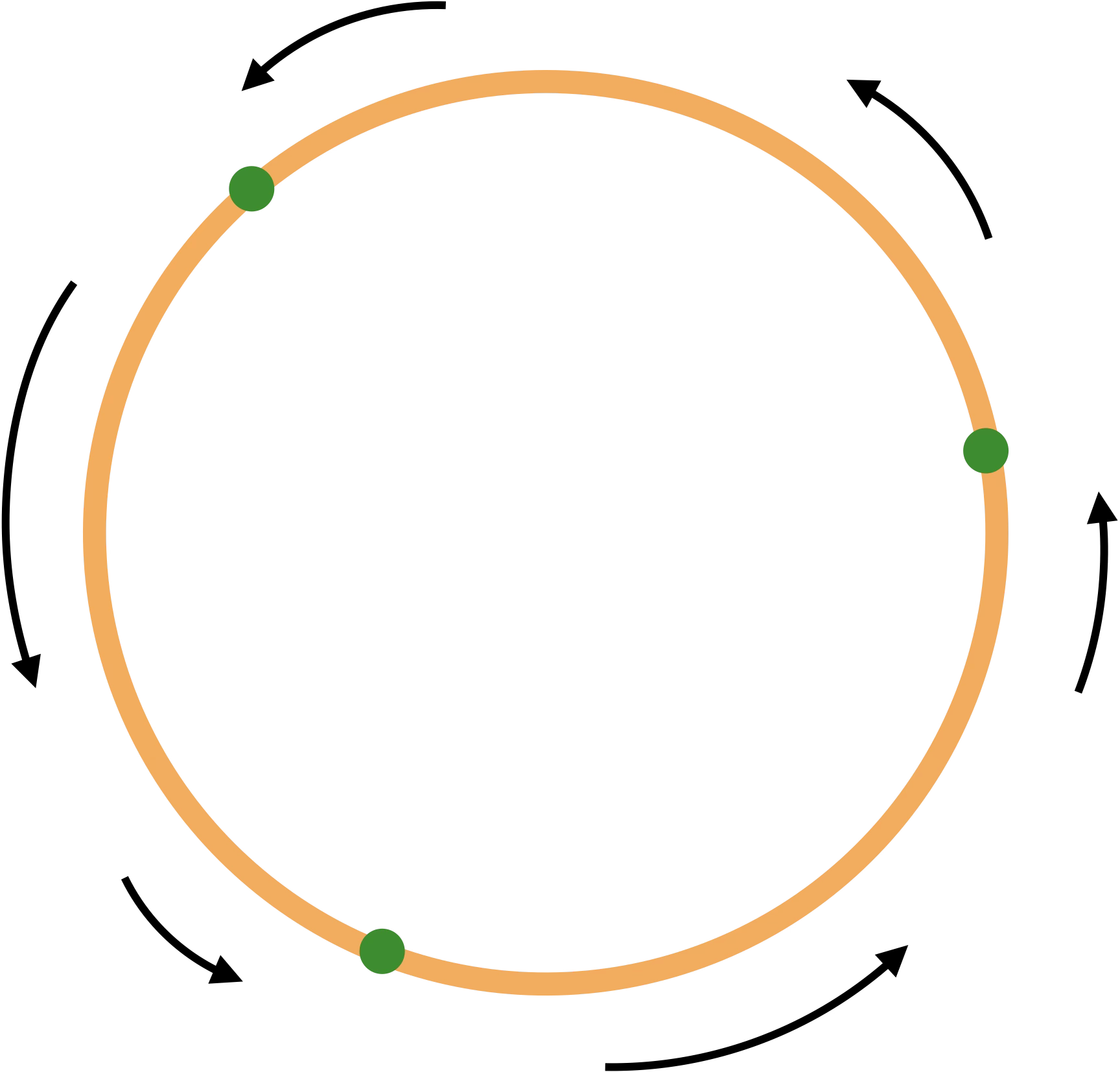}
    \caption{Representation of the dynamics on the phase boundary along the spatial circle on the plane at Euclidean time slices $\sigma_0^*=-\frac{1}{q\alpha}$ for $q=3$. The green dots are the degenerate fixed points in this case. So the fixed points are attractive from one side and repulsive on the other.}
    \label{fig:pb_arcs}
\end{figure}

From the zeros of the flow equation of $\frac{d\Theta_{\textrm{pb}}}{dx}=\alpha(1+\cos{q\Theta})=0$ we get $q$ degenerate fixed points at $\Theta=\frac{(2n^\pm+1)\pi}{q}\ \mathrm{with}\ n=0,\ldots,q-1$. For a given branch $n^\pm$, the coordinate limits satisfy
\begin{equation}
\lim_{x^\pm\to-\infty}\Theta_{\mathrm{pb}}^\pm
=\frac{(2n^\pm-1)\pi}{q},
\qquad
\lim_{x^\pm\to+\infty}\Theta_{\mathrm{pb}}^\pm
=\frac{(2n^\pm+1)\pi}{q}.
\end{equation}
So, each branch covers the interval between two consecutive
fixed points (see fig.[\ref{fig:pb_arcs}]). The fixed points on branch $n^\pm$ are reached as
$x^\pm\to+\infty$.  They can also be reached from branch $n^\pm+1$ but in the limit
$x^\pm\to-\infty$. In this case, the branch points of the map are at $\sigma^0=0\ \textrm{and}\ -\frac{2}{q\alpha}$ at $x^{\pm}=0$.

\subsection{Entanglement Entropy }\label{CFTEE}
%\textcolor{red}{Needs to add bridging sentences}\\
After having discussed the map in detail, we will now exploit the branch structure of the map to understand the time evolution of the EE \cite{FO20_1}, directly from the Hamiltonian-adapted coordinates. 

In the heating phase, the energy density shows peaks at the unstable fixed points whereas the late time behaviour of $EE$ depends on the location of the end points of the interval w.r.t. the unstable fixed points. There are four distinct cases of interest as discussed in \cite{FO20_1}. We rewrite them below but now in terms of $P_n$ and $S_n$s\footnote{m runs from 1 to q for both $P_n$ and $S_n$}:
\begin{itemize}
    \item If each of the two end points are on $P_n$ and $P_{n+m}$: includes one or more pairs of fixed points.
    \item If one end is on $P_n$ and another on $S_{n}$: includes one stable fixed point.
    \item If one end is on $S_{n-1}$ and another on $P_{n}$: includes only one unstable fixed point.
     \item If two end points are exactly on two different fixed points.
    
\end{itemize}

Below we will try to understand these different cases for the EE in our deformed coordinates: (let us first discuss only the chiral sector)

\begin{figure}
    \centering
    \makebox[\textwidth][c]{%
    \fbox{%
    \begin{minipage}{1.2\textwidth}%
        \begin{subfigure}{0.32\linewidth}%
            \includegraphics[width=\linewidth]{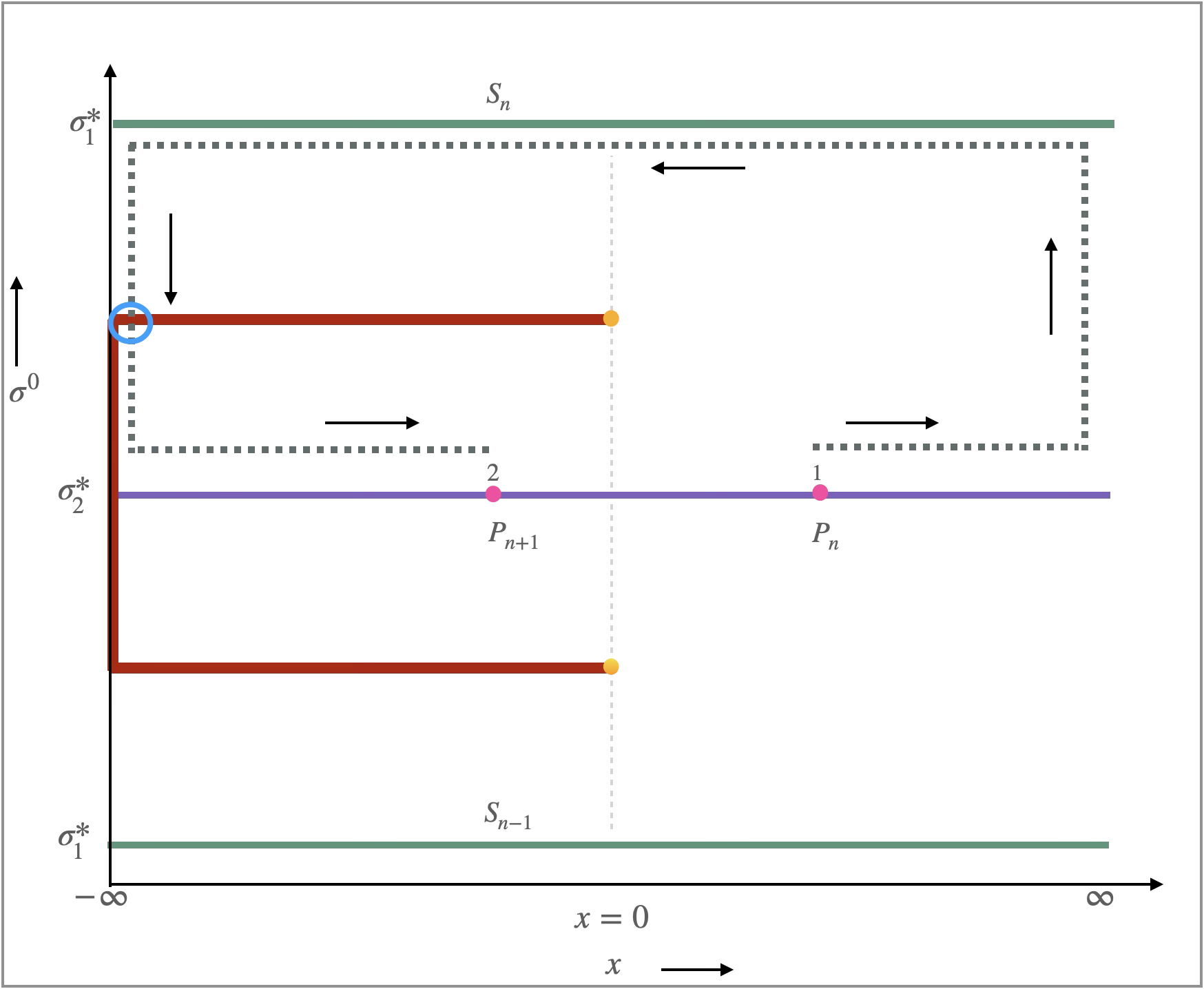}%
        \end{subfigure}%
        \hfill%
        \begin{subfigure}{0.32\linewidth}%
            \includegraphics[width=\linewidth]{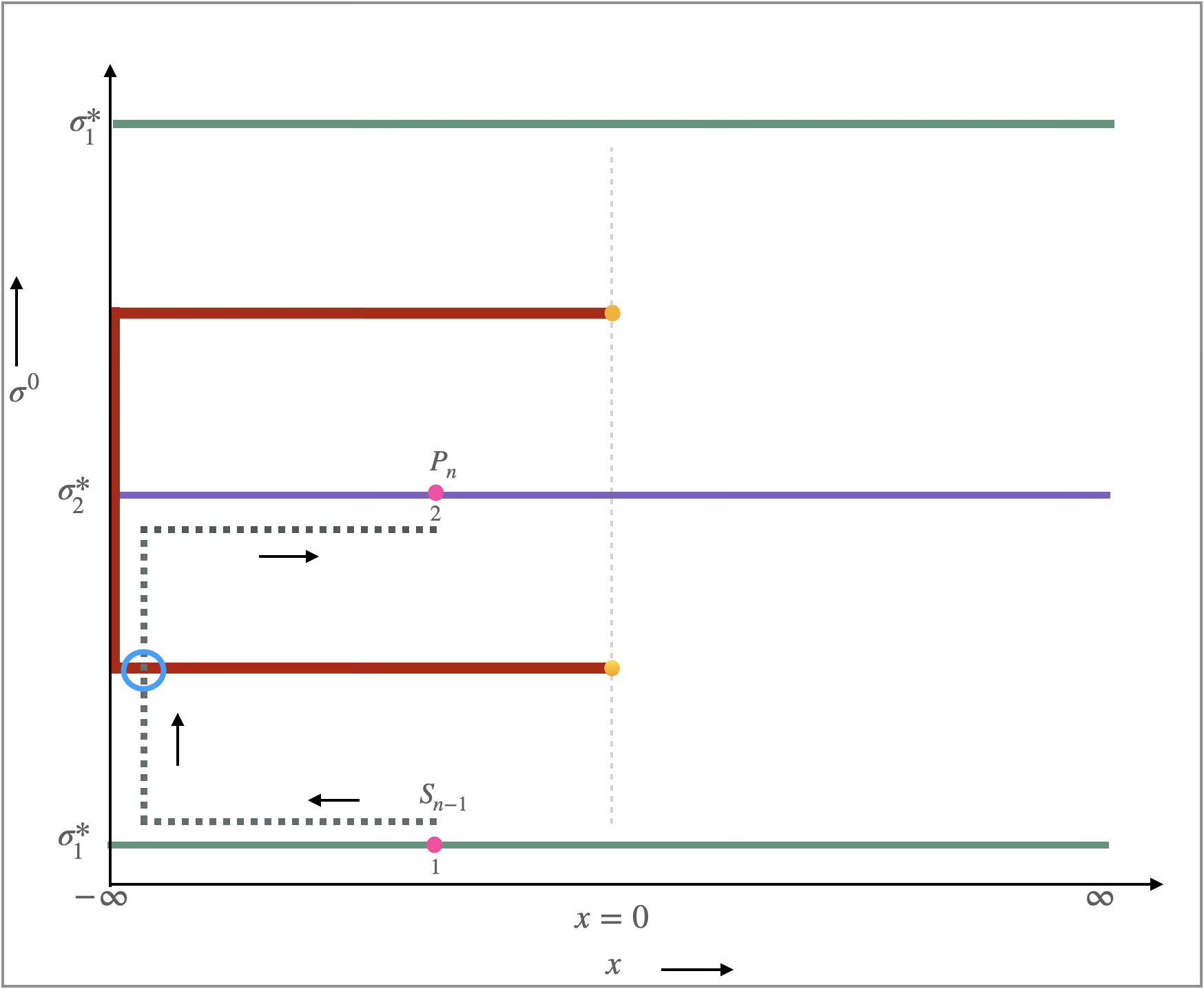}%
        \end{subfigure}%
        \hfill%
        \begin{subfigure}{0.32\linewidth}%
            \includegraphics[width=\linewidth]{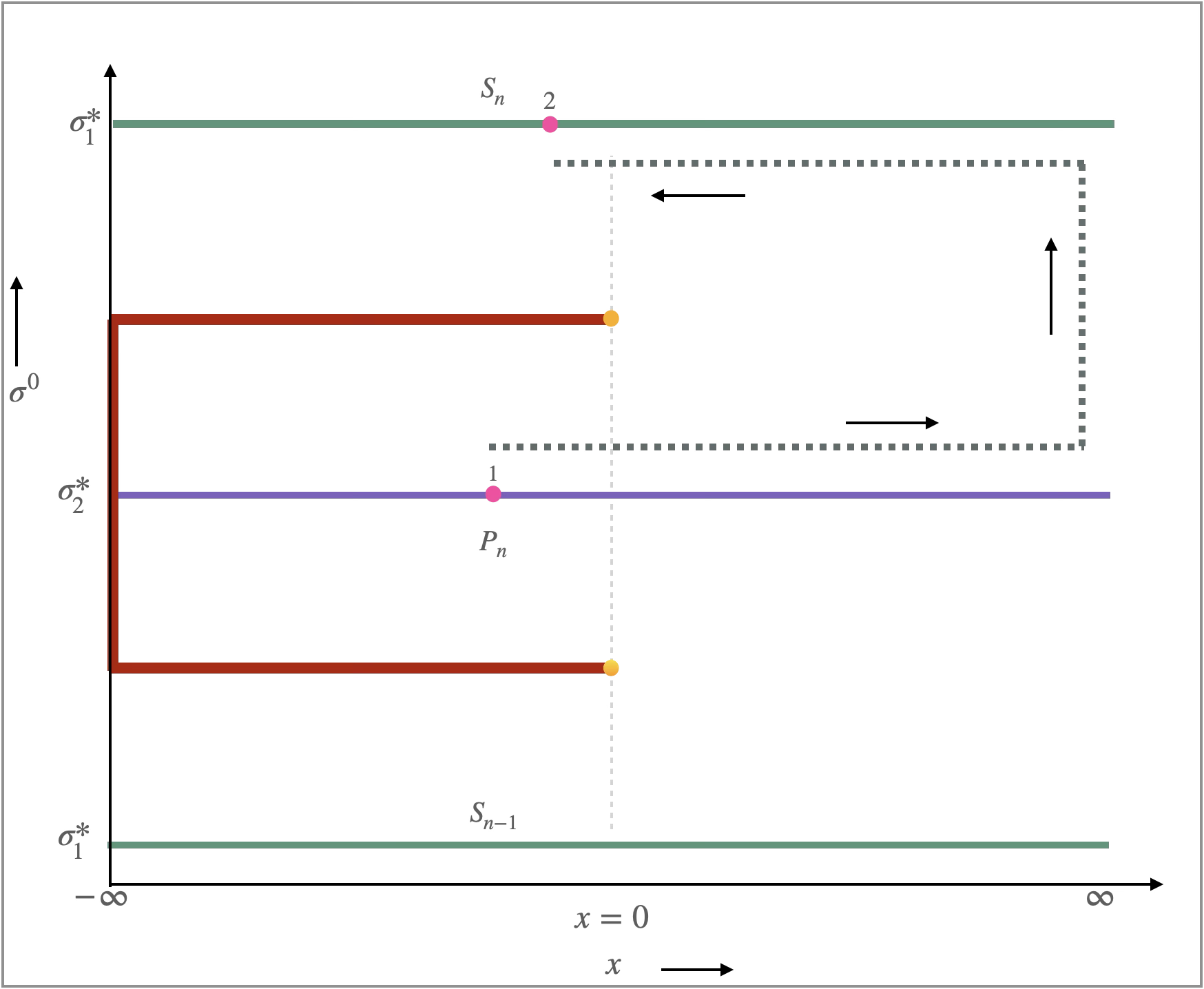}%
        \end{subfigure}%
    \end{minipage}%
    }}
   \caption{The three cases of Section~\ref{CFTEE} on the Hamiltonian-adapted cylinder, for the chiral sector. The purple line is the $P$-line, and the green lines are the $S$-line, which is drawn twice since $\sigma^0$ is periodic. The red curve is the branch cut, which joins the two branch points (yellow) at $x=0$ through $x=-\infty$. The dotted curve is the interval, with its endpoints marked 1 and 2. Left: endpoints on $P_n$ and $P_{n+1}$ (Case 1 with $m=1$). Middle: $S_{n-1}$ and $P_n$ (Case 2). Right:  $P_n$ and $S_n$(Case 3). The interval crosses the cut (blue circle) in Cases 1 and 3, but not in Case 2.}
\label{fig:EEcases}
    %\label{fig:placeholder}
\end{figure}

$\zeta_t$ in the deformed coordinate chart at time $t$ is a cylinder with Euclidean circle $\sigma^0$ and  $x\in (-\infty,+\infty)$. All the q unstable fixed points corresponding to the original coordinates get mapped to $-\infty$ and the stable ones to $+\infty$.
It has two special lines at $\sigma_1^*$ (P-line) and  $\sigma_2^*$(S-line); in terms of the original arc, going from one P-arc to S-arc means going along the P-line towards $+\infty$ and then entering the S-line from the direction of $+\infty$ upon crossing it and walking towards $-\infty$, crossing it to again enter the P-line. There are also two special points at $x=0$; the infinity branch point at $\sigma^0=\pi/k$ and the zero branch point at $\sigma^0=(2\phi+\pi)/k$. We will consider a branch cut connecting these two points passing through the P-Line (or S-line) through $x=-
\infty$ (See figure \ref{fig:EEcases}). The entanglement entropy is given by \footnote{$\epsilon_{x,i}$s are the cutoff in the Hamiltonian adapted coordinates. It is related to the uniform cut-off($\epsilon$) on the original circle as $\epsilon_{x,i}=\frac{q}{k}\cosh{kx_i}\ \epsilon$}

\begin{equation}
    S^{chiral}_{EE}=\frac{c}{6} \log\frac{|F_P^1-F_P^2|}{\sqrt{\epsilon_{x_1}\epsilon_{x_2}}\sqrt{|F_P^{1'}F_P^{2'}|}}
\end{equation}
Now suppose in the original coordinate frame we consider the first case: i.e. end points $F_P^1$ and $F_P^2$ on different P-arcs, say $P_n$ and $P_{n+m}$ respectively. We find that the  $F_P^2$ acquires an additional phase factor $e^{2\pi i m/q}$ each time it crosses the branch cut (Fig.~\ref{fig:EEcases}). Therefore, the EE becomes 
\begin{equation}
\resizebox{\linewidth}{!}{$
\begin{aligned}
S^{\text{chiral}}_{EE} &= \frac{c}{6} \log\frac{|F_P^1 - e^{\frac{2\pi i m}{q}} F_P^2|}{\sqrt{\epsilon_{x_1}\epsilon_{x_2}}\sqrt{|F_P^{1'}F_P^{2'}|}} \\
&= \frac{c}{6} \log \left|e^{i \frac{\psi_P(t+x_1)}{q}} - e^{i\left(\frac{2\pi m}{q} + \frac{\psi_P(t+x_2)}{q}\right)}\right| - \frac{c}{12} \log \sqrt{\epsilon_{x_1}\epsilon_{x_2}} \left|\frac{\sqrt{\mathcal{D}}\sin\phi}{\cos \phi + \cosh k(t+x_1)} \frac{\sqrt{\mathcal{D}}\sin\phi}{\cos \phi + \cosh k(t+x_2)}\right|
\end{aligned}
$}
\end{equation}

%\begin{align}
%&S^{chiral}_{EE}=\frac{c}{6} \log\frac{|F_P^1-e^{\frac{2\pi i m}{q}} F_P^2|}{\sqrt{\epsilon_{x_1}\epsilon_{x_2}}\sqrt{|F_P^{1'}F_P^{2'}|}}\\\nonumber&=\frac{c}{6} \log \left|e^{i \frac{\psi_P(t+x_1)}{q}}-e^{i\left(\frac{2\pi m}{q}+ \frac{\psi_P (t+x_2)}{q}\right)}\right| -\frac{c}{12} \log \sqrt{\epsilon_{x_1}\epsilon_{x_2}} \left|\frac{\sqrt{\mathcal{D}}\sin\phi}{\cos \phi +\cosh k(t+x_1)} \frac{\sqrt{\mathcal{D}}\sin\phi}{\cos \phi +\cosh k(t+x_2)}\right|
 %     \nonumber&=&
%\end{align}
since, $|F_P'|=\frac{\sqrt{\mathcal{D}}\sin\phi}{(\cos \phi +\cosh k(t+x))} $.  At late times both $\psi_p\rightarrow\phi$ (more precisely, $\psi_P(t+x)\rightarrow\phi -2\sin\phi\ e^{-k(x+t)} $ and we have $c/6 \log(2|\sin {\pi m/q}|)$ from the first term and $|F'_P|\simeq 2\sqrt{\mathcal{D}}\sin\phi\ e^{-k(t+x)}$ from the second, leading to the final expression at late time to be

\begin{equation}
    S^{chiral}_{EE}= \frac{c}{6} \log\left(\frac{ 1}{\sqrt{\epsilon_{x_1}\epsilon_{x_2}} \sqrt{\mathcal{D}}}\frac{| \sin {\frac{\pi m}{q}|}}{|\sin \phi|} \right) + \frac{c k}{12}  (2 t + x_1+x_2) 
\end{equation}
for $m\neq 0$. For $m=0$ both endpoints are on the same line (no branch-cut crossing ) and in the late-time limit both the first term and second term have the same $e^{-k t}$ fall-off rate which cancels against each other to give a finite number
\footnote{The numerator: $\left| e^{i\psi_1/q} - e^{i\psi_2/q} \right| \simeq \frac{|\psi_1 - \psi_2|}{q}= \frac{2\sin\phi}{q}\, e^{-kt} \left| e^{-kx_1} - e^{-kx_2} \right|$ and therefore the EE saturates to $\frac{c}{6}\log\!\left(\frac{2}{\sqrt{\epsilon_{x_1}\epsilon_{x_2}}q\sqrt{\mathcal{D}}}\left|\sinh\frac{k(x_1-x_2)}{2}\right|\right)$}, 
and hence the EE saturates.

Now suppose we consider the second case: i.e. when one of the end points is on nth $P$ arc  and another on nth $S$ arc in the original coordinate frame . One can immediately see that along this path from $P$ to $S$ one needs to walk towards  $+\infty$ along the $P$-line, entering  in $S$-line, without encountering the branch cut ever. Therefore, no extra phase picked up, and the EE becomes:
\begin{equation}
    S_{EE}^{chiral}= \frac{c}{6} \log\frac{|F_P^1-F_S^2|}{\sqrt{\epsilon_{x_1}\epsilon_{x_2}}\sqrt{|F^{1'}_P F^{2'}_S|}}\rightarrow \frac{c}{6}\log\!\left(\frac{2}{q\sqrt{\epsilon_{x_1}\epsilon_{x_2}}\sqrt{\mathcal{D}}}\left|\cosh\frac{k(x_1-x_2)}{2}\right|\right)
\end{equation}
as $\psi_P(t+x)\rightarrow\phi -2\sin\phi\ e^{-k(x+t)} $ and $\psi_S(t+\theta)\rightarrow\phi+2\sin\phi\ e^{-k(\theta+t)} $ in the late time. Thus, we see that the EE saturates for this case too in the late time limit.

%For the third case too we need to cross the branch cut as we go from $S_{n-1}$ to $P_n$, 
Let us now consider the third case: ie. when one of the interval end points is on (n-1)th $S$ arc  and another on nth $P$ arc in the original coordinate frame. Following the correct walking pointers we can again convince ourselves that this time for going from $S$- line to $P$-line in our new frame, we will need to cross the branch cut, and therefore an additional phase picked up by the path, and the EE at late time  grows as:
\begin{equation}
    S_{EE}^{chiral}= \frac{c}{6} \log\frac{|F_S^1-e^{\frac{2\pi i}{q}}F_P^2|}{\sqrt{\epsilon_{x_1}\epsilon_{x_2}}\sqrt{|F^{1'}_S F^{2'}_P|}}\rightarrow \frac{c}{6}\log\!\left(\frac{1}{\sqrt{\epsilon_{x_1}\epsilon_{x_2}}\sqrt{\mathcal{D}}}\frac{\sin\frac{\pi}{q}}{\sin\phi}\right)+\frac{ck}{12}\left(2t+x_1+x_2\right)
\end{equation}
and there is a linear growth. Note that $x\to-\infty$(or $+\infty$) are the unstable (or stable) fixed points for the
chiral sector while they are the opposite i.e. stable(or unstable) for the anti-chiral sector, since the latter depends on
$x^-=t-x$. Therefore while computing the anti-chiral contribution of the EE, we will need to choose the branch cut on the right side $x\to+\infty$ in the
anti-chiral sector, and the second and the third cases are exchanged and both will grow at late time.

\begin{table}[t]
\centering
\small
\renewcommand{\arraystretch}{1.5}
\begin{tabular}{@{}lcccc@{}}
\toprule
 & Endpoints & Chiral & Anti-chiral & Late-time $S_{EE}$ \\
\midrule
Case 1, $m=0$    & $P_n$, $P_n$      & saturates & saturates & $\frac{c}{3}\log\frac{2\,|\sinh(k\Delta x/2)|}{q\sqrt{\epsilon_{x_1}\epsilon_{x_2}}\sqrt{\mathcal D}}$ \\
Case 1, $m\neq0$ & $P_n$, $P_{n+m}$  & grows     & grows     & $\simeq\frac{ck}{3}\,t$ \\
Case 2           & $P_n$, $S_n$      & saturates & grows     & $\simeq\frac{ck}{6}\,t$ \\
Case 3           & $S_{n-1}$, $P_n$  & grows     & saturates & $\simeq\frac{ck}{6}\,t$ \\
\bottomrule
\end{tabular}
\caption{Late-time behaviour of the entanglement entropy in the heating phase for the first three cases, with Case 1 split into two subcases: endpoints on the same $P$-arc ($m=0$) or on $P$-arcs $m$ apart ($m\neq0$). For each case, the chiral and antichiral contributions are listed separately, together with the resulting total entanglement entropy at late times. Here $\Delta x=x_1-x_2$.}
\label{tab:EEcases}
\end{table}

\vskip 0.2cm

Case 4: When the two endpoints are sitting exactly on the fixed points. Since the fixed points are at $x\rightarrow\pm \infty$, it is easier to study with the regulated fixed point at $x_i^c$ and finally take $x_i^c\to\pm\infty$. For $1\ll t\ll|x_i^c|$, each endpoint contributes $\mp\frac{ck}{12}(t+x_i^c)$ to the chiral sector and $\pm\frac{ck}{12}(t-x_i^c)$ to the anti-chiral sector. For example, for two consecutive unstable fixed points,
%These can be further divided into three sub-cases
%\begin{itemize}
  %  \item When the end points are exactly on two consecutive unstable fixed points,
\begin{align}
S_{EE}^{chiral}|_{t\rightarrow\infty}\sim  -\frac{c\ k }{12} (2 t +x_1^c+x_2^c)\ ,
\ \ 
   \ \ \ S_{EE}^{anti-chiral}|_{t\rightarrow\infty}\sim  \frac{c\ k }{12} (2 t -x_1^c-x_2^c)
\end{align}
Note that, for the chiral sector, this matches with the linear decrease in \cite{FO20_1}
for endpoints at unstable fixed points. Together with the anti-chiral sector, whose fixed points have the opposite stability, the two contributions cancel, and the total entanglement entropy is time independent.
$$S_{EE}\simeq\frac{ck}{6}\big(|x_1^c|+|x_2^c|\big)+\text{const}$$
This is true for all such cases with interval end points on the fixed points. 
Two remarks are in order. First, the divergence as $x_i^c\to\mp\infty$ is an artifact of imposing the cutoff in the deformed coordinate $x$, which stretches without bound near a fixed point. With a fixed cutoff on the original circle, the entanglement entropy remains finite.
Second, the expressions above assume that the endpoints approach the fixed points before the late-time limit is taken, i.e. $t\ll|x_i^c|$. If instead $t\to\infty$ at fixed $x_i^c$, the end points are generic points and Cases 1–3 apply.
We have not pursued this particular case, or its bulk description, further in this paper.
% \subsection{Conformal map corresponding to the Heating phase}
% \paragraph{Correct analytic continuation}
\section{Holographic dual description} \label{holodualSec}
% We now discuss the holographic description of the \(sl_q(2,\mathbb{R})\)-deformed CFT. Assuming that the CFT admits a semiclassical holographic dual described by Einstein gravity, we investigate the corresponding bulk geometries. We use the conformal maps derived in the last section. In particular, we compute the associated Bañados functions from the Schwarzian derivatives of these maps and substitute them into the general Bañados metric. This allows us to study the bulk geometry in the Hamiltonian-adapted frame across the three phases of the deformed CFT.
% Bañados metrics are the most general solutions of the vacuum Einstein equation with negative cosmological constant, subject to the Brown-Henneaux boundary condition. In the Fefferman-Graham gauge, it takes the following form
% \begin{equation}\label{banadoslt}
% ds^2=
% \frac{dy^2}{y^2}
% +\mathcal{L}_+(x^+)(dx^+)^2
% +\mathcal{L}_-(x^{-})(dx^-)^2
% -\left(\frac{1}{y^2}+y^2\mathcal{L}_+(x^+)\mathcal{L}_-(x^-)\right)\,dx^+dx^-.
% \end{equation}
% The functions \(\mathcal L_\pm(x^\pm)\) are determined by the conformal maps introduced in the previous section through their Schwarzian derivatives,
% \begin{equation}\label{banadosfactor2}
% \mathcal L_\pm(x^\pm)=-\frac12\{F_\pm(x^\pm),x^\pm\}.
% \end{equation}
% We use this relation to construct the corresponding Bañados geometries in the three phases of the deformed CFT. 

In this section, we discuss the holographic bulk description of the \(sl^{(q)}(2,\mathbb{R})\)-deformed CFT. Assuming that the CFT has a semiclassical holographic dual description in terms of Einstein gravity, we derive the corresponding bulk geometries. We use the conformal maps from the previous section to do this. We calculate the Bañados functions using the Schwarzian derivatives of these maps and plug them into the general Bañados metric. This way, we find the geometry in terms of the  Hamiltonian-adapted coordinates for each of the three phases of the deformed CFT.
Bañados metrics are the most general solutions of the vacuum Einstein equation with negative cosmological constant, with the Brown-Henneaux boundary condition. In the Fefferman-Graham gauge, it takes the following form
\begin{equation}\label{banadoslt}
ds^2=
\frac{dy^2}{y^2}
+\mathcal{L}_+(x^+)(dx^+)^2
+\mathcal{L}_-(x^{-})(dx^-)^2
-\left(\frac{1}{y^2}+y^2\mathcal{L}_+(x^+)\mathcal{L}_-(x^-)\right)\,dx^+dx^-.
\end{equation}
The functions \(\mathcal L_\pm(x^\pm)\) are determined by the conformal maps introduced in the previous section. They are related to the  Schwarzian derivatives as
\begin{equation}\label{banadosfactor2}
\mathcal L_\pm(x^\pm)=-\frac12\{F_\pm(x^\pm),x^\pm\}.
\end{equation}
We use this relation to construct the corresponding Bañados geometries in the three phases of the deformed CFT. 

In order to study the horizons, geodesics and the defect in these geometries, we will also oftentimes find it convenient to map the Bañados metric \eqref{banadoslt} to the Poincaré patch,
\begin{equation}\label{poincare}
    ds^2=\frac{du^2-dX^+dX^-}{u^2}.
\end{equation}

Using the bulk diffeomorphism maps, 
\begin{eqnarray}\label{liftbulk}
X_{+} &\rightarrow& \hat{F}(x_{+})+\frac{2y^{2}(\hat{F}'(x_{+}))^{2}\hat{F}''(x_{-})}
{4\hat{F}'(x_{+})\hat{F}'(x_{-})-y^{2}\hat{F}''(x_{+})\hat{F}''(x_{-})}\\ \nonumber
X_{-} &\rightarrow& \hat{F}(x_{-})+\frac{2y^{2}(\hat{F}'(x_{-}))^{2}\hat{F}''(x_{+})}
{4\hat{F}'(x_{+})\hat{F}'(x_{-})-y^{2}\hat{F}''(x_{+})\hat{F}''(x_{-})}\\ \nonumber
u &\rightarrow& y\,\frac{4(\hat{F}'(x_{+})\hat{F}'(x_{-}))^{3/2}}
{4\hat{F}'(x_{+})\hat{F}'(x_{-})-y^{2}\hat{F}''(x_{+})\hat{F}''(x_{-})}
\end{eqnarray}
for a given boundary conformal transformation $\hat{F}$ with $\mathcal L_\pm=-\frac12\{\hat F,x^\pm\}$.
Note that $\hat F$ is not unique, since any Möbius transformation of it gives the same Bañados function. One might therefore be tempted to use the map $F$ itself. 
However, $F$ is the coordinate $\zeta$ on the plane\eqref{eq:HTheta}, so on the physical slices, after the analytic continuation $\sigma^0\to\sigma^0+it$, it becomes a pure phase, $F^\pm=e^{i\Theta^\pm}$.  Using $F$ directly in \eqref{liftbulk} would therefore give complex $X^\pm$ and $u$. 
%The Euclidean Poincaré patch built on $\zeta$ therefore does not continue to the Lorentzian Poincaré patch.
One should instead use the Cayley transform
\begin{equation}\label{eq:cayley}
   \chi=i\,\frac{1-F}{1+F}=\tan\frac{\Theta}{2}, 
\end{equation}
which maps the unit circle to the real line and 
in particular $\zeta^\pm=e^{i\Theta^\pm}$ to $X^\pm=T\pm X=\tan\frac{\Theta^\pm}{2}$.
While $\chi$ is complex at generic values of $\sigma^0$, it is real on the two physical slices P and S. Moreover, since the Cayley transformation is itself a Möbius transformation, the Bañados factors are unchanged, $\{\chi^\pm,x^\pm\}=\{F^\pm,x^\pm\}$.
%,and the map lifts to an isometry of the bulk. 
Therefore, throughout this paper we use $\hat F=\chi$ in \eqref{liftbulk}.
%the Lorentzian Poincaré light-cone coordinates of the boundary point i
%However, $F$ is the coordinate $\zeta$ on the plane, where the Euclidean time is the radial direction and the time reflection is the inversion $\zeta\to1/\bar\zeta$ about the unit circle.
\subsubsection*{Branch points and the branch locus}
As noted earlier in Section~\ref{Sec:Slq2r}, %\eqref{eq:HTheta},
the function $F=\tilde F^{1/q}$ is multivalued in the Hamiltonian-adapted coordinates $z$, with branch points at the zero and the pole of $\tilde F$. Encircling a branch point multiplies $F$ by $e^{2\pi i/q}$, which shifts the labels $n^\pm=0,\dots,q-1$ in \eqref{eq:HTheta} by one. These $q$ values label the $q$ sectors (the segments of Section~\ref{Sec:Slq2r}, each containing one $P$ and one $S$ arc), and the map $F\to e^{2\pi i/q}F$ takes one to the next. Since $F=\frac{i-\chi}{i+\chi}$ by \eqref{eq:cayley}, this map acts on $\chi$ as\footnote{This follows by substituting the inverse into $i\,\frac{1-e^{2\pi i/q}F}{1+e^{2\pi i/q}F}$ and multiplying the numerator and denominator by $e^{-i\pi/q}$.}
\begin{equation}\label{mobiusMap}
    \chi'=\frac{\cos\frac{\pi}{q}\,\chi+\sin\frac{\pi}{q}}{-\sin\frac{\pi}{q}\,\chi+\cos\frac{\pi}{q}} .
\end{equation}
Its fixed points, $\chi=\pm i$ ($\chi'=\chi$ implies $\chi^2=-1$), are the images of $F=0$ and $F=\infty$, i.e.\ of the two branch points. Being a Möbius transformation, \eqref{mobiusMap} leaves the Bañados functions unchanged, so all $q$ sectors carry the same pair of $P$ and $S$ metrics. Through \eqref{liftbulk} it lifts to an isometry of the Poincaré metric \eqref{poincare}, and the natural extension of the two branch points into the bulk is the curve left fixed by this isometry, the branch locus.

To find it, we use \eqref{liftbulk} in Euclidean signature, with $\chi=X+i\tau_E$, $X^+\to\chi$ and $X^-\to-\bar\chi$, and with $\hat F$ replaced by \eqref{mobiusMap}. The bulk Poincaré coordinates $(u,\chi)$ then transform as
\begin{align}
    &\chi'=\frac{\big(\cos\frac{\pi}{q}\,\chi+\sin\frac{\pi}{q}\big)\big(\cos\frac{\pi}{q}-\sin\frac{\pi}{q}\,\bar\chi\big)-\sin\frac{\pi}{q}\cos\frac{\pi}{q}\,u^2}{\Big|\cos\tfrac{\pi}{q}-\sin\tfrac{\pi}{q}\,\chi\Big|^2+\sin^2\!\tfrac{\pi}{q}\,u^2},\nonumber\\ &u'=\frac{u}{\Big|\cos\tfrac{\pi}{q}-\sin\tfrac{\pi}{q}\,\chi\Big|^2+\sin^2\!\tfrac{\pi}{q}\,u^2}.
\end{align}
A bulk fixed point requires $u'=u$ and $\chi'=\chi$. For $u\neq0$, the first condition gives
$$\Big|\cos\tfrac{\pi}{q}-\sin\tfrac{\pi}{q}\,\chi\Big|^2+\sin^2\!\tfrac{\pi}{q}\,u^2=1,$$
and the second then has the unique solution\footnote{Expanding both conditions gives
$$2\cos\tfrac{\pi}{q}\,X=\sin\tfrac{\pi}{q}\,\big(u^2+X^2+\tau_E^2-1\big),\qquad 2\sin\tfrac{\pi}{q}\,X=-\cos\tfrac{\pi}{q}\,\big(u^2+X^2+\tau_E^2-1\big).$$
Multiplying the first by $\cos\frac{\pi}{q}$, the second by $\sin\frac{\pi}{q}$, and adding gives $X=0$, and then either equation gives $u^2+\tau_E^2=1$.} for $q>1$
$$X=0,\qquad u^2+\tau_E^2=1 .$$
So the two fixed points on the boundary, $(X,\tau_E)=(0,\pm1)$, extend into the bulk as the geodesic joining them: this is the branch locus. In Lorentzian signature, $\tau_E=-iT$, it becomes the timelike curve $X=0$, $u^2-T^2=1$.

\subsection{Metric corresponding to the heating phase}
As mentioned in the previous section, the Hamiltonian-adapted coordinate map has two Lorentzian continuations around the Euclidean slices at $\sigma_1^*=\phi/k$ and $\sigma_2^*=(\phi+\pi)/k$. At these points, the Schwarzian derivative, the Bañados factor, and the corresponding metric become real. In Appendix~\ref{kswa} we show that the KSW conditions are saturated throughout the regular coordinate ranges only at these specific values of $\sigma^0$. Continuing around these two points gives two distinct Bañados factors.

\begin{equation}\label{banadosfactorhp}
\mathcal{L}^{\mathrm{P/S}}_{\pm}=\frac{\mathcal{D}}{4}\left(q^2+ \frac{(q^2-1)\sin^2\phi}{(\cosh (kx^{\pm})\pm\cos \phi)^2}\right)
\end{equation}
If we substitute this into (\ref{banadoslt}), we get the metric in Hamiltonian-adapted coordinates. For $(q=1)$, the functions are constant, and the metric takes the AdS-Rindler form.
When $(q>1)$, the metric is explicitly time-dependent. But as $x^\pm \to \infty $, both functions approach $\frac{k^2}{4}$, so the geometry becomes AdS-Rindler in the asymptotic region.

The Bañados functions for the P and S regions satisfy \[\mathcal{L}^{\mathrm{P/S}}\left(x^\pm+\frac{i\pi}{k}\right)=\mathcal{L}^{\mathrm{S/P}}(x^\pm).\] If we identify $\beta_{\mathrm{eff}}=\frac{2\pi}{q\sqrt{\mathcal{D}}}$, the two functions are related by an imaginary time shift of $(i\beta_{\mathrm{eff}}/2)$. This is similar to the analytic continuation that relates two complementary AdS-Rindler wedges. For $(q>1)$, the physical boundary circle has $q$ alternating P and S intervals (see Fig. [\ref{fig:heating_arcs}]), so each bulk region is a patch of a single global geometry.

One can also think of an auxiliary quotient description due to the  $q$-fold structure of the map . As we can see from the boundary map (\ref{eq:HTheta}), a discrete rotation $\Theta \to \Theta+\frac{2\pi}{q}$ sends $P_n$ to $P_{n+1}$ and $S_n$ to $S_{n+1}$. If we take the global $\mathrm{AdS}_3$ and identify the $q$ sectors using the $\mathbb{Z}_q$ symmetry, we get one complementary P/S pair on the quotient boundary. The resulting bulk geometry is locally $\mathrm{AdS}_3$ with a deficit angle $\delta_q=2\pi\left(1-\frac{1}{q}\right)$ along the fixed rotation axis. However, this quotient picture is different from the smooth covering-space geometry of the original CFT.

\subsubsection*{Time evolution of the horizon}
We now turn to the causal horizons associated with the individual P/S regions. These horizons characterize the portion of the bulk spacetime accessible to the observers restricted to the corresponding boundary regions.

In the heating phase, the Hamiltonian flow confines the boundary observer \footnote{By \textit{observer} what we really mean is that its clock in H and the region it can access is the causal diamond  ($\mathcal{D}(A)$) of $A$, where $A$ is an arc between adjacent fixed points of the heating phase flow.} between two fixed points. To understand the corresponding bulk causal horizons, consider a subregion with endpoints $(T,X)=(0,\pm a)$ in Poincaré coordinates\footnote{On the boundary of the global AdS boundary the fixed points are at $\tau=0$ and $\Theta_\pm=\pm\frac{\phi}{q}$ which corresponds to $T=0$ and $X=\pm \tan\left(\frac{\Theta_\pm}{2}\right)\equiv \pm a_P$ on the boundary of the Poincaré metric.}. Its causal diamond has past and future tips at  $(T, X)=(\mp a,0)$ respectively. Under the Hamiltonian evolution, the observer's worldline asymptotes to these tips as $t \to \mp \infty$.

Hence, a bulk event at $(T, X,u)$ can communicate to the observer if it lies in the causal past of the future tip. The boundary of this region defines the future causal horizon
\[\mathcal H^+:\qquad (T-a)^2=X^2+u^2,\qquad T\leq a.\]

Similarly, the past horizon is defined as the boundary of
\[\mathcal H^-:\qquad (T+a)^2=X^2+u^2,\qquad T\geq-a.\]
The intersection of these two null hypersurfaces is
\[T=0,\qquad X^2+u^2=a^2,\]
which is the RT geodesic between these endpoints in the vacuum $\mathrm{AdS}_3$. 
\begin{figure}[H]
    \centering
    \includegraphics[width=0.95\textwidth]{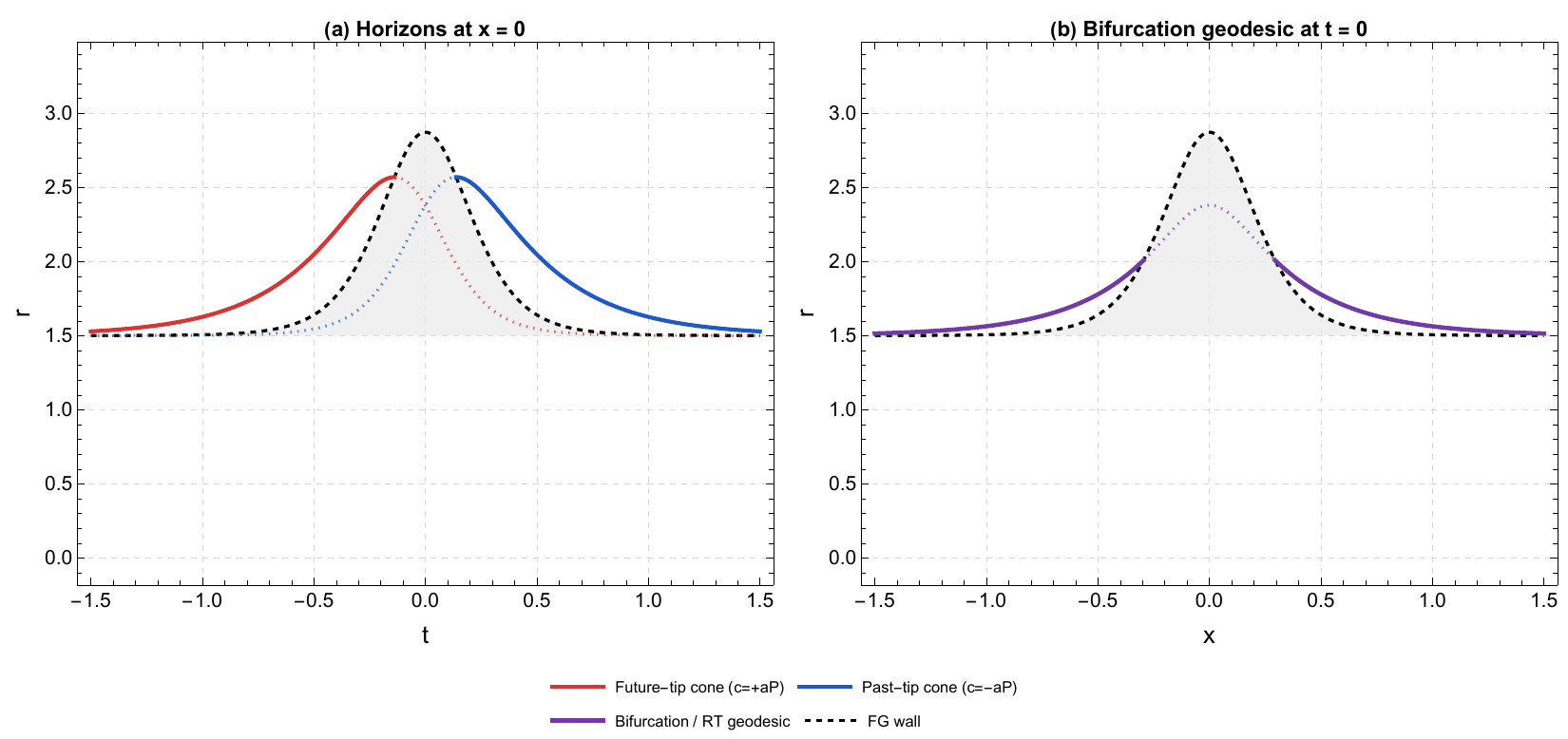}
    \caption{Causal horizons and bifurcation geodesic associated with the P
region in the heating phase, for $q=3$, $\mathcal D=1$ and
$\phi=2\pi/3$.
\textbf{(a)} Radial profiles of the future (red) and past (blue)
horizon branches at fixed $x=0$, obtained from the Roberts
transformation \cite{Roberts:2012aq}.
\textbf{(b)} Bifurcation geodesic (purple), corresponding to the
intersection of the future and past horizons, on the $t=0$
slice. In the smooth AdS$_3$ covering geometry, this is the
RT geodesic associated with the P interval.
In both panels, the black dashed curve denotes the
Fefferman--Graham (FG) coordinate wall,
$r_{\mathrm{FG}}^4=\mathcal L_+\mathcal L_-$, and the shaded
region lies below this wall.
Solid colored curves represent the portions that are in the outer-FG-chart and in the causal domain. The dotted curves indicate the remaining algebraic extensions of the horizon equations.
The passage through the FG wall reflects a breakdown of the coordinate chart.}
    \label{fig:heating_horizon}
\end{figure}
To express these horizons in the Hamiltonian-adapted coordinates, we use the Roberts transformation. The detailed derivation is presented in Appendix~ \ref{heating horizon derivation}. For the P/S regions, the resulting horizons take the common form 
\begin{equation}
   \left[r^2-Q_{a,+}(x^+)Q_{a,+}(x^-)\right]
\left[r^2-Q_{a,-}(x^+)Q_{a,-}(x^-)\right]=0,
\qquad a=P,S, 
\end{equation}
where $x^\pm=t\pm x$ and $Q_{\mathrm{P/S},m}(z)=\frac{\sqrt{\mathcal{D}}}{2(\cosh{kz}\pm \cos{\phi})}\left[q\sinh{kz}\pm \sin{\phi}\cot{\left(\frac{ \Theta_{h,\mathrm{P/S}}-m\delta_{\mathrm{P/S}}}{2}\right)}\right]$ with $\delta_{\mathrm{P}}=\frac{\phi}{q},\ \delta_{\mathrm{S}}=\frac{\pi-\phi}{q}$ and $m=\pm$.
For $q=1$, both branches reduce to $r=\frac{\sqrt{\mathcal{D}}}{2}$. For $q>1$, the horizon surfaces have non-trivial dependence on Hamiltonian-adapted coordinates, which is illustrated in Fig.[\ref{fig:heating_horizon}].

\subsubsection{Global embedding}\label{globalembed}
%\textcolor{red}{Needs to add bridging sentences}\\
Until now, the holographic description was restricted to individual patches corresponding to the P and S arcs. Each arc has a different Hamiltonian-adapted coordinate chart.
However, we can embed all patches in global $\mathrm{AdS}_3$ to understand how these regions are related. 

\begin{figure}[H]
    \centering
    \includegraphics[width=0.65\textwidth]{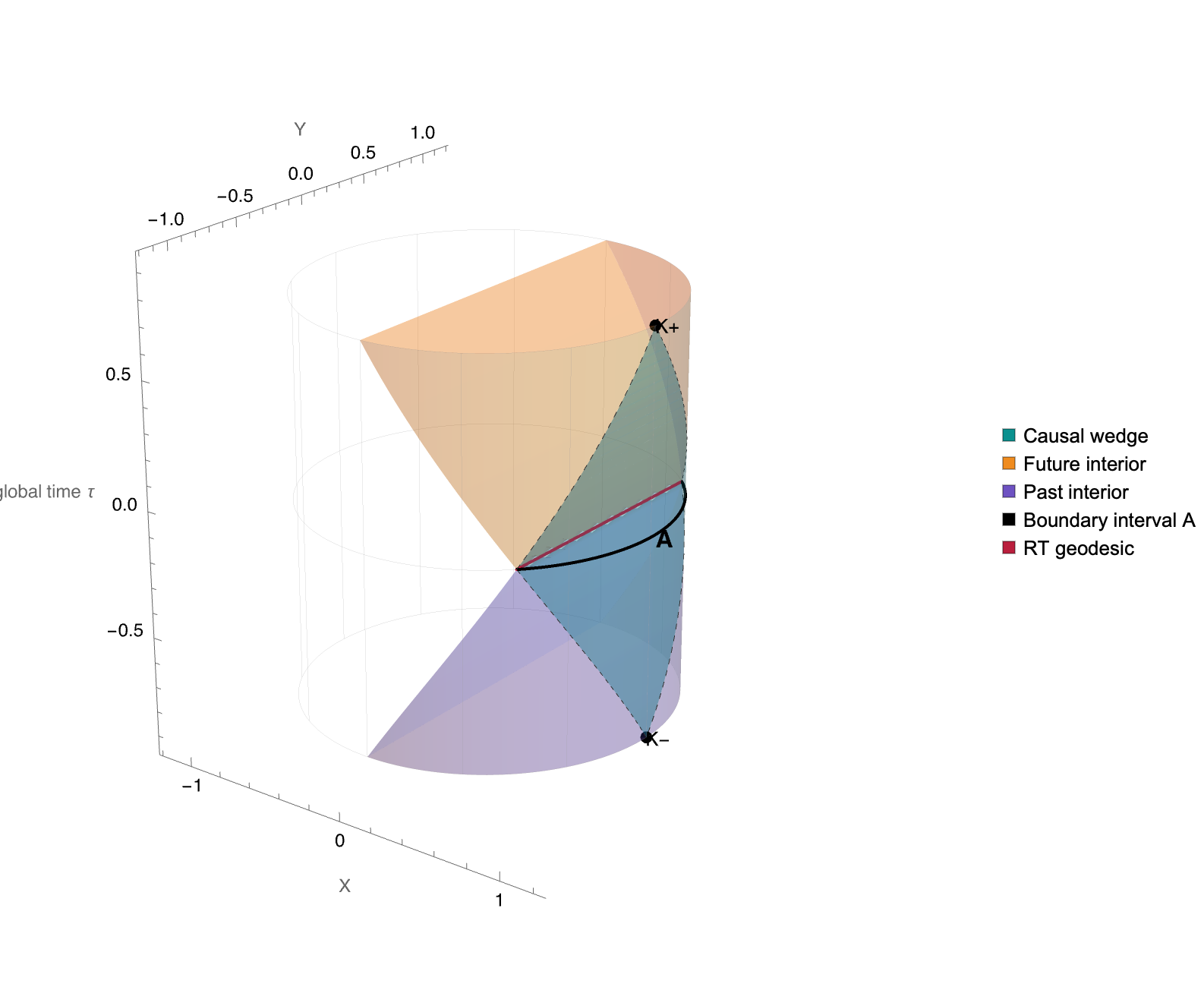}
    \caption{Causal wedge, future interior and past interior corresponding to a boundary subregion.}
    \label{causal3d}
\end{figure}

We can study the causal wedges associated with all the $2q$ boundary arcs in a common geometry.
We can then identify the past and future causal horizons corresponding to each arc. We can also distinguish portions of the bulk as exteriors, future and past interiors, and causally disconnected regions relative to each arc. We will use the embedding coordinates to discuss the global geometry.

As we know, global $\mathrm{AdS}_3$ is a hyperboloid 
\[-Y_0^2-Y_1^2+Y_2^2+Y_3^2=-1\] in $\mathbb{R}^{2,2}$ with the metric $\mathrm{ds}^2=-dY_0^2-dY_1^2+dY_2^2+dY_3^2$. After substituting 
 \begin{align*}
   Y_0 &=\cosh{\rho}\ \cos{\tau} \\
   Y_1 &=\cosh{\rho}\ \sin{\tau} \\
   Y_2 &=\sinh{\rho}\ \cos{\Theta} \\
   Y_3 &=\sinh{\rho}\ \sin{\Theta} 
 \end{align*}
we get the global $\mathrm{AdS}_3$ metric, $\mathrm{ds}^2=-\cosh^2{\rho}\ d\tau^2+d\rho^2+\sinh^2{\rho}\ d\Theta^2$. 

The conformal boundary is at $\rho \to \infty$. A boundary point is described in the embedding space as $P(\tau,\ \Theta)=(\cos{\tau},\ \sin{\tau},\ \cos{\Theta},\ \sin{\Theta})$.For a boundary point $K$, a null hypersurface in the bulk is defined as \[\mathcal{H}_K=\{Y:\ Y \cdot K=0\}\]. We can use this to find the causal horizons of a boundary interval. 

% in the embedding space, we need to find the tips of the causal diamond corresponding to the boundary interval. 

If the upper and lower tips of the causal diamond are $K_\pm$, two bulk null hypersurfaces are $\mathcal{H}^\pm=\{\ Y:\ Y \cdot K_\pm=0\}$. Appropriate portions of these hypersurfaces are the future and past horizons of the interval's causal wedge. In smooth $\mathrm{AdS}_3$, the horizons intersect along the extremal geodesic between the endpoints of the interval \cite{Hubeny_2012}. 

For $-a<\tau<a$, any bulk point $Y$ is in
\begin{align*}
    \mathrm{causal\ wedge\ if}\qquad Y\cdot K_+>0,\ Y \cdot K_->0,\\
    \mathrm{future\ interior\ if}\qquad Y\cdot K_+<0,\ Y \cdot K_->0,\\
    \mathrm{past\ interior\ if}\qquad Y\cdot K_+>0,\ Y \cdot K_-<0,\\
    \mathrm{causally\ disconnected\ if}\qquad Y\cdot K_+<0,\ Y \cdot K_-<0.
\end{align*}
We will do this construction for each of the P and S arcs. This gives a common description of their causal wedges and their relative positions in the bulk. In section \ref{Sec:ExactEE} we use this global picture to compute entanglement entropy when the endpoints of a subregion lie on different arcs of the boundary.
\subsection{Bulk geometry for the non-heating phase}

In the non-heating phase, the Schwarzian factor of Ba\~{n}ados metric can be written as,\\
% $\zeta^{\pm}=F (x_{\pm})= e^{-i\Theta^{\pm}}$\\
% where $\Theta^{\pm}=$
\begin{equation}\label{banadosfactorNhp}
    \mathcal{L}_{\pm}=-\frac{\mathcal{\varepsilon}}{4}\left(q^2- \frac{(q^2-1)\sinh^2\mu}{(\cos (q\sqrt{\mathcal{\varepsilon}}x^{\pm})-\cosh \mu)^2}\right)
\end{equation}

$\mathcal{L}_{\pm}(x^{\pm})$ is periodic with period $T=\frac{2\pi}{q\sqrt{\epsilon}}$. In such periodic cases, one can construct the monodromy matrix ($M$) to check the horizon structure\cite{Martinec:1998wm, Sheikh_Jabbari_2016}. If $\tr{M}_{\pm}>2$, the monodromy matrix corresponds to {\it hyperbolic} {\it class} and in this case the geometry may admit an event horizon depending on its global structure.
When $\tr{M}_{\pm}<2$, the monodromy matrix corresponds to the so called {\it elliptic} {\it class}. In this case, the metric is horizonless if both chiral and anti-chiral sectors belong to the elliptic class \cite{Sheikh_Jabbari_2016}. 
%Finally, if $\tr{M}_{\pm}=2$, the matrix falls into the {\it parabolic} {\it class}. In the last case, it indicates the horizonless geometry trivially if $M$ is identity matrix.\\
Now we need the solutions of the Hill equation to construct the Monodromy matrix.
The Hill equation can be written as \cite{Hochstadt_1967},
\begin{eqnarray}
    \psi''-\mathcal{L}_{\pm}\psi= 0
\end{eqnarray}
where $\mathcal{L}_{\pm}$ is the Schwarzian factor.
%As $\mathcal{L}_{\pm}$ is by construction a Schwarzian, it has a closed-form solution. 
One of the solutions can be explicitly written as $\psi_{1}=\frac{1}{\sqrt{F'}}$ and $\psi_{2}=\frac{F}{\sqrt{F'}}$. Where $F$ is the boundary conformal transformations and in non heating phase $F$ is given by 
%F_{\pm}=\tan{\frac{\Theta^{\pm}_{\textrm{nh}}}{2}}$ where
%\begin{equation}
%\Theta^{\pm}_{\textrm{nh}}=\frac{ 2\arctan\Bigl(\coth\tfrac{\phi}{2}\,\tan\tfrac{k x}{2}\Bigr)}{q}+\frac{2\pi n^{\pm}}{q}
%\label{eq:Theta2nh}
%\end{equation}
%\begin{equation}
%\Theta^{\pm}_{\textrm{nh}}(x^{\pm}+T)= \Theta^{\pm}_{\textrm{nh}}(x^{\pm})+\frac{2\pi}{q} \label{eq:Theta2nh2} \end{equation}
$F_{\pm}=\tan{\frac{\Theta^{\pm}_{nh}(x^{\pm})}{2}} $  where $\Theta^{\pm}_{nh}(x^{\pm})$ is written in \ref{eq:Theta2}.
%\begin{equation}
%\Theta^{\pm}_{\textrm{nh}}(x^{\pm})=\frac{ 2\arctan\Bigl(\coth\tfrac{\phi}{2}\,\tan\tfrac{k x}{2}\Bigr)}{q}+\frac{2\pi n^{\pm}}{q}
%\label{eq:Theta2nh}
%\end{equation}
$F$ is not periodic under $T$, rather it mapped to the next sheet as $\Theta^{\pm}_{\textrm{nh}}(x^{\pm}+T)= \Theta^{\pm}_{\textrm{nh}}(x^{\pm})+\frac{2\pi}{q} \label{eq:Theta2nh2}$.
%Now $F$ has a periodicity of period $\frac{2\pi}{q\sqrt{\epsilon}}$. So the solutions of the Hill equation have the same periodicity $\frac{2\pi}{q\sqrt{\epsilon}}$ i.e, $ \psi_{1/2}(x)= \psi_{1/2}(x+\frac{2\pi}{q\sqrt{\epsilon}})$\\
%$\psi_{1}=\frac{\cos}{\sqrt{\theta'}}$
The solutions can be explicitly written as 
\begin{equation}
\psi_{1}=\frac{\cos\left({\Theta^{\pm}_{\textrm{nh}}}/2\right)}{\sqrt{\left(\Theta '^{\pm}_{\textrm{nh}}/2\right)}}, \quad \psi_{1}=\frac{\sin\left({\Theta^{\pm}_{\textrm{nh}}}/2\right)}{\sqrt{\left(\Theta '^{\pm}_{\textrm{nh}}/2\right)}}
\end{equation}

Now the solutions at $x^{\pm}+T$ can be written as
\begin{eqnarray}
    \psi_{1}(x^{\pm}+T)=%\frac{\cos{(\frac{\Theta^{\pm}_{\textrm{nh}}}{2}+\frac{\pi}{q})}}{\sqrt{(\Theta '^{\pm}_{\textrm{nh}}/2)}}
     \psi_{1}(x^{\pm})\cos\left(\frac{\pi}{q}\right)-\psi_{2}(x^{\pm})\sin\left(\frac{\pi}{q}\right)\\
    \psi_{2}(x^{\pm}+T)=%\frac{\sin{(\frac{\Theta^{\pm}_{\textrm{nh}}}{2}+\frac{\pi}{q})}}{\sqrt{(\Theta '^{\pm}_{\textrm{nh}}/2)}} = 
    \psi_{1}(x^{\pm})\sin\left(\frac{\pi}{q}\right)+\psi_{2}(x^{\pm})\cos\left(\frac{\pi}{q}\right)
\end{eqnarray}
Same equations hold for the derivatives of the solutions, $\psi'_{1}$ and $\psi'_{2}$. Now we want to write these solutions and the derivative of the solutions at $x^{\pm}=0$ and it can be written in the following matrix form.
\begin{eqnarray}
{\begin{pmatrix}
\psi_{1}(T) & \psi_{2}(T) \\
\psi'_{1}(T) & \psi'_{2}(T)
\end{pmatrix}} =
  \begin{pmatrix}
\psi_{1}(0) & \psi_{2}(0) \\
\psi'_{1}(0) & \psi'_{2}(0)
\end{pmatrix} 
{\begin{pmatrix}
 \cos\left(\frac{\pi}{q}\right)& \sin\left(\frac{\pi}{q}\right) \\
-\sin\left(\frac{\pi}{q}\right) & \cos\left(\frac{\pi}{q}\right)
\end{pmatrix}}  
\end{eqnarray} 
So the monodromy matrix in the non heating phase is given by,
\begin{eqnarray}
  M_{\pm}= {\begin{pmatrix}
 \cos\left(\frac{\pi}{q}\right)& \sin\left(\frac{\pi}{q}\right) \\
-\sin\left(\frac{\pi}{q}\right) & \cos(\frac{\pi}{q})
\end{pmatrix}}  
\end{eqnarray} 
%where $T$ is the periodicity of $\mathcal{L}_{\pm}$.\\
Thus, $\tr(M_{\pm})=2\cos(\frac{\pi}{q})$. 
%For q=1 \quad $|\tr(M)|=2$ and |M| is the identity matrix. So it belongs to trivial parabolic monodromy. But 
Hence for $q\ge2$, we have $\tr(M_{\pm})<2$. Thus this belongs to elliptic class, which implies that the bulk metric corresponding to non-heating phase has no horizon.

\subsection{Bulk geometry  for the phase boundary}\label{Sec:PBbulk}

 On the physical slice, the Bañados factor is
\begin{equation}\label{banadosfactorpb}
    \mathcal{L}_{\pm}(x^{\pm})=\frac{(q^2-1)\alpha^2}{(1+q^2\alpha^2(x^{\pm})^2)^2}
\end{equation}
which falls off as $\mathcal L_\pm\simeq\frac{q^2-1}{q^4\alpha^2t^4}$, at late time ($t\to\infty$) and fixed $x$, and the metric becomes AdS Poincaré.
From \eqref{banadosfactorhp} we see that this is a limiting case of the heating phase: as $\varphi\to\pi$, and therefore $\mathcal D\to0$, the heating phase Bañados factors (for $\alpha>0$) on the P slice ($\mathcal{L}^P_{\pm}$) becomes \eqref{banadosfactorpb} whereas vanishes for S slice $\mathcal{L}^S_{\pm}\rightarrow0$\footnote{Both terms in $\cosh kx+\cos\phi$ vanish at the same rate: $\cosh kx+\cos\phi\simeq\frac{\mathcal D}{2\alpha^2}(1+q^2\alpha^2x^2)$, while $\sin^2\phi\simeq\mathcal D/\alpha^2$ and the factors of $\mathcal D$ cancel.}.
Correspondingly, from the boundary map itself \eqref{eq:HTheta} we see that in the phase boundary, when viewed as a limiting case of the heating phase, the P-arcs grow to fill the sectors between consecutive degenerate fixed points whereas the length of the S-arcs shrink\footnote{Since from \eqref{eq:HTheta}, for $\eta=-1$, $\psi_{\rm S}=\pi-2\arctan\big(\cot\frac\phi2\tanh\frac{kx}{2}\big)$, as $\phi\to\pi$, $\cot\frac\phi2\to0$, so the entire S-arc reduces to a point $\psi_{\rm S}=\pi$.}. %We comment on the late-time entanglement entropy in the quotient picture in Section~\ref{sec:pbcomments}.
Further comments on the late time EE in the phase boundary in the quotient picture are given in Section \ref{pbnhpcommentSec}.

\section{Entanglement Entropy from the bulk  description}
\label{twopicture}
In this section we derive the entanglement entropy from the bulk geometry using two descriptions. In Section~ \ref{Sec:ExactEE} we use the smooth global $\mathrm{AdS}_3$ description and compute the EE from the length of bulk geodesics. In Section~ \ref{quotientdesciption} we present an alternative, quotient description of the bulk geometry. In this description, $q$ identical sectors are identified, and their effect is carried by a conical defect. Later, we use it in Section~ \ref{shockwavedesc} to understand
%give a geometric understanding of 
the late-time growth of EE. Throughout the sections, we mainly focus on the heating phase while making some comments on the other two phases. Also for simplicity we set $\phi=\frac{\pi}{2}$, i.e. $\alpha=0$.

\subsection{EE in the global description}\label{Sec:ExactEE}
%While understanding the global geometry as a whole, that is the geometries corresponding to the $2q$ arcs together it is best to understand it in the global coordinates. 

In this section, we compute EE using an embedding space description of the global geometry introduced in Section~ \ref{globalembed}. Using the HRT prescription \cite{Ryu_2006, Hubeny_2007}, we know that the leading-order entanglement entropy of a boundary interval is \(S=\frac{c}{6}L.\) where $L$ is the regulated geodesic length of the corresponding bulk geodesic. 

% We will use elements of Section \ref{globalembed} 
In the following, we will compute lengths of geodesics joining points from different arcs. In each case, we will be able to identify the portions of the geodesic that lie in the causal wedge of a given arc. We will also find where it crosses the causal horizons. This will help us identify which portions of the geodesics contribute to the time dependence at late times. We will discuss three cases separately and compare the results with the CFT results of Section~ \ref{CFTEE} for each case.

In the embedding coordinates, a spacelike geodesic between two boundary points $A$ and $B$ can be parametrized in terms of proper length $\lambda$ \cite{Dyer:2017zef}  as \[Y^M(\lambda)=\frac{e^{-\lambda}P_A^M+e^{\lambda}P_B^M}{N}\] where $N=\sqrt{-2 P_A \cdot P_B}$ is the normalization factor and $M=0, \ldots , 3$. The length between two such points is $L=\lambda_B-\lambda_A$. 
The geodesic crosses the null hypersurface of a boundary interval with causal tips at $K_\pm$, at $\lambda_\pm=\frac{1}{2}\log\left(-\frac{P_A \cdot K^\pm}{P_B\cdot K^\pm}\right)$.  Also, we need to regulate the geodesic lengths near the boundary. Let $\epsilon_A$ and $\epsilon_B$ be the global cutoffs at the two endpoints.
So the regulated length between  two boundary points is
\begin{equation}\label{embeddinglength}
L=\lambda_B-\lambda_A=\log \left(\frac{-2P_A \cdot P_B}{\epsilon_A\epsilon_B}\right).
\end{equation}
The global cut-off is related to the physical UV cutoff $\epsilon$ through \footnote{The global radial cutoff is
% $\epsilon_{\mathrm{global}}$ is related to the geodesic cut-off as
$\cosh{\rho_c}=\frac{1}{\epsilon_{\mathrm{global}}}$. In the Hamiltonian-adapted coordinate, at the point $(t,x)$, it is related to the UV cut-off in physical system as $\epsilon_{\mathrm{global}(t,x)=\epsilon\frac{\cosh{kx}}{\sqrt{\cosh{kx^+\cosh{kx^-}}}}}.$ At late times, with fixed $x$,  the relation becomes $\epsilon_{\mathrm{global}}\simeq 2 \epsilon\ \cosh{kx}e^{-kt}$.} \[\epsilon_{\mathrm{global}}(t,x)=\epsilon\, \frac{\cosh{kx}}{\sqrt{\cosh{kx^+\cosh{kx^-}}}}.\]

% when describing the boundary points, we need to introduce a cut-off ($\epsilon_{\mathrm{global}}$), which corresponds to $\lambda_c=\log \left(\frac{N}{\epsilon}\right)$. 

%\begin{itemize}
    %\item 
    First we consider the scenario when both the end points lie on the same arc (for example we can take it to be the $P_0$ arc). The arc is $(-a,a)$ and the tips of its causal diamond are $K_\pm=P(\pm a,0)$. We take the two boundary points at $\Theta^{\pm}_{1,2}$ with $-a<\Theta^{\pm}_{1,2}<a$. For these points, $Y(\lambda) \cdot K_{\pm}>0$, so the geodesic remains within the causal wedge of $\mathrm{P}_0$. For two boundary points \(P_A \cdot P_B=2 \sin{(\Delta\Theta^+/2)}\sin{(\Delta\Theta^-/2)}\). Then, using Eq.~\eqref{embeddinglength} and the cutoff relation, the contribution from one chiral sector is
    \begin{equation}
        S=\frac{c}{6}\log \left[\frac{2}{\epsilon}\sin{\frac{|\Theta^+_2-\Theta_1^+|}{2}\sqrt{\frac{\cosh{k(t+x_1)}\ \cosh{k(t+x_2)}}{\cosh{kx_1}\cosh{kx_2}}}}\right]
    \end{equation}
where $\Theta^\pm_i=\frac{2}{q}\arctan \tanh{\frac{k(t\pm x_i)}{2}}$ as defined in \eqref{eq:HTheta}. Similarly, there will be a contribution from the other sector as well. In the late time limit $t \gg x_i$, the time-dependent factors between the angular separation and the cutoff factors cancel exactly. Adding both chiral contributions, we get
\begin{equation}\label{EEcase1}
    S=\frac{c}{3}\log \left[\frac{2}{q\epsilon}\frac{\sinh{\frac{k(x_2-x_1)}{2}}}{\sqrt{\cosh{kx_1}\cosh{kx_2}}}\right]=\frac{c}{3}\log \left[\frac{2}{k}\frac{\sinh{\frac{k\Delta x}{2}}}{\sqrt{\epsilon_{x_1}\epsilon_{x_{2}}}}\right]
\end{equation}
where $\epsilon_{x_{i}}=\frac{q\epsilon}{k}\cosh{kx_i}$. This is exactly what we get from the CFT computation in this case, as discussed in section \ref{CFTEE}.\\

Also, we can see that with the identification $\beta_{\mathrm{eff}}=\frac{2\pi}{k}$, \eqref{EEcase1}
has the same functional form as the Calabrese-Cardy result of an interval of length $\Delta x$ on an infinite line at temperature $T=\frac{k}{2\pi}$. However, the effective cutoff $\epsilon_{x_i}$ depends on position. This is an artifact of the fact that the bulk metric for the patch approaches a static AdS Rindler geometry at late times.
%\item 
Now we consider the case when the endpoints of the subregion are at different P arcs. For concreteness, let's take the arcs to be $\mathrm{P}_0$ and $\mathrm{P}_1$ and, for simplicity, take the endpoints to be the midpoints of the arcs. For the $\mathrm{P}_0$ arc, the center is at $x=0$ in its Hamiltonian-adapted coordinates. In the global coordinate with\footnote{Here, we are using the identity that $2 \arctan(\tanh{x/2})=\arctan(\sinh{x})$.} it is located at $(\tau=\mathtt{u}(t), \Theta=0)$ where  $\mathtt{u}(t)=\frac{1}{q}\arctan \sinh{kt} $. The corresponding boundary point and the causal tips are 
\begin{equation}
 A=P(\mathtt{u}(t),0) \qquad  K^{\mathrm{P}_0}_\pm=P\left(\pm a\left(=\frac{\phi}{q}\right),0\right). \end{equation}

% $$ while the tips of the corresponding causal diamonds are at $$.
Similarly, the center of $\mathrm{P}_1$ would also be at $x=0$ in its coordinate chart, but in global coordinates $\Theta$ moves to $4a$. So, in the embedding coordinates the point is $B=P(\mathtt{u}(t),4a)$ and the causal diamond tips are at $K^{\mathrm{P}_1}_\pm=P(\pm a, 4a)$. Since, for both the end points $x=0$, the cut-off surfaces are at $\epsilon_{A/B}=\epsilon \sech{kt}$. 
Now, we determine if the geodesic crosses the horizon of the two arcs.
Solving $Y(\lambda) \cdot K^{\mathrm{P}_0/\mathrm{P}_1}_\pm=0$ and find that there are four zeroes at 
\[
\resizebox{\linewidth}{!}{$\displaystyle
\lambda^{\mathrm{P}_0}_+=-\lambda^{\mathrm{P}_1}_+=\frac12\log\left[\frac{\sin^2\frac{a-\mathtt{u}}{2}}{\sin\frac{5a-\mathtt{u}}{2}\,\sin\frac{3a+\mathtt{u}}{2}}\right],\qquad
\lambda^{\mathrm{P}_0}_-=-\lambda^{\mathrm{P}_1}_-=\frac12\log\left[\frac{\sin^2\frac{a+\mathtt{u}}{2}}{\sin\frac{5a+\mathtt{u}}{2}\,\sin\frac{3a-\mathtt{u}}{2}}\right]
$}
\]

% \[\lambda^{\mathrm{P}_0}_+=-\lambda^{\mathrm{P}_1}_+=\frac{1}{2}\log\left[\frac{\sin^2{\frac{a_P-\mathtt{u}}{2}}}{\sin{\frac{5a_P-\mathtt{u}}{2}\sin{\frac{3a_P+\mathtt{u}}{2}}}}\right],\qquad  \lambda^{\mathrm{P}_0}_-=-\lambda^{\mathrm{P}_1}_-=\frac{1}{2}\log\left[\frac{\sin^2{\frac{a_P+\mathtt{u}}{2}}}{\sin{\frac{5a_P+\mathtt{u}}{2}\sin{\frac{3a_P-\mathtt{u}}{2}}}}\right].\]
For $\mathtt{u}>0$, the crossing points satisfy 
\begin{equation}
\lambda_+< \lambda_-<-\lambda_-<-\lambda_+   
\end{equation}

These points divide the whole geodesic into five portions and compute the corresponding lengths. For each of the pieces, the lengths are given in Table~\ref{tab:ex2pieces}.
\begin{table}[h]
\centering
\small
\renewcommand{\arraystretch}{2.4}
\resizebox{\linewidth}{!}{
\begin{tabular}{@{}c c l l@{}}
\toprule
\ & $\lambda$ range & Region & Length \\
\midrule
1. & $[\lambda_A,\ \lambda_+]$ & $\mathrm{P}_0$'s exterior &
$\ell_{\rm ext}=\dfrac12\log\dfrac{4\sin^2\frac{\delta}{2}\,\sin^2 2a\,\cosh^2 kt}{\sin\frac{5a-\mathtt{u}}{2}\,\sin\frac{3a+\mathtt{u}}{2}\;\epsilon^2}$ \\
2. & $[\lambda_+,\ \lambda_-]$ & $\mathrm{P}_0$'s future interior &
$\ell_{\rm int}=\dfrac12\log\dfrac{\sin^2\frac{a+\mathtt{u}}{2}\,\sin\frac{5a-\mathtt{u}}{2}\,\sin\frac{3a+\mathtt{u}}{2}}{\sin^2\frac{\delta}{2}\,\sin\frac{5a+\mathtt{u}}{2}\,\sin\frac{3a-\mathtt{u}}{2}}$ \\
3. & $[\lambda_-,\ -\lambda_-]$ & outside both the wedges and interiors &
$\ell_{\rm mid}=\log\dfrac{\sin\frac{5a+\mathtt{u}}{2}\,\sin\frac{3a-\mathtt{u}}{2}}{\sin^2\frac{a+\mathtt{u}}{2}}$ \\
4. & $[-\lambda_-,\ -\lambda_+]$ & $\mathrm{P}_1$'s future interior & $\mathrm{same\ as\ \ell_{\mathrm{int}}}$ \\
5. & $[-\lambda_+,\ \lambda_B]$ & $\mathrm{P}_1$'s exterior & $\mathrm{same\ as\ \ell_{\mathrm{ext}}}$\\ 
\bottomrule
\end{tabular}}
\caption{The five pieces of the geodesic from the centre of $\mathrm{P}_0$ to the centre of $\mathrm{P}_1$, with $\delta=a-\mathtt{u}(t)$. The total length is $L=2\ell_{\rm ext}+2\ell_{\rm int}+\ell_{\rm mid}$.}
\label{tab:ex2pieces}
\end{table}
At late times, only the two $\ell_{\mathrm{int}}$ grow with time, and all the other pieces saturate to constant values, and the total length becomes
\(L=2kt+2\log\frac{\sin{\pi/q}}{\epsilon}\) and the leading growth of EE is therefore \[S \simeq \frac{ck}{3}t\]
This also agrees with the CFT computation.\\
Finally, we consider the third scenario, where one endpoint (A) is on the P arc, and the other (B) is on the S arc. For concreteness, we take one endpoint on $\mathrm{P}_0$ and the other on $\mathrm{S}_0$, with both endpoints at the center of their arcs. 

% As we discussed, $\Theta_{\mathrm{S}}^\pm=\pm 2a_P-\Theta_{\mathrm{S}}^\pm$. 
As we discussed, the S arc has the opposite orientation to P in the global description.
So B moves down, towards the bottom tip of its causal diamond. At time $t$ the center of $\mathrm{S}_0$ is at $(\tau=-\mathtt{u}(t), \Theta=2a)$. And, the causal tips are at $K^{\mathrm{S}}_\pm=P(\pm a_, 2a)$. Again, following the same method as above, we can divide the geodesic into five segments using the four crossing points. The lengths of each portion are listed in Table~ \ref{tab:ex3pieces}
\begin{table}[h]
\centering
\small
\renewcommand{\arraystretch}{2.4}
\resizebox{\linewidth}{!}{
\begin{tabular}{@{}c c l l@{}}
\toprule
\ & $\lambda$ range & Region & Length \\
\midrule
1. & $[\lambda_A,\ \lambda_+]$ & $\mathrm{P}_0$'s exterior &
$\ell_{\rm ext}=\dfrac12\log\dfrac{4\sin\frac{\delta}{2}\,\sin \delta\,\sin{(a+\mathtt{u})}\cosh^2 kt}{\sin\frac{3a+\mathtt{u}}{2}\,\epsilon^2}$ \\
2. & $[\lambda_+,\ \lambda_-]$ & $\mathrm{P}_0$'s future interior &
$\ell_{\rm int}=\dfrac12\log\dfrac{\sin\frac{a+\mathtt{u}}{2}\,\sin\frac{3a+\mathtt{u}}{2}}{\sin\frac{a-\mathtt{u}}{2}\,\sin\frac{3a-\mathtt{u}}{2}}$ \\
3. & $[\lambda_-,\ -\lambda_-]$ & outside both the wedges and interiors &
$\ell_{\rm mid}=\log\dfrac{\sin\frac{3a-\mathtt{u}}{2}}{\sin\frac{a+\mathtt{u}}{2}}$ \\
4. & $[-\lambda_-,\ -\lambda_+]$ & $\mathrm{S}_0$'s past interior & $\mathrm{same\ as\ \ell_{\mathrm{int}}}$ \\
5. & $[-\lambda_+,\ \lambda_B]$ & $\mathrm{S}_0$'s exterior & $\mathrm{same\ as\ \ell_{\mathrm{ext}}}$\\ 
\bottomrule
\end{tabular}}
\caption{The five pieces of the geodesic from the centre of $\mathrm{P}_0$ to the centre of $\mathrm{S}_0$, with $\delta=a-\mathtt{u}(t)$. The total length is $L=2\ell_{\rm ext}+2\ell_{\rm int}+\ell_{\rm mid}$.}
\label{tab:ex3pieces}
\end{table}
At late times, again only the interior pieces grow in time, giving $L=kt+\log \frac{2\sin{(\pi/q)}}{q\epsilon^2}$. As a result, the leading EE growth is
\[S\simeq\frac{c}{6}kt.\]
This result also agrees with the CFT computation.

In \cite{Hartman_2013}, the authors identified growing extremal surface length in a genuine black hole interior. However, in this case, although there are no black holes in the bulk geometry, the causal horizons of each arc give us a natural way to divide the bulk geodesic into different sections. When the endpoints of the interval lie on different arcs, the geodesic crosses these horizons and goes beyond the causal wedge. Interestingly, we find that only these portions contribute to the late-time growth of EE, while the other portions approach constant values. This gives a simple geometric way to understand the different EE growth rates for intervals in the same arc and on different arcs.

\subsection{An alternative approach: the quotient picture}\label{quotientdesciption}
In the previous section, we computed the bulk entanglement entropy from the embedding of the various patches inside global AdS$_3$.
%In the global AdS$_3$, each of the $q$ P-arcs and the $q$ S-arcs, has their causal wedge and horizon. The RT geodesic between two end points of an interval can be split into different pieces according to the different wedges and the horizon interior it crosses and that its late time growth comes entirely from the pieces lying behind the horizons (Table~\ref{tab:ex2pieces}).

In this section we present an alternative "quotient" description, in which we restrict to a single $q$ P-arc and $q$ S-arc. 
%are identified to one $P$-arc and one $S$-arc. 
The dual bulk picture has a single pair of P and S metrics, along with a conical defect, which is sourced by two heavy operators at the branch points, and the effect of the remaining $q-1$ sectors is carried by the defect. This description gives a simple geometric picture of the late-time growth of the entanglement entropy.  This single-patch or quotient description can be best described from the boundary. 

To understand why this description works, note that the deformed Hamiltonian acts as a Möbius transformation on $\tilde F=\zeta^q$. So instead of the CFT on the $\zeta$-plane, which covers the $\tilde F$-plane $q$ times, one can equivalently consider $q$ copies of the CFT on a single $\tilde F$-plane, on which the time evolution acts simply as a Möbius transformation. Then the price that one pays for this is the addition of twist operators inserted at the branch points $\tilde F=0$ and $\tilde F=\infty$ that keep track of these $q-$copies. These are heavy operators with weight $h_{\sigma_q}=\bar h_{\sigma_q}=\frac{c}{24}\big(q- \frac1{q}\big)$ or $h_q=\bar h_q=\frac{c}{24}\big(1-\frac1{q^2}\big)$ per copy that sources the conical defect with deficit angle $\delta_q=2\pi\big(1-\frac1q\big)$ \cite{Benjamin:2020mfz} in the bulk.
We discuss this in more detail in the Appendix~\ref{appendixPIComparison}, where we explain how this description can be understood in the boundary path integral formalism of Calabrese-Cardy \cite{Calabrese:2004eu}.

Up to the backreaction of the defect, the bulk is then the Bañados geometry built on $\tilde F$ rather than on $F$. Note that the Floquet time evolution is an isometry of this Bañados geometry, whereas the defect does not remain invariant under it. So in the quotient description, the time dependence is carried only through the defect. We will make this point more precise later in Section~\ref{sec:timeboost}. In the heating phase, in this quotient description, the geometry is AdS–Rindler with a conical defect along the defect worldline. Geodesics of intervals whose endpoints lie on either side of a fixed point cannot avoid the effects of the defect, and this is where the late-time growth of the entanglement entropy comes from.

In what follows we first show in Section~\ref{Sec:EffectiveShockwave} that the boosted defect acts effectively as a shockwave along the Rindler horizon at late times, which reproduces the growth rate of the entanglement entropy. In Section~\ref{comparison}, we compare this effective description with the exact result of Section~\ref{Sec:ExactEE}. Finally, in Section~\ref{pbnhpcommentSec}, we show that on the phase boundary the defect is only translated, which leads to a logarithmic growth instead.

%These heavy operators above, of weight $h_q=\bar h_q=\frac{c}{24}\big(1-\frac1{q^2}\big)$.
%(see Appendix~\ref{appendixPIComparison}).

%In the single-sheet description, the bulk is just one sheet, with one $P$ and one $S$ metric, together with a conical defect along the same geodesic. The effect of the remaining $q-1$ sheets is now carried entirely by the defect, which is sourced by two heavy operators at the branch points. the heavy operators have weight $h_q=\bar h_q=\frac{c}{24}\big(1-\frac{1}{q^2}\big)$, and the defect has deficit angle $\delta_q=2\pi\big(1-\frac1q\big)$.
%\textcolor{blue}{Up to the backreaction of the defect, the $P$ and $S$ metrics are the right and left wedges of a single AdS–Rindler geometry, and the Floquet evolution acts on them as a Rindler boost.} The shockwave analysis below uses this description, with the defect treated as a source on the Rindler background.
\subsection{Late time limit: effective AdS Rindler shockwave geometry}\label{shockwavedesc}

In this section, we present a geometric understanding of the late-time growth in entanglement entropy from the quotient description of Section~\ref{quotientdesciption}. For simplicity, let us consider $\alpha=0$, i.e.,  $\phi=\pi/2$, so that $\sigma^*_1=\pi/(2k)$ and $\sigma^*_2=3\pi/(2k)$, where, $k=q\sqrt{\mathcal{D}}$. At late time and for fixed $x$,  the $P$ and $S$ Bañados metrics both approach the same AdS Rindler geometry,
\begin{equation}\label{latetimeRindler}
ds^2
=
-(r^2-k^2)dt^2
+
\frac{dr^2}{r^2-k^2}
+
r^2dx^2
\end{equation}
since the second term of
$$\mathcal L_\pm=-\frac12\{F,x^\pm\}=\frac{k^2}{4}+\frac{k^2}{4}\Big(1-\frac{1}{q^2}\Big)\mathrm{sech}^2(kx^\pm),\qquad x^\pm=t\pm x $$
decays and $\mathcal L_\pm\to k^2/4$, where, $r$ is related to $y$ used earlier in \ref{banadoslt} as $r=1/y+k^2y/4$.

Now, in the quotient picture, two heavy operators source a conical defect in the bulk, whose locus is the geodesic (the branch locus of Section~\ref{holodualSec}).
%We want to show that this defect when viewed from the late time slice (where $P$ and $S$ are the the the metric \eqref{latetimeRindler}) is effectively described by a shockwave, to linear order in the deficit angle. The entanglement entropy computation in then becomes a geodesic length computation on this shockwave geometry. 
We want to show that, at late times, when both the $P$ and $S$ metrics approach \eqref{latetimeRindler}, this defect is effectively described by a shockwave: its stress tensor is dominated by a single null component, which backreacts on the AdS–Rindler background as a shock on the horizon. The entanglement entropy computation then becomes a geodesic length computation in this shockwave geometry. This effective description reproduces the growth rate of the entanglement entropy exactly, while its time-independent part differs from the exact result, as we show in Section~\ref{comparison}.
In what follows, we first locate the defect in Kruskal coordinates, then show that in the Rindler frame it appears increasingly boosted toward the horizon, compute the resulting shock, and compute the geodesic crossing it. Finally, we compare this effective description with the exact result.
%%In what follows, we first locate the defect in Kruskal coordinates, then show that in the Rindler frame it appears increasingly boosted toward the horizon, compute the resulting shock, and finally compute the geodesic crossing it. 
%In Section~7.5 we compare with the exact result of Section~\ref{…}: the two agree to first order in the deficit angle, while at integer $q$ the nonlinear corrections, encoded in the holonomy of the defect, modify only the time-independent terms.

We begin by writing the late time $P$ or $S$ metric in Kruskal coordinates. To this end, we introduce the tortoise coordinate
\begin{equation}
dr_*=\frac{dr}{r^2-k^2},
\end{equation}
therefore, $r_*=\frac{1}{2k}\log\frac{r-k}{r+k}$ is chosen, such that, $r_*\to0$ at $r\to\infty$.
The Kruskal coordinates may be chosen as
\begin{equation}
u=-e^{-k(t-r_*)},
\qquad
v=e^{k(t+r_*)}.
\end{equation}
we write, 

\begin{equation}
ds^2
=
-\frac{4\,du\,dv}{(1+uv)^2}
+k^2
\left(
\frac{1-uv}{1+uv}
\right)^2
dx^2.
\label{eq:localized-shock-full}
\end{equation}

The branch points, on the other hand, are mapped to $\pm i$ under the Cayley map. We can see this by writing the Cayley coordinate \eqref{eq:cayley}:
\begin{equation}
    \chi
    =
    i\frac{1-F}{1+F}.
    \label{eq:cayley_chi}
\end{equation}
The zero and pole of $F$ are mapped to
\begin{equation}
    F=0
    \quad\Longrightarrow\quad
    \chi=+i,
    \qquad
    F=\infty
    \quad\Longrightarrow\quad
    \chi=-i.
\end{equation}
Now it is useful to introduce another Cayley coordinate, built from $\widetilde F$ instead of $F$,\footnote{Since $F=\widetilde F^{1/q}$, the chain rule for the Schwarzian gives
\begin{equation*}
\{F,x\}=\{\widetilde F,x\}+\frac{1-q^{-2}}{2}\left(\frac{\widetilde F'}{\widetilde F}\right)^2 .
\end{equation*}
On the $P$-line, $\widetilde F=e^{i\psi}$ with $\psi'=k\,\mathrm{sech}(kx)$, and $\{\widetilde F,x\}=-k^2/2$. The Ba\~nados functions therefore split as
\begin{equation*}
\mathcal L^{(F)}_\pm=\mathcal L^{(\widetilde F)}_\pm+\frac{k^2}{4}\Big(1-\frac{1}{q^2}\Big)\mathrm{sech}^2(kx^\pm),
\qquad
\mathcal L^{(\widetilde F)}_\pm=\frac{k^2}{4}.
\end{equation*}
%The first term $\mathcal L^{(\widetilde F)}_\pm$ is the 
The second term is the stress tensor of the heavy operators at the branch points, and it vanishes at late times for fixed $x$. The late-time geometry is therefore the Ba\~nados geometry generated by the first term $\widetilde F$, i.e.\  AdS--Rindler, and its Poincar\'e patch is the one built on the Cayley transform of $\widetilde F$. In that patch the Floquet evolution is a M\"obius map, i.e.\ a Rindler boost. In the patch built on $\chi$, by contrast, the evolution acts as $F\to\big(\mathcal M(F^q)\big)^{1/q}$ with $\mathcal M$ M\"obius, which is not an isometry for $q>1$,
%so no Rindler or Kruskal chart exists there and 
there the Floquet time is not a Killing time (not a Killing flow) of the exact metric.}
\begin{equation}
    \widetilde\chi
    =
    i\frac{1-\widetilde F}{1+\widetilde F}
    =
    \tan\!\big(q\arctan\chi\big).
    \label{eq:cayley_chitilde}
\end{equation}
Since $F$ and $\widetilde F=F^q$ have the same zero and pole, the branch points are again mapped to $\pm i$,
\begin{equation}
    F=0
    \quad\Longrightarrow\quad
    \widetilde\chi=+i,
    \qquad
    F=\infty
    \quad\Longrightarrow\quad
    \widetilde\chi=-i.
\end{equation}
Writing the Euclidean boundary coordinate as
\begin{equation}
    \widetilde\chi=\widetilde X+i\widetilde\tau_E,
\end{equation}
the two branch points are located at
\begin{equation}
    (\widetilde X,\widetilde\tau_E)=(0,+1),
    \qquad
    (\widetilde X,\widetilde\tau_E)=(0,-1).
    \label{eq:branch_points_cayley}
\end{equation}
The corresponding Euclidean Poincar\'e metric is (with $\ell=1$)
\begin{equation}
    ds^2
    =
    \frac{d\widetilde Z^2+d\widetilde X^2+d\widetilde\tau_E^2}{\widetilde Z^2},
    \label{eq:Euclidean_Poincare_defect}
\end{equation} 
 The map between the Kruskal coordinates of the late-time geometry and this Poincar\'e patch is
\begin{align}
    \widetilde Z
    &=
    \frac{1+uv}{(1-uv)\cosh \xi+v-u},
    \nonumber\\
    \widetilde{\mathcal T}
    &=
    \frac{u+v}{(1-uv)\cosh \xi+v-u},
    \nonumber\\
    \widetilde X
    &=
    \frac{(1-uv)\sinh \xi}{(1-uv)\cosh \xi +v-u}.
    \label{eq:Poincare_Kruskal_map}
\end{align}

where, $\xi\equiv kx$ and  $\widetilde\tau_E$ is analytically continued to $\widetilde\tau_E=-i\widetilde{\mathcal T}$.
At the asymptotic boundary $uv=-1$ it gives
\begin{equation}\label{tildechi}
   \widetilde\chi= \widetilde X\pm\widetilde{\mathcal T}=\tanh\frac{k(x\pm t)}{2},
\end{equation}
on the $P$-line,\footnote{Analytically continued $\widetilde\chi$ is the light-cone coordinates $\widetilde X\pm\widetilde{\mathcal T}$ on the Lorentzian Poincare metric. On the $P$- and $S$-lines these are
\begin{equation*}
P:\ \ \widetilde X\pm\widetilde{\mathcal T}=\tanh\frac{k(x\pm t)}{2},
\qquad
S:\ \ \widetilde X\pm\widetilde{\mathcal T}=\coth\frac{k(x\pm t)}{2}.
\end{equation*}

At the asymptotic boundary $uv=-1$ (where $r_*=0$), substituting $(u,v)=(-e^{-kt},\,e^{kt})$ (right exterior) and $(u,v)=(e^{-kt},\,-e^{kt})$ (left exterior) into \eqref{eq:Poincare_Kruskal_map} reproduces exactly these two expressions. Thus the $P$-line ($|\widetilde X|<1$) is the right Rindler wedge, the $S$-line ($|\widetilde X|>1$) is the left Rindler wedge.
}
The defect locus in the Poincare coordinate is the bulk geodesic joining the two branch points,\footnote{In the global description, the $\mathbb Z_q$ generator $F\to e^{2\pi i/q}F$ acts on $\chi$ as an elliptic Möbius map with fixed points $\chi=\pm i$. It extends to a bulk isometry whose fixed set is the geodesic joining $\pm i$. This is the exact defect, the edge along which the $q$ sheets are glued. In the quotient description, the branch points become two heavy operators of weight $h_q=\frac{c}{24}(1-\frac{1}{q^2})$ at $\tilde\chi=\pm i$. The defect becomes a point particle created and absorbed by these operators, so it moves along the geodesic joining them, which extremizes its worldline action. In Poincaré coordinates this geodesic is the semicircle $\tilde X=0$, $\tilde Z^2+\tilde\tau_E^2=1$, and its continuation $\tilde\tau_E=-i\tilde{\mathcal T}$ is the timelike worldline $\tilde X=0$, $\tilde Z^2-\tilde{\mathcal T}^2=1$.}

\begin{equation}
    \widetilde X=0,\quad \widetilde Z^2+\widetilde\tau_E^2=1
    \qquad\longrightarrow\qquad
    \widetilde X=0,\quad \widetilde Z^2-\widetilde{\mathcal T}^2=1 .
    \label{eq:defect_geodesic}
\end{equation}
In what follows we use the quotient description, in which this geodesic is the worldline of a point particle of mass $\mu_q=\frac{1}{4G_N}\big(1-\frac1q\big)$. This is the mass that produces the conical deficit $\delta_q=8\pi G_N\mu_q=2\pi\big(1-\frac1q\big)$\cite{Deser:1983tn,Benjamin:2020mfz}.

Since $\widetilde X=0\Leftrightarrow x=0$ for $uv\neq1$, the map \eqref{eq:Poincare_Kruskal_map} at $x=0$ gives
\begin{equation}
    \widetilde Z^2-\widetilde{\mathcal T}^2
    =
    \frac{(1+u)(1-v)}{(1-u)(1+v)},
\end{equation}
and  the $\widetilde Z^2-\widetilde{\mathcal T}^2=1$ gives
\begin{equation}
    u=v
    \label{eq:unboosted_defect_uv}
\end{equation}
A convenient parametrization is
\begin{equation}
    u=v=\tan\frac{\lambda}{2},
    \qquad
    \lambda\in\Big(-\frac{\pi}{2},\frac{\pi}{2}\Big),
    \qquad
    ds^2=-d\lambda^2,
    \label{eq:defect_param}
\end{equation}
where, the parameter $\lambda$ is proper time. 
\subsubsection{Time evolution as a boost}\label{sec:timeboost}
The entanglement entropy of an interval at time $t$ is given by the length of the extremal geodesic anchored at the interval’s endpoints at that time, i.e. on the boundary slice $\tau=kt$. In this computation the endpoints move with $t$, while the defect stays at $u=v$. Since $\partial_t=k(v\,\partial_v-u\,\partial_u)$ is a Killing vector of the AdS–Rindler background, time evolution by the deformed Hamiltonian $H$ in \eqref{q} acts as the boost.
$$u\to e^{-kt}u,\qquad v\to e^{kt}v,$$
By the time-translation symmetry of the background, we can shift the whole configuration back by $t$. This brings the interval endpoints back to $\tau=0$ and moves the defect to $u=e^{2kt}v$, without changing the geodesic length. Evolving the state forward is therefore equivalent to boosting the defect backward, with the endpoints held fixed. The entanglement entropy at time $t$ can thus be computed with the endpoints held at $\tau=0$ and the defect boosted backward by $kt$.
Equivalently,\footnote{ The time reflection $(u,v)\to(-v,-u)$ is an isometry that leaves the $\tau=0$ slice, and hence the endpoints, unchanged, and maps $u=e^{2kt}v$ to $u=e^{-2kt}v$. This places the shock on $u=0$
%, the convention used below.
} the boosted defect can be written as
$$u(\lambda)=e^{-kt}\tan\frac{\lambda}{2},\qquad v(\lambda)=e^{kt}\tan\frac{\lambda}{2},\qquad x=0,\qquad\text{i.e.}\qquad u=e^{-2kt}v ,$$
which approaches the horizon $u=0$ at late times.
\begin{figure}
    \centering
    \includegraphics[width=0.5\linewidth]{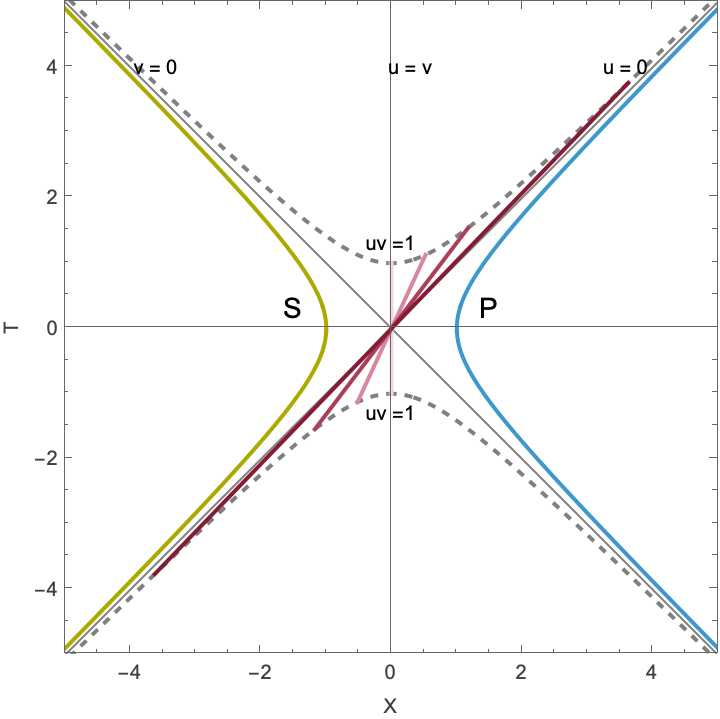}
    \caption{The defect worldline in the quotient-description, on the slice $x=0$, in the Kruskal coordinates \eqref{eq:localized-shock-full} of the late-time AdS–Rindler geometry. The solid hyperbolas $uv=-1$ are the asymptotic boundaries: the $P$-line on the right and the $S$-line on the left. The dashed hyperbolas $uv=1$ bound the patch, and the diagonals $u=0$ and $v=0$ are the horizons. The worldline $u=e^{-kt}\tan\frac\lambda2$, $v=e^{kt}\tan\frac\lambda2$ is shown for $kt=0,\,0.5,\,1,\,2$, from light to dark. At $kt=0$ it is the segment $u=v$ through the bifurcation point. Under the boost it rotates toward the horizon $u=0$, which it approaches at late times. It runs from the past region to the future region and never enters the $P$ or $S$ wedge.}
    \label{fig:placeholder}
\end{figure}
\subsubsection{The large-boost limit}
At late time $t$, for fixed $v$, $u=e^{-2kt}v\to0$: the worldline approaches the null generator $u=0$, $x=0$ of the horizon. We can compute the momentum and the stress tensor of the defect particle along
$$u(\lambda)=e^{-kt}\tan\frac{\lambda}{2},\qquad v(\lambda)=e^{kt}\tan\frac{\lambda}{2},\qquad x=0.$$
%using the local mass of the defect.
Momentum:
\begin{align}\label{momentum}
p^v=\frac{\mu_q}{2}e^{kt}\sec^2\frac{\lambda}{2},\qquad
p^u=\frac{\mu_q}{2}e^{-kt}\sec^2\frac{\lambda}{2},\\\nonumber
p_u=-\mu_q\,e^{kt}\cos^2\frac{\lambda}{2} \qquad p_v=-\mu_q\,e^{-kt}\cos^2\frac{\lambda}{2},
\end{align}
where, $\mu_q$ is the local mass of the defect and for writing the second line we have used $1+uv=\sec^2\frac{\lambda}{2}$, $g_{uv}=-2/(1+uv)^2=-2\cos^4\frac{\lambda}{2}$.
We note that, $p_u$ grows as $e^{kt}$ while $p_v$ decays as $e^{-kt}$. Therefore, the particle becomes effectively null and moves along the horizon generator.

%Lowering the indices gives
%$$p_u=-\mu_q\,e^{kt}\cos^2\frac{\lambda}{2},\qquad p_v=-\mu_q\,e^{-kt}\cos^2\frac{\lambda}{2}.$$
Stress Tensor:
For a point particle of mass $\mu_q$,

$$T_{\mu\nu}(y)=\mu_q\int d\lambda\;\frac{\dot x_\mu\,\dot x_\nu}{\sqrt{-g}}\;\delta^3\big(y-x(\lambda)\big),\qquad \sqrt{-g}=\frac{2(1-uv)}{(1+uv)^3},$$
Integrating $\lambda$ and using \eqref{momentum},
\begin{align}
&T_{uu}=\frac{\mu_q\,e^{kt}}{1-uv}\,\delta\big(u-e^{-2kt}v\big)\,\delta(\xi),\ 
T_{uv}=\frac{\mu_q\,e^{-kt}}{1-uv}\,\delta\big(u-e^{-2kt}v\big)\,\delta(\xi),\\\nonumber\ 
&T_{vv}=\frac{\mu_q\,e^{-3kt}}{1-uv}\,\delta\big(u-e^{-2kt}v\big)\,\delta(\xi),
\end{align}
            As we can see, only $T_{uu}$ grows, while $T_{uv}, T_{vv}$ decay with time $t$. Since, at late times and and fixed $v$, $u=e^{-2kt}v\to0$, so, $1-uv\rightarrow 1$ and $T_{uu}\approx \mu_q\,e^{kt} \delta(u)\delta(\xi)$.
The boosted defect therefore acts as a null source on the horizon $u=0$, $\xi=0$, with strength $\mu_q\,e^{kt}$.
%begin{equation}
 %   \mathcal P(t)=\mu_q\,e^{kt}.
%\end{equation}
%sourcing the shockwave geometry. 
The effective late-time description of the boosted defect is the shock geometry it sources. 
\subsubsection{The effective shockwave geometry}\label{Sec:EffectiveShockwave}

The effective late-time geometry is
\begin{equation}
ds^2=-\frac{4\,du\,dv}{(1+uv)^2}+\left(\frac{1-uv}{1+uv}\right)^2d\xi^2+4h(\xi)\,\delta(u)\,du^2,
\label{eq:shock_metric}
\end{equation}
Geodesics crossing $u=0$ experience a discontinuity in the $v$ coordinates\cite{Dray:1984ha},
$$v\rightarrow v +h(\xi),$$
where, the shift profile $h(\xi)$ is obtained by solving Einstein eq. with \eqref{eq:shock_metric} and $T_{uu}$ in late time, by solving (eq 3.9 in \cite{Roberts:2014isa}),
\begin{equation}
h''(\xi)-h(\xi)=-4\pi G_N\,\mu_q\,e^{kt}\,\delta(\xi),
\end{equation}
or, 
\begin{equation}\label{shift}
h(\xi)=2\pi G_N\,\mu_q\,e^{kt}\,e^{-|\xi|}.
\end{equation}
%The late time limit of entanglement entropy computation e.g. for case 2 or 3 in appendix B (S-P or P-S), reduces to to the computation of extremized geodesic length of the interval from left to the right boundary, $\tau=0$ across the shockwave at $u=0$.
%This is essentially the same computation as in \cite{RS, RR}. For the completion, let us outline the main arguments below.
At late times, the entanglement entropy of an interval say e.g. with one endpoint on the $S$-line and the other on the $P$-line (Section~\ref{CFTEE}) reduces to the length of the extremal geodesic connecting the left and right boundaries at $\tau=0$, crossing the shock at $u=0$. This is essentially the same computation as in \cite{Roberts:2014isa,Shenker:2013pqa,Roberts:2014ifa,Reynolds:2016pmi}. For completeness, we outline it below.

Consider one endpoint on the $S$-line (left wedge), at $(r,\tau,\xi)=(r_\infty,0,\xi_1)$, and the other on the $P$-line (right wedge), at $(r_\infty,0,\xi_2)$, where $r_{\infty}$ is $r\rightarrow\infty$ in the shockwave geometry \eqref{eq:shock_metric}. The geodesic connecting them consists of two segments, glued by the shift $v\to v+h(\xi)$: one of length $l_1$ from the $S$ endpoint to $(u,v,\xi)=(0,\,v+h(\xi'),\,\xi')$, and one of length $l_2$ from $(u,v,\xi)=(0,\,v,\,\xi')$ to the $P$ endpoint. Since we use both $(r,\tau,\xi)$ coordinates for the boundary points and $(u,v,\xi)$ for the crossing point it is convenient to use embedding coordinates,\footnote{The embedding coordinates satisfy $-Y_{-1}^2-Y_0^2+Y_1^2+Y_2^2=-1$, with $Y\cdot Y'=-Y_{-1}Y'_{-1}-Y_0Y'_0+Y_1Y'_1+Y_2Y'_2$. In terms of $(r,\tau,\xi)$ and $(u,v,\xi)$,
\begin{align*}
(Y_{-1},Y_0,Y_1,Y_2)
&=\Big(\tfrac{r}{k}\cosh \xi,\ \pm\sqrt{\tfrac{r^2}{k^2}-1}\,\sinh\tau,\ \tfrac{r}{k}\sinh \xi,\ \pm\sqrt{\tfrac{r^2}{k^2}-1}\,\cosh\tau\Big)\\
&=\Big(\tfrac{1-uv}{1+uv}\cosh \xi,\ \tfrac{u+v}{1+uv},\ \tfrac{1-uv}{1+uv}\sinh \xi,\ \tfrac{v-u}{1+uv}\Big),
\end{align*}
with the upper (lower) sign in the right (left) wedge.} in which the geodesic distance $l$ between two points $Y$ and $Y'$ is given by $\cosh l=-Y\cdot Y'$.
At large $r_\infty$ this gives
\begin{equation}
l_1=\log\frac{2r_\infty}{k}+\log\big[\cosh(\xi'-\xi_1)+v+h(\xi')\big],\qquad
l_2=\log\frac{2r_\infty}{k}+\log\big[\cosh(\xi'-\xi_2)-v\big].
\end{equation}
Next step is to extremize\cite{Roberts:2014ifa,Reynolds:2016pmi}, $l_1+l_2$ w.r.t. $v$ and $\xi'$. For simplest case, let us  consider the symmetric configuration $\xi_1=\xi_2=0$, for which extremization over $v$ gives $v=-h(\xi')/2$. Thus,
\begin{equation}
l_1+l_2=2\log\frac{2r_\infty}{k}+2\log\Big[\cosh \xi'+\tfrac12 h(\xi')\Big].
\end{equation}
With $h(\xi')=h(0)\,e^{-|\xi'|}$, extremizing over $\xi'$ gives $e^{2|\xi'|}=1+h(0)$, and finally
\begin{equation}
l_1+l_2=2\log\frac{2r_\infty}{k}+\log\big(1+h(0)\big),\qquad h(0)=2\pi G_N\,\mu_q\,e^{kt}.
\end{equation}
The entanglement entropy, $S=\frac{c}{6}(l_1+l_2)$, is therefore
\begin{equation}\label{eqshockEE}
S=\frac{c}{3}\log\frac{2r_\infty}{k}+\frac{c}{6}\log\big(1+2\pi G_N\mu_q\,e^{kt}\big),
\end{equation}
where the first term is the usual time-independent UV-divergent contribution. At late times,
\begin{equation}
S\simeq\frac{ck}{6}\,t+\text{const},
\end{equation}
reproducing the slope $ck/6$ of Cases~2 and~3 in Section~\ref{CFTEE}.
\subsubsection{Remarks on the effective description}\label{comparison}
In the previous subsection, we called the shockwave geometry an effective description because, to obtain it, we allowed only the leading late-time component of the stress tensor of the defect, $T_{uu}\simeq\mu_q e^{kt}\delta(u)\delta(\xi)$, to backreact on the AdS–Rindler background, producing the shock. In this subsection we ask how good this approximation is.

The comparison is clearest in terms of holonomies. In the global description, going once around the branch locus, takes one sector to the next, so the defect holonomy is given by the Möbius transformation \eqref{mobiusMap},
\begin{equation}
    M=\begin{pmatrix}\cos\frac{\pi}{q} & \sin\frac{\pi}{q}\\[2pt] -\sin\frac{\pi}{q} & \cos\frac{\pi}{q}\end{pmatrix},
\end{equation}

which is elliptic, since $\mathrm{tr}\,M=2\cos\frac{\pi}{q}<2$ for every integer $q>1$.
The single-patch/ quotient description is the quotient of this geometry by the $\mathbb Z_q$ generated by $M$, in which points related by $M$ are identified and the branch locus becomes the conical defect. Going once around the defect therefore amounts to applying $M$ once, so $M$ is also the holonomy around the defect.

It is useful to rewrite $M$ in terms of the deficit angle $\delta_q$. Since $\frac{2\pi}{q}=2\pi-\delta_q$, $M$ is equivalently a rotation by the deficit angle, $M_\delta=-M^{-1}$, which reads
\begin{equation}
M_\delta=\begin{pmatrix}\cos\frac{\delta_q}{2} & \sin\frac{\delta_q}{2}\\[2pt] -\sin\frac{\delta_q}{2} & \cos\frac{\delta_q}{2}\end{pmatrix}.
\end{equation}

For $q>1$ the deformed Hamiltonian is not a boost of the full geometry. Recall, however, that at late times the geometry approaches the AdS–Rindler background generated by $\tilde F$ (see Section~\ref{quotientdesciption}). The deformed Hamiltonian acts on $\tilde F$ as a Möbius transformation, and is therefore an isometry of this background. It generates the Rindler boost $u\to e^{-kt}u$, $v\to e^{kt}v$ of Section~\ref{sec:timeboost}
, whose matrix, in the basis where it is diagonal\footnote{By \eqref{tildechi}, $\tilde\chi=\tanh\frac{kx^+}{2}$ on the $P$-line, and time evolution shifts $x^+\to x^++t$. In terms of $\frac{1+\tilde\chi}{1-\tilde\chi}=e^{kx^+}$, this shift becomes a rescaling by $e^{kt}$, so the boost is diagonal in this basis. Since this change of basis fixes the branch points $\tilde\chi=\pm i$ (same as in \eqref{mobiusMap}), as does $M_\delta$, it commutes with $M_\delta$, and $M_\delta$ has the same matrix in both bases.}, is $B=\mathrm{diag}\big(e^{\eta/2},e^{-\eta/2}\big)$, with $\eta=kt$.
Since this boost is not an isometry of the defect, it moves the defect, and its holonomy becomes
\begin{equation}
M_\eta=B\,M_\delta\,B^{-1}=\begin{pmatrix}\cos\frac{\delta_q}{2} & e^{\eta}\sin\frac{\delta_q}{2}\\[2pt] -e^{-\eta}\sin\frac{\delta_q}{2} & \cos\frac{\delta_q}{2}\end{pmatrix}.
\end{equation}

%%Note that the boosted holonomy is still elliptic: its trace, $2\cos\frac{\delta_q}{2}$, is unchanged by the boost.

%%It is nevertheless useful to separate it as
%%$$M_\eta=\begin{pmatrix}1 & \frac{\delta_q}{2}\,e^{\eta}\\[2pt] 0 & 1\end{pmatrix}+\begin{pmatrix}\cos\frac{\delta_q}{2}-1 & e^{\eta}\big(\sin\frac{\delta_q}{2}-\frac{\delta_q}{2}\big)\\[2pt] -e^{-\eta}\sin\frac{\delta_q}{2} & \cos\frac{\delta_q}{2}-1\end{pmatrix}.$$
%The first matrix is the holonomy of a shock, a shift rather than a rotation, and it is what the effective description keeps. The second contains everything the effective description misses.
%%Written in this form, the first matrix is exactly the holonomy of a shock, with trace $2$. It is what the effective description keeps: the leading late-time component $T_{uu}$, which creates the shock. The second matrix contains everything the effective description misses. Its lower entry decays at late times, its upper entry grows as $e^{\eta}$, and its diagonal entries are unchanged by the boost.

The boosted holonomy has two features that determine how the effective description compares with the exact one.
First, time enters only through the boost: it multiplies the upper entry of $M_\eta$ by $e^{\eta}$ and the lower one by $e^{-\eta}$, and leaves the diagonal unchanged. At late times the growing part is therefore the upper entry, and this is the entry the effective description keeps. Indeed, the shock \eqref{eq:shock_metric} glues the two sides of the horizon $u=0$ by the shift $v\to v+h(\xi)$ \cite{Dray:1984ha}. Written, like \eqref{mobiusMap}, as a matrix acting on $v$ by $v\to\frac{av+b}{cv+d}$, this shift is
$$\begin{pmatrix}1 & h(\xi)\\ 0 & 1\end{pmatrix},\qquad h(\xi)=2\pi G_N\mu_q\,e^{kt}e^{-|\xi|}\propto e^{\eta},$$
by \eqref{shift}. This is a shift rather than a rotation: its only nonzero off-diagonal entry is the upper one. 
%Both holonomies therefore grow by the same factor $e^{\eta}$, and we expect the entanglement entropy to grow at the same rate in both descriptions.
In both descriptions, the growth therefore comes from the same boost factor $e^{kt}$, and we expect the entanglement entropy to grow at the same rate.
Second, $M_\eta$ is still elliptic: its trace, $2\cos\frac{\delta_q}{2}$, is unchanged by the boost and differs from $2$, the trace of a shift. No boost turns the defect into a shock, so the effective description is not exact.\footnote{Formally, the two holonomies agree to first order in $\delta_q$. Since $\delta_q\ge\pi$ for integer $q$, this is not a controlled approximation.} Since in both descriptions the growth comes from the same boost factor $e^{\eta}$, however, the difference can only appear in the time-independent part of the entanglement entropy.

To check this, we compare the two results for the configuration computed in Section~\ref{Sec:EffectiveShockwave}, with one endpoint on the $S$-line and the other on the $P$-line, both at $\xi=0$, for which the effective result \eqref{eqshockEE} is available in closed form. The exact result of Section~\ref{Sec:ExactEE}, equivalently Cases~2 and~3 of Section~\ref{CFTEE}, gives \footnote{Here, the cut-off is the cutoff in terms of the Hamiltonian adapted coordinate at $x=0$, i.e. $\epsilon_{(x=0)}$.}
\begin{equation}\label{eq:exact_symmetric}
S_{\rm exact}=\frac{c}{3}\log\frac{2}{k\epsilon}+\frac{c}{6}\log\Big[q^2\cosh^2(kt)\,\sin\frac{\frac{\pi}{2}-\psi_{\rm P}(t)}{q}\,\sin\frac{\frac{\pi}{2}+\psi_{\rm P}(t)}{q}\Big]
\end{equation}

with $\psi_{\rm P}(t)=2\arctan\big(\tanh\frac{kt}{2}\big)$, as in \eqref{eq:HTheta} at $\phi=\pi/2$. \footnote{A consistency check: the argument of the logarithm equals $1$ for $q=1$, so the entropy is time-independent in the undeformed CFT. }
%Note that, the first term in the above eq is $2\cosh(kt)$ for $q=2$. 
On the other hand, with $r_\infty=1/\epsilon$ and $2\pi G_N\mu_q=\delta_q/4$, the effective result \eqref{eqshockEE} reads
\begin{equation}
   S_{\rm shock}=\frac{c}{3}\log\frac{2}{k\epsilon}+\frac{c}{6}\log\Big(1+\frac{\delta_q}{4}\,e^{kt}\Big). 
\end{equation}
At late times, using $\psi_{\rm P}(t)\simeq\frac{\pi}{2}-2e^{-kt}$ (Section~\ref{CFTEE}), the argument of the time-dependent logarithmic term  in $S_{\rm exact}$ becomes $\frac{q}{2}\sin\frac{\delta_q}{2}\,e^{kt}+\cos\frac{\delta_q}{2}+O(e^{-kt})$, and is $\frac{\delta_q}{4}e^{kt}+1$ in $S_{\rm shock}$. Both grow as $e^{kt}$ 
 with different coefficients and different constant terms \footnote{The factor $q$ in the exact coefficient is the familiar prefactor of geodesic lengths in a conical-defect geometry with opening $1/q$ \cite{Asplund:2014coa, Balasubramanian:2014sra}.}. 
Pulling out the growing terms in each respectively,
$$\log\Big(\frac{q}{2}\sin\frac{\delta_q}{2}\,e^{kt}+\cos\frac{\delta_q}{2}\Big)=kt+\log\Big(\frac{q}{2}\sin\frac{\delta_q}{2}\Big)+\log\Big(1+\frac{2\cos\frac{\delta_q}{2}}{q\sin\frac{\delta_q}{2}}\,e^{-kt}\Big),$$
$$\log\Big(\frac{\delta_q}{4}\,e^{kt}+1\Big)=kt+\log\frac{\delta_q}{4}+\log\Big(1+\frac{4}{\delta_q}\,e^{-kt}\Big),$$
where the last terms vanish at late times. Therefore 

\begin{equation}
    S_{\rm exact}\simeq\frac{ck}{6}\,t+\frac{c}{3}\log\frac{2}{k\epsilon}+\frac{c}{6}\log\Big(\frac{q}{2}\sin\frac{\delta_q}{2}\Big),\qquad S_{\rm shock}\simeq\frac{ck}{6}\,t+\frac{c}{3}\log\frac{2}{k\epsilon}+\frac{c}{6}\log\frac{\delta_q}{4}.
\end{equation}

The two descriptions thus give the same growth at every $q>1$, and differ only in the time-independent part:
\begin{equation}
    S_{\rm exact}-S_{\rm shock}\simeq\frac{c}{6}\log\frac{q\sin(\pi/q)}{\pi(1-1/q)} .
\end{equation}

This difference decreases monotonically with $q$
%, from $\frac{c}{6}\log\frac{4}{\pi}\approx0.04\,c$ at $q=2$ 
to zero as $q\to\infty$.
\subsubsection{Comments on phase boundary and non-heating phase}\label{pbnhpcommentSec}

%\textbf{Phase boundary}\\
%From \eqref{banadosfactorhp}, the phase boundary is a limiting case of the heating phase, i.e. as $\phi\rightarrow\pi$ and therefore $\mathcal{D}\rightarrow 0$ the banados factor (for $\alpha>0$) becomes
%\begin{align}&\mathcal{L}_{\pm}^P \rightarrow \frac{(q^2-1)\alpha^2}{(1+q^2\alpha^2 (x^\pm)^2)^2}\\&\mathcal{L}_{\pm}^S \rightarrow 0    \end{align}
%Moreover, the length of $P$ metric increases while that of $S $ shrinks to zero as we consider this limit. 
As discussed before in Section \ref{Sec:PBbulk}, the phase boundary can be understood as a limiting ($\phi\to\pi$) case of the heating phase. When seen from the boundary, the S-arcs shrink to  points and the P-arcs grows to cover the whole part between two degenerate fixed points. Similarly, in the bulk quotient picture at late time, as one approaches the phase boundary from the heating phase: the P region of the heating phase AdS-Rindler metric \eqref{latetimeRindler} (sharing a bifurcation point with the S-region and with an effective temperature $k/2\pi$), ultimately becoming the whole of the AdS-Poincare, while S region shrinks to one of the fixed points and the effective temperature goes to zero (since,  $k\rightarrow0$). 

Let us now discuss what happens to the defect in this limit. Unlike the heating phase scenario, where the boosted defect, as seen from the late time slice, produced the effective shockwave description, in the phase boundary late time AdS–Poincaré slice (with Poincaré horizons and zero effective temperature) the defect just get translated rather than being boosted.
One more important point we must note before moving forward: on the phase boundary, we cannot set $\alpha=0$, as was done in the case of heating phase for simplicity\footnote{In the heating phase, the late time slice is AdS-Rindler for every $\phi$, i.e. irrespective of the choice of $\alpha$. There the $\alpha=0$ choice only simplified the position of the defect, which was passing through the bifurcation point.}, since here $\alpha=2\beta$. 
This can be argued cleanly from the boundary. Consider the $P$-line \eqref{eq:HTheta}: here the Cayley coordinate of the quotient picture is $\tilde\chi=\tan\frac{\psi_P}{2}=\tan\frac\phi2\tanh\frac{kx}{2}$, and thus the Floquet evolution $x^\pm\to x^\pm+t$(from \eqref{mobiusmapAppendix} with $a(t)$, $b(t)$ in \eqref{FloquetParametersAppendix}) is the Möbius transformation 
\begin{equation}
    \tilde\chi\;\to\;\frac{\cosh\frac{kt}{2}\,\tilde\chi+\tan\frac\phi2\sinh\frac{kt}{2}}{\cot\frac\phi2\sinh\frac{kt}{2}\,\tilde\chi+\cosh\frac{kt}{2}}
\end{equation}
with fixed points at $\tilde\chi=\pm\tan\frac\phi2$. This in $\phi\rightarrow\pi$ limit 
%ince $\tan\frac\phi2\sinh\frac{kt}{2}\to q\alpha t$ and $\cot\frac\phi2\sinh\frac{kt}{2}\to0$
becomes,
\begin{equation}
    \tilde\chi\;\to\;\tilde\chi+q\alpha t
\end{equation}
which is a time translation rather than a boost as in the heating phase.
Now since $\tilde X\pm\tilde{\mathcal T}=\pm\tilde\chi^\pm$\eqref{tildechi}, in the bulk this is the Poincaré time translation $\tilde{\mathcal T}\to\tilde{\mathcal T}+q\alpha t$ at fixed $\tilde X$ and $\tilde{Z}$.
%\footnote{Note that the $SL(2,\mathbb R)$ matrix of the Floquet evolution changes from hyperbolic, with trace $2\cosh\frac{kt}{2}>2$, to parabolic, with trace $2$, as $\phi\to\pi$.}.
Just as in the heating phase, in this case too the defect follows same bulk geodesic $\tilde X=0$, $\tilde Z^2-\tilde{\mathcal T}^2=1$ as that of \eqref{eq:defect_geodesic}\footnote{This is because it does not depend on the value of $\alpha$ since the heavy operators are at $\tilde F=0,\infty$ for every $\phi$ and the Cayley map sends them to $\tilde\chi=\pm i$.}.

 In the late time, the boundary interval endpoints at fixed $(x_i,t)$, sit on the late-time Poincaré slice $\tilde{\mathcal T}=q\alpha t$, at $\tilde X=q\alpha x_i$ and using the \eqref{liftbulk} at the cutoff $\tilde Z=q\alpha\epsilon$. To analyze the defect from this late time slice we define $\tilde{\mathcal T}'=\tilde{\mathcal T}-q\alpha t$, therefore on at observer's time $t$, the two endpoints sit at $\tilde{\mathcal T}'=0$. Thus the defect geodesic becomes
$$\tilde X=0,\qquad \tilde Z^2-(\tilde{\mathcal T}'+q\alpha t)^2=1$$
which at $\tilde{\mathcal T}'=0$ is located at $\tilde Z=\sqrt{1+q^2\alpha^2t^2}\simeq q\alpha t$.
Now let us come to the EE. Since there is no S region in this phase, we can only have both the interval endpoints in P boundary. This means there can be two cases: one in which the interval includes one (or more ) degenerate fixed points and another containing none. In the first situation geodesics needs to loop around this defect line (number of loops equals the number of fixed points) see Appendix~\ref{appendixPIComparison} and has EE 
$$S\simeq\frac{2c}{3}\log t.$$
If the interval contains no fixed points then it does not loop around the defect and hence saturates at late time \footnote{In this case the end point coming closer at the rate $\sim t^{-2}$ cancels the cut off contribution in geodesic length $\rho\simeq\log\frac{2}{\epsilon_{\rm global}}$, where $\epsilon_{\rm global}=\epsilon\sqrt{\psi'_{\rm pb}(x^+)\psi'_{\rm pb}(x^-)}\simeq\frac{2\epsilon}{q\alpha t^2}$ }. 

Finally, we note that in the non-heating phase, since there are no fixed points (equivalently no horizons in the bulk description), the time evolution under the $np$ Hamiltonian covers the entire global AdS$_3$. Thus the quotient description does not exist in this case. 
%The quotient desciption exists only when there exists $q$ patches which are causally disconnected from each other 

%For a late time floquet observer the boundary transformation $\tilde\chi^\pm=q\alpha x^\pm$ to the hamiltonian adapted coordinates at late time $t$, the defect geodesic  (using the bulk diffeomorphism map \eqref{robertsmap}:  
%$\tilde{\mathcal T}=q\alpha t,\qquad \tilde X=q\alpha x,\qquad \tilde Z=q\alpha y$)
%becomes,
%\begin{equation}
%    x=0,\qquad y^2-t^2=\frac{1}{(q\alpha)^2}.
%\end{equation}
%which results in the logarithmic growth of the EE
%$$S\simeq\frac c3\log t$$for each chiral and anti chiral sector.

%In the late time, at fixed $x$, the Thus, we see from the \eqref{banadoslt}

% \subsection{In heating phase}
% \subsection{In non-heating phase}
% \subsection{On phase boundary}

\section{Conclusion}
In this paper, we study the dual bulk description of $sl^{(q)}(2, \mathbb{R})$  CFTs, generalizing the analysis of \cite{Das:2024vqe} to the case of $q>1$. The dual geometries are described in the Hamiltonian-adapted frame, in which the coordinate time is the time generated by the system Hamiltonian. for $q=1$, these geometries are time independent (global AdS for the $np$ phase and AdS Poincare for the $pb$), reflecting the $sl(2,\mathbb{R})$ symmetry of the vacuum. The most interesting case was the heating phase. In the $hp$, the Hamiltonian has two fixed points, which causally restricts Heisenberg evolution of operators in each side of the fixed points into two separate causal wedges. In the bulk, this results in two AdS Rindler exteriors as the bulk dual of each patch. In this paper, we were able to complete the analysis of \cite{Das:2022pez}, showing how the AdS Rindler description, exactly reproduces the OTOC computation in the $sl(2, \mathbb{R})$ case as presented in \cite{Das:2022jrr}. Further, we also explain why the CFT OTOC shows the exponential growth only when both operators involved in the OTOC are on the same side of the fixed points. The main point is that, in the same side configuration of the operators, the bulk description involves a geodesic computation across the shockwave shifted worm hole, similar to the Robert Stanford OTOC analysis in thermal CFT$_2$ \cite{Roberts:2014ifa}. However, for the case of opposite side configuration of operators, the bulk computation would translate to a geodesic computation in one AdS Rindler exterior, completely bypassing the  effect of the shock wave which lies entirely in the other exterior.

For the $q>1$ case, the bulk duals are time dependent Ba\~{n}ados geometries. In the $hp$, there are $2q$ fixed points, and the Hamiltonian adapted frame describes the dynamics inside region bounded by nearest neighbour fixed points. In the bulk, the geometry associated with each such patch describes the bulk causal wedge of the subregion restricted between the two fixed points.  This geometry is time dependent in each wedge which asymptotes to an AdS Rindler metric at late times, which accounts for the saturation of the vacuum $EE$ of an interval, restricted to lie between two such nearest-neighbour fixed points. 

We are also able to reproduce the late time temporal growth of the $EE$, for the case when the interval straddles  multiple fixed points. This is done in two ways. In the first method, we use the structure of the embedding of the bulk causal wedges inside the global AdS$_3$, to compute the time dependence of the $EE$ and explicitly match it with the CFT answer. In the second method, we exploit the fact that the  $sl^{(2)}(2,\mathbb{R})$, while locally isomorphic to $q=1$, involves a further  quotienting of the spatial circle into $q$ distinct pairs of regions. Each of these $q$ regions may be reached by crossing a branch cut of the coordinate map relating the usual cylinder coordinates to the Hamiltonian-adapted coordinate system. The existence of the fixed points in the $hp$ and $pb$, causally decouples them into $q$ distinct dynamical {\it replicas}, with fields in each sector related via the branch cut. Thus one can equivalently restrict to a single sector, with a twist operator inserted at the branch point. Correlation functions across the sectors can be computed equivalently in this single sector, in the presence of the twist operator. In the bulk, the twist operator corresponds to a heavy operator moving along a geodesic, which is the bulk extension of the branch point in the boundary. At late time, we show how the effective computation translates into a geodesic computation in the presence of a shockwave caused by the twist operator, reproducing the late time growth of the vacuum $EE$. 

In our paper, we have only analyzed $EE$ in the $sl^{(q)}(2,\mathbb{R})$ deformed CFTs. It would be nice to study other observables involving higher point functions such as the OTOC in these systems. Recently,  OTOCs in time dependent states have been studied in 2D CFTs\cite{Hampton:2026qhq}. It would be nice to apply this for the $q>1$ deformed CFTs and understand the fixed point dependence of this observable in this case and reproduce this from a bulk analysis. $sl^{(q)}(2,\mathbb{R})$ deformed CFTs  are interesting because they provide us with simple models where one can analyze properties of time-dependent states. It would be interesting to see, if our methods generalize to other non equilibrium systems. An interesting class of Floquet CFTs are those where the drive is random, instead of being periodic. It would be interesting to see how to understand some of these results from a bulk analysis, perhaps along the lines of \cite{Antonini:2023hdh}. Recently, there have been discussions in the literature on the connection between quantization in the $hp$ of the $sl^{q}(2,\mathbb{R})$ CFTs and recent discussions of the brickwall models of AdS$_{3}$ Rindler horizons\cite{Das:2024mlx}. It would be nice to see if these connections can be generalized to the $q>1$ case.  We hope to be able to address some of these questions in the near future.

%of the bulk description of $sl(2,\mathbb{R})$ deformed CFTs to $q>1$, while also completing the bulk explanation of the position-dependent OTOC behaviour in $q=1$ case. Our main result for the $q>1$ case, is an explicit metric description of the time evolved vacuum of 2D CFT under the $sl(2,R)$ deformed Hamiltonian in each dynamical phase. We then reproduce the CFT result for the $EE$ from the bulk picture. In the $hp$, there are $2q$ fixed points and the bulk description in the Hamiltonian-adapted frame, describes the region inside the bulk causal wedge, anchored across two neighbouring fixed points in the boundary CFT. In each such patch, the bulk metric asymptotes at late time to the static AdS Rindler geometry. This accounts for the late time saturation of the  $EE$ for an interval lying inside two fixed points. We also showed that the late-time linear growth in time of the $EE$, when the interval straddles a fixed point occurs due to the bulk geodesic crossing across a conical defect. One can also understand it from an effective shock wave computation in a single patch due to the presence of a twist operator. We then also discuss the $EE$ computation in the other two phases and show how they match the boundary answer. 

\section*{Acknowledgements}
We thank Diptarka Das, Suchetan Das and Ronak Soni for useful discussions related to this work. We would also like to thank Suchetan Das for crucial discussions during initial stages of the paper. RR would also like to thank the organizers of `Student Talks on Trending Topics in Theory' ($ST^4$ 2026), during which period some part of the work was done.
%We thank $\cdots$ for comments on the draft.
SP would like to thank Infosys Foundation for partial support at CMI. RR would like to acknowledge financial support under the DST INSPIRE Fellowship (IF240152) from the Department of Science and Technology (DST), Government of India.

\section*{AI Disclosure Statement}
%\paragraph{Disclosure of AI use.}
During the preparation of this work, the authors used Anthropic's Claude Opus 5.5, which suggested the choice of Euclidean slices about which to perform the analytic continuation. This choice was extremely important to the analysis. Claude also suggested many relevant references for other analyses. The authors used OpenAI's ChatGPT to help prepare the plots in Appendix~\ref{kswa}, Fig.~\ref{causal3d} and visualize many features of the bulk geometry, and check intermediate calculations and the reasoning behind them. The authors developed the physical interpretation, carried out the calculations, and wrote the manuscript. The authors take full responsibility for the scientific content of the paper.
\appendix
\section{OTOC in \texorpdfstring{$q=1$} Case}\label{otoc}
\begin{figure}[h]
    \centering
    \includegraphics[width=0.95\textwidth]{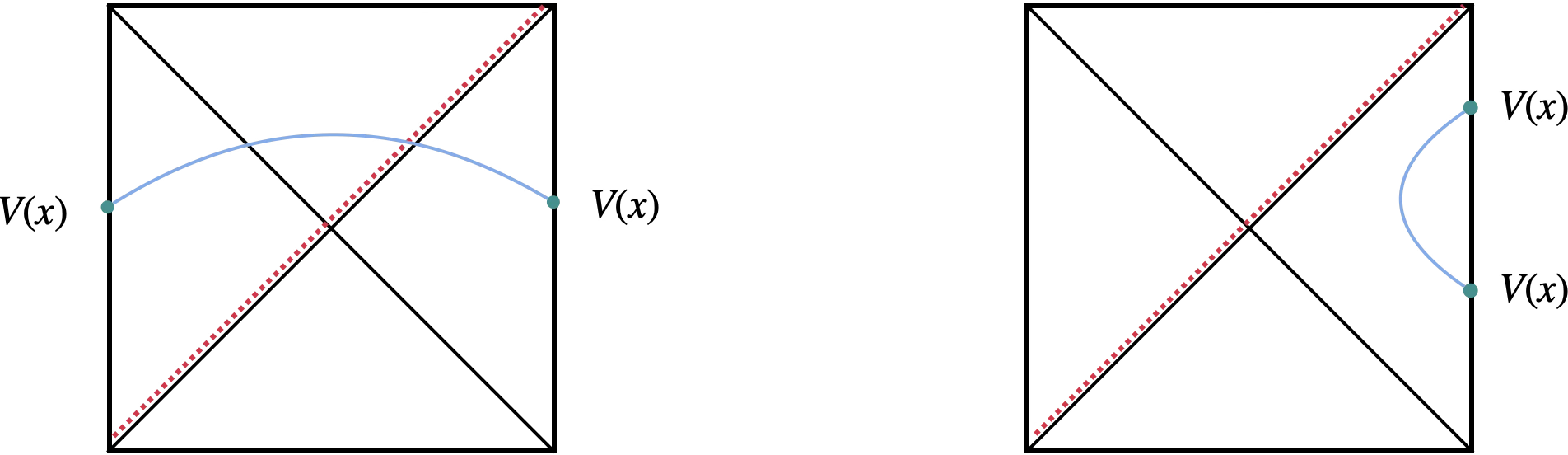}
    \caption{A schematic description of the OTOC computation in the two cases. The left figure corresponds to the case when two operators are on different sides of the fixed points while the right panel corresponds to the situation when both the operators are inside two fixed points.}
\end{figure}
The bulk dual of the  $sl(2,\mathbb R)$-Floquet CFTs, i.e. the $q=1$ case of the maps of Section~\ref{Sec:Slq2r}, was studied earlier in \cite{Das:2022pez,Das:2024vqe}\footnote{More specifically, two different methods were used. In \cite{Das:2022pez}, the duals of the $hp$, $pb$ and $nh$ were found to be different AdS$_2$ slicings of pure AdS$_3$, with AdS$_2$ black-hole, Poincaré AdS$_2$ and global AdS$_2$ slices, respectively. On the other hand, in \cite{Das:2024vqe}: AdS–Rindler, Poincaré AdS$_3$ and global AdS$_3$, respectively. The two descriptions are equivalent and are related by small diffeomorphism.}. 
As mentioned in the introduction, it was shown in \cite{Das:2022jrr} that the vacuum OTOC in the $hp$ grows exponentially at late times when both operators lie on the same side of a fixed point, and does not grow when they lie on opposite sides. The Lyapunov exponent of this same-side growth was reproduced from  the bulk description in \cite{Das:2022pez}. In this appendix, we try to provide a complete understanding of OTOC in $hp$ of $q=1$ with a much simpler argument. More specifically, using coordinate mapping to the Hamiltonian-adopted frame, we show that the full cross ratio of same-side OTOC \cite{Das:2022jrr} can be exactly matched  and it is same as that of the AdS–Rindler shockwave computation of \cite{Roberts:2014ifa}. 
%\textcolor{red}{Needs to add bridging sentences}\\
For $q=1$, the deformed Hamiltonian is built from the global generators $L_0$ and $L_{\pm1}$, and therefore leaves the vacuum invariant. The Bañados functions of Section~\ref{holodualSec} are then constant, $\mathcal L_\pm=\mathcal D/4$. Using the bulk diffeomorphism of Section~\ref{holodualSec}, one can lift the boundary conformal transformations to the bulk. After the analytic continuation to real time used in Section 2, the resulting metrics are:
%%Using the standard Roberts transform, one can now lift this conformal transformation to a bulk coordinate system with $s$ being the time in the new metric. We then analytically continue to real time $s\rightarrow it$, to get the Lorentzian metrics. The resulting expression of the metrics are given below:

\begin{eqnarray}
    {\bf nh\; (\mathcal{D} < 0)}: \;&& ds^2 = -(\rho^2 +\mathcal{|D|}) dt^2 + \frac{d\rho^2}{\rho^2 +\mathcal{|D|}} + \rho^2dx^2\ \nonumber \\
   {\bf pb\; (\mathcal{D} = 0)}: \;&& ds^2 = \frac{-dt^2 + dx^2 +dy^2}{y^2} \nonumber \\
   {\bf  hp\; (\mathcal{D} >0)}: \;&& ds^2 = -(r^2 -\mathcal{D}) dt^2 + \frac{dr^2}{r^2 -\mathcal{D}} + r^2dx^2
\end{eqnarray}

As we see, from the metrics above, the bulk metric for the $nh$ is global $AdS$, while for the $pb$ and $hp$ it is $AdS$ Poincare and $AdS$ Rindler respectively. The $AdS$ Rindler patch covers the bulk causal wedge associated with the boundary region between two fixed points. Thus there are only two such patches associated to the two sides of the fixed points. 
%\textcolor{cyan}{(the corresponding two causal wedges  share a common bifurcation point unlike $q\neq1$)}. 
When both operators $V$ and $W$ lie on the same side of the fixed points, the bulk computation of the $OTOC$ reduces to a two-point computation of the $V$ operator  in the  extended $AdS$ Rindler perturbed by a shock wave geometry caused by the back reaction of the large blue shifted energy ($E_W \sim e^{\sqrt{\mathcal{D}} t} $) of the $W$ particle near the $AdS$ Rindler horizon, exactly as in \cite{Roberts:2014ifa}. Mathematically, we can express this as follows:
\begin{eqnarray}
    \langle V_{u}(0)W_{u}(t)V_{u}(0)W_{u}(t)\rangle|_{\textrm{AdS Rindler}} &=&\langle W_{u}(t)V_{d}(0)V_{u}(0)W_{u}(t)\rangle|_{\textrm{extended AdS Rindler}} \nonumber  \\
     &=&\langle V_{d}(0)V_{u}(0)\rangle|_{\textrm{shock wave in  AdS Rindler}}\nonumber
\end{eqnarray}
$u$ and $d$ label the two sides of the fixed point and their bulk duals. On the other hand, when the two operators lie on different sides, the computation in the bulk reduces to a two point computation of $V$ operators in one patch, while the shock wave exists in the other patch. Thus the shock wave has no effect on the two point function and so the $OTOC$ does not develop any non trivial time dependence.  

\begin{eqnarray}
    \langle V_{d}(0)W_{u}(t)V_{d}(0)W_{u}(t)\rangle|_{\textrm{AdS Rindler}} &=&\langle W_{u}(t)V_{d}(0)V_{d}(0)W_{u}(t)\rangle|_{\textrm{extended AdS Rindler}} \nonumber  \\
     &=&\langle V_{d}(0)V_{d}(0)\rangle|_{\textrm{shock wave in AdS Rindler patch ($u$)} }\nonumber
\end{eqnarray}

We now want to show how the OTOC result of \cite{Das:2022jrr}, when written in terms of the Hamiltonian-adapted coordinate, matches exactly to  that of \cite{Roberts:2014ifa}.
%To see that the same-side computation is exactly that of \cite{Roberts:2014ifa}, we rewrite the boundary result of \cite{Das:2022jrr} in the Hamiltonian-adapted coordinate.
%This same-side versus opposite-side behaviour follows more directly from the boundary computation of~\cite{Das:2022jrr}, which we briefly recall. 
Again for simplicity we consider $\alpha=0$\footnote{Note that \cite{Das:2022jrr} uses a different $\alpha$ parameter and that should not be confused with $\alpha$ used here.}. 
%$\alpha=0$ here corresponds to $\alpha\to\infty$ in \cite{Das:2022jrr}, where fixed points approach $x_c=\pm\pi/2$.
Since $H$ is built from global generators, it acts on the plane as a Möbius map, $\zeta\to (a(t)\zeta+b(t))/(b*(t)\zeta+a^*(t))$\footnote{ The expressions of $a(t),b(t)$ 
%where, $c(t)=b^*(t)$ and $d(t)=a^*(t)$ 
are given in \eqref{FloquetParametersAppendix}, with the substitution of $q=1$.}, with fixed points $\zeta=\pm i$, i.e. $x_c=\pm \pi/2$. The OTOC is therefore a conformally transformed four-point function, which at large $c$ is determined by the cross ratio $\eta$~\cite{Roberts:2014ifa}. Note that the two $W$’s are placed at $x_c=0$ and evolved in time, while the two $V$’s are placed at position $x_c$ as in \cite{Roberts:2014ifa}. At late times,
$$\eta\;\simeq\;-4\,e^{-\sqrt{\mathcal{D}}\,t}\,\frac{1+\tan\frac{x_c}{2}}{1-\tan\frac{x_c}{2}}\;\sin\frac{\epsilon_1-\epsilon_2}{2}\,\sin\frac{\epsilon_3-\epsilon_4}{2},$$
where the $\epsilon_i$ are the Euclidean regulators that order the operators, taken  as $\sqrt{\mathcal D}\,t\to\sqrt{\mathcal D}\,t+i\epsilon_i$ and $\eta$ is given up to an overall phase which goes to 1 as $\epsilon_i\rightarrow0$. The OTOC grows exponentially only if $\eta$ crosses the real axis at a value larger than one, moving onto the second sheet, and this happens only when the operators lie between the two fixed points, $-\pi/2\le x_c\le \pi/2$. This region is exactly one AdS–Rindler patch. On the $\tau_c=0$ slice, the Hamiltonian-adapted coordinate is related to $x_c$ by
$$e^{\sqrt{\mathcal{D}}\,x}=\frac{1+\tan\frac{x_c}{2}}{1-\tan\frac{x_c}{2}},$$
which maps the arc between the fixed points onto $x\in\mathbb R$, with the fixed points at $x=\pm\infty$ (at $q=1$, $\alpha=0$ and $t=0$ from the map of Section \ref{Sec:Slq2r} we get $\tan\frac{x_c}{2}=\tanh\frac{\sqrt{\mathcal D}\,x}{2}$). In terms of $x$, 
\begin{equation}
    \eta\simeq-4\,e^{-\sqrt{\mathcal{D}}(t-x)}\sin\frac{\epsilon_1-\epsilon_2}{2}\sin\frac{\epsilon_3-\epsilon_4}{2}
\end{equation}
and the crossing condition of eq. (3.7) of \cite{Das:2022jrr} simplifies to $t=x$. These are exactly the late-time cross-ratio and the crossing conditions for the thermal OTOC of~\cite{Roberts:2014ifa} at inverse temperature $2\pi/\sqrt{\mathcal{D}}$. This ensures that bulk shock wave computation in our case becomes identical to that of the shockwave computation presented in~\cite{Shenker:2013pqa,Roberts:2014ifa}. 
Thus, the exponential growth found in \cite{Das:2022jrr} matches exactly with the Roberts–Stanford shockwave in the AdS–Rindler patch as seen by the Hamiltonian-adapted observer.

%When the two operators lie on opposite sides of the fixed points, they are in different patches, $u$ and $d$, and $\eta$ does not cross onto the second sheet, consistent with the absence of growth.
The OTOC computation for $q>1$ is different because the deformed Hamiltonian no longer keeps the vacuum invariant. Therefore, even if simply think in the quotient description of Section~\ref{quotientdesciption}, where the late-time background is again AdS–Rindler, has an extra contribution of a conical defect. The bulk dual picture of OTOC for $q\neq 1$ is then no longer a simple coordinate transformation of the thermal one and we leave its computation for future work.

\section{Comparison of dynamics in two frames}\label{AppEE}
In the Floquet CFT literature, e.g. \cite{FO19,FO20_1,FO20_2,FO22}, the dynamics are studied in plane coordinates. That is, under Floquet Hamiltonian evolution, in Euclidean time `s' any point on the plane transforms as 
\begin{equation}\label{mobiusmapAppendix}
    \zeta \rightarrow \zeta_s= \left(\frac{a(s) \zeta^q +b(s)}{b^*(s) \zeta^q +a^*(s)}\right)^{1/q}
\end{equation}
where,
$a(s),
b(s)$ depends on the drive parameters $\alpha, \beta$ and $s$ and $a(t)$,
$b(t)$ upon analytic continuation of $s\rightarrow i t $. Quantities like EE, energy density in different phases are finally expressed in terms of the original cylinder coordinate $x_c$ and time $t$.

However, in the Hamiltonian-adapted coordinates, the Floquet time evolution  's' corresponds to $\zeta_s=F(z+s)$. 

For example, if we consider the heating phase, the $\zeta_s, \bar{\zeta}_s$ are 

\begin{equation}
   \zeta_s=F(z+s)= \left(-\frac{\alpha}{2\beta}+\frac{\sqrt{\mathcal{D}}}{2\beta} \tan \frac{q\sqrt{\mathcal{D}}(\sigma^0+s+i x)}{2} \right)^{\frac{1}{q}}; 
\end{equation}
and \begin{equation}
   \bar{\zeta}_s=F(\bar{z}+s)= \left(-\frac{\alpha}{2\beta}+\frac{\sqrt{\mathcal{D}}}{2\beta} \tan \frac{q\sqrt{\mathcal{D}}(\sigma^0+s-i x)}{2} \right)^{\frac{1}{q}}; 
\end{equation}
which in Lorentzian time t becomes 
\begin{equation}
   \zeta_t^{\pm}=F(x^{\pm})= \left(-\frac{\alpha}{2\beta}+\frac{\sqrt{\mathcal{D}}}{2\beta} \tan \frac{q\sqrt{\mathcal{D}}(\sigma^0+i x^{\pm})}{2} \right)^{\frac{1}{q}}; 
\end{equation}

where, $x^{\pm}=t\pm x$.  

%and are exactly what 
This is precisely what we obtain by substituting $a(t)$,$b(t)$   directly into Eq.~ \eqref{mobiusmapAppendix}. The following are the explicit expressions \cite{FO20_1} of $a(t)$ and $b(t)$
\begin{equation}\label{FloquetParametersAppendix}
a(t)
=
-\cosh\!\left(
\frac{\pi q t\sqrt{\mathcal D}}{L}
\right)
-i\,\frac{\alpha}{\sqrt{\mathcal D}}\,
\sinh\!\left(
\frac{\pi q t\sqrt{\mathcal D}}{L}
\right),
\qquad
b(t)
=
-i\,\frac{2\beta}{\sqrt{\mathcal D}}\,
\sinh\!\left(
\frac{\pi q t\sqrt{\mathcal D}}{L}
\right)    
\end{equation}

This shows that the Mobius transformation generated by the deformed Hamiltonian acts as a time translation in the Hamiltonian-adapted coordinates.

Let us try to understand our setup in two coordinate systems more clearly. 

In the old chart, for $s=0$ in (B.1), we have $a(s)=1,\ b(s)=0$, and hence, $\zeta_s=(\zeta^q)^{1/q}$. The $q$-th root splits the circle into $q$ sectors: the flow fixed points $\zeta^q=e^{\pm i \phi}$ subdivide each into a primary (P-arc) and a secondary (S-arc), giving $2q$ arcs and 
$2q$ fixed points. This structure is $s$-independent as the drive acts within each of these arcs. 

In the new deformed frame, on the other hand, we have everything in terms of $\tilde {F}=\frac{\cos (\frac{k}{2} (\sigma^0+ix)-\phi)}{\cos (\frac{k}{2} (\sigma^0+ix))}$ and $F=\tilde{F}^{1/q}$. Each of $q$ P-arcs and $q$ S-arcs gets mapped to two particular values of $\sigma^0$s: $\sigma_1^*=\phi/k$ and $\sigma_2^*=(\phi+\pi)/k$, with fixed points at $x=\pm \infty$, corresponding to physical slice at $\tau=0$ in the original cylinder coordinates.\footnote{ This is because, for $\zeta=e^{2\pi \omega /L}$, we can write $|\zeta|=e^{2\pi \tau_c/L }$, and therefore, at $\tau_c=0$ corresponds to the unit circle on plane as $|\zeta|=1$. Also, since $\tau_c=\frac{L}{2 \pi q}\log |\tilde {F}|$, the $\tau_c=0$ slice corresponds to $|\tilde{F}|=1$.} The $|\tilde{F}|=1$ at these values of $\sigma^0$ with $\tilde{F_P}=e^{2i \tan^{-1}(\tan \frac{\phi}{2}\tanh \frac{k \theta}{2})}$ and $\tilde{F_S}=e^{2i \tan^{-1}(\tan \frac{\phi}{2}\coth \frac{k \theta}{2})}$ . There are two other special points where at $\theta=0$, $\tilde{F}=0$ at $\sigma^0=(2\phi+\pi)/k$ and $\tilde{F}=\infty$ at  $\sigma^0=\pi/k$, which are also equivalently the branch points for $F$. 
Turning on $s$ (or $t$) corresponds to writing $\tilde{F_P}=e^{i\psi_{P}(x^{\pm})}$ and $\tilde{F_S}=e^{i\psi_S(x^{\pm})}$ with $\psi_P=2 \tan^{-1}(\tan \frac{\phi}{2}\tanh \frac{k x^{\pm}}{2})$ and $\psi_S= 2 \tan^{-1}(\tan \frac{\phi}{2}\coth \frac{k x^{\pm}}{2})$.

Note that there are two kinds of fixed points: one stable and one unstable for each pair.\footnote{These can be identified by analyzing the value of $\zeta'$ at the fixed points: ie unstable if $\zeta'>1$ and stable if $\zeta'<1$ \cite{FO20_1}.} One can check that, eg., $-\phi/q$ is the unstable fixed point whereas $\phi/q$ is the stable fixed point.

\section{Path-integral interpretation of the quotient description}\label{appendixPIComparison}

In Section \ref{twopicture}, we showed the entanglement entropy computations in two different ways: the global description \ref{Sec:ExactEE} and an alternative, quotient description in \ref{quotientdesciption}, in which the $q$ sectors are identified and their effect is carried by a conical defect. In this subsection we show that this alternative has a simple origin on the boundary, using the path-integral formalism of Calabrese-Cardy \cite{Calabrese:2004eu,Calabrese:2009qy}.  

In the path-integral formalism, vacuum correlation functions are computed by the Euclidean path integral of the CFT on the $\zeta$-plane, $\zeta=e^{\omega}$. The deformed Hamiltonian, however, does not act simply on $\zeta$ itself. As \eqref{mobiusmapAppendix} shows, it acts as a Möbius transformation on $\tilde F=\zeta^q$. The map $\zeta\to\tilde F$ divides the $\zeta$-plane into $q$ angular sectors(segments), each of which covers the whole $\tilde F$-plane once. Each point of the $\tilde F$-plane therefore has $q$ preimages,
$$\zeta_n=e^{2\pi in/q}\,\tilde F^{1/q},\qquad n=0,\dots,q-1,$$
one on each sector, and going once around the branch point $\tilde F=0$ takes each one to the next.
In each sector we can define a field on the $\tilde F$-plane, $\phi_n(\tilde F)=\phi\big(\zeta_n(\tilde F)\big)$. Since the action is local, it splits into a sum over sectors, and the path integral over $\phi$ on the $\zeta$-plane becomes a path integral over $q$ fields on a single $\tilde F$-plane,
$$Z=\int_{\mathcal C_q}\prod_{n=0}^{q-1}[D\phi_n]\;\exp\Big(-\sum_{n=0}^{q-1}S[\phi_n]\Big),$$
where $\mathcal C_q$ denotes the gluing conditions that encode the continuity of $\phi$: along a cut from $\tilde F=0$ to $\tilde F=\infty$, the lower edge of $n$ is glued to the upper edge of $n+1$, $\phi_n(x,0^-)=\phi_{n+1}(x,0^+)$, with $n+q\equiv n$. Viewed as a theory of $q$ copies of the CFT on one $\tilde F$-plane, these gluing conditions can be traded for branch-point twist fields at the two ends of the cut~\cite{Calabrese:2004eu,Cardy:2007mb,Calabrese:2009qy},
$$Z\;\propto\;\big\langle\sigma_q(0)\,\tilde\sigma_q(\infty)\big\rangle_{\mathcal C^{\otimes q}},$$
and correlation functions on the $\zeta$-plane become correlators in the $q$-copy theory with these twist fields inserted, each operator placed in the copy corresponding to its sector. The left-hand side is the global description, one CFT on the full $\zeta$-plane. The right-hand side is the quotient description, $q$ copies of the CFT on a single $\tilde F$-plane, with twist fields at the branch points.
This is precisely the construction of~\cite{Calabrese:2004eu,Calabrese:2009qy}, with our $q$ playing the role of their replica number $n$. 
%This is the construction of~\cite{CalabreseCardy2004,CalabreseCardy2009}, with $q$ in the role of the replica number. Two things differ here. The $q$-sheeted plane is the physical one, so $q$ is fixed by the deformed Hamiltonian and cannot be analytically continued. And the deformed Hamiltonian acts as a Möbius transformation on $\widetilde F$ but not on $\zeta$, which makes the single-sheet description natural for time evolution.
The only difference is that here $q$ is a fixed integer, set by the deformed Hamiltonian, and cannot be analytically continued. Accordingly, the entanglement entropies computed in Section~\ref{twopicture} require a separate replica index, independent of $q$.

The weights of these twist fields can be computed from the stress tensor. The vacuum stress tensor vanishes on the $\zeta$-plane, $\langle T(\zeta)\rangle=0$, whereas the stress tensor of each copy on the $\tilde F$-plane is
$$\langle T_n(\tilde F)\rangle=\frac{c}{12}\,\{\zeta_n,\tilde F\}=\frac{c}{24}\Big(1-\frac1{q^2}\Big)\frac{1}{\tilde F^2},$$
due to the Schwarzian of the map $\zeta_n(\tilde F)=e^{2\pi in/q}\tilde F^{1/q}$.
Since $\langle T_n(\tilde F)\rangle$ is the same for every copy, $\langle T_{\rm tot}(\tilde F)\rangle=\sum_{n=0}^{q-1}\langle T_n(\tilde F)\rangle=\frac{c}{24}\big(q-\frac1q\big)\frac{1}{\tilde F^2}$. From the conformal Ward identity, the twist fields at the two branch points, $\tilde F=0$ and $\tilde F=\infty$, have weights
$$h_{\sigma_q}=q\,h_q=\frac{c}{24}\Big(q-\frac1q\Big),\qquad h_q=\frac{c}{24}\Big(1-\frac1{q^2}\Big),$$
where $h_q$ is the weight per copy~\cite{Calabrese:2004eu,Lunin:2000yv}. The anti-holomorphic sector gives the same, $\bar h_q=h_q$. We note that $h_q$ is the weight, per copy, of the heavy operators that source the conical defect in the quotient description of Section~\ref{quotientdesciption}.
The total weight $h_{\sigma_q}$ has a natural algebraic counterpart, which also explains why the vacuum is nontrivial for $q>1$. We define the modes
$$\hat L_j=\frac1q\,L_{qj}+\frac{c}{24}\Big(q-\frac1q\Big)\delta_{j,0},$$
which satisfy a Virasoro algebra with central charge $c_{\rm eff}=qc$ \cite{Borisov:1997nc,Caputa:2022zsr}. The original vacuum is a highest-weight state of this algebra, $\hat L_j|0\rangle=0$ for $j>0$, with
$$\hat L_0|0\rangle=h_{\sigma_q}|0\rangle,\qquad h_{\sigma_q}=\frac{c_{\rm eff}}{24}\Big(1-\frac1{q^2}\Big)=\frac{c}{24}\Big(q-\frac1q\Big).$$
For $q>1$, however, the vacuum is not invariant under the global subalgebra generated by $\hat L_{0,\pm1}$: unlike for $q=1$, it is not annihilated by $\hat L_{-1}=\frac1q\,L_{-q}$, since $L_{-q}|0\rangle\neq0$.
%%For $q>1$ it is not invariant under the global subalgebra generated by $\hat L_{0,\pm1}$, since $\hat L_{-1}|0\rangle=\frac1q\,L_{-q}|0\rangle\neq0$.
In terms of these modes, the holomorphic part of the deformed Hamiltonian \eqref{q} reads
$$\alpha L_0+\beta\big(L_q+L_{-q}\big)=q\Big[\alpha\hat L_0+\beta\big(\hat L_1+\hat L_{-1}\big)\Big]-q\,\alpha\,h_{\sigma_q},$$
i.e., up to a constant, it is a global $sl(2,\mathbb R)$ generator of the embedded algebra. 
%The vacuum evolved by the deformed Hamiltonian is therefore a heavy primary of this algebra acted on by global conformal transformations. States of this form appear as $SL^{(q)}(2,\mathbb R)$ coherent states in~\cite{caputa,ge}, where they are treated as a family of states rather than as the time evolution generated by a fixed Hamiltonian.
The vacuum evolved by the deformed Hamiltonian is therefore a primary state of the $\hat L_j$ algebra, with weight $h_{\sigma_q}$, acted on by a global conformal transformation of that algebra, i.e. by a Möbius transformation of $\tilde F=\zeta^q$, as in \eqref{mobiusmapAppendix}. States of this form appear as $SL^{(q)}(2,\mathbb R)$ coherent states in~\cite{Caputa:2022zsr}, where they are treated as a family of states rather than as the time evolution generated by a fixed Hamiltonian.
This has a direct holographic interpretation. A primary of weight $h$ below the black-hole threshold, $h<c_{\rm eff}/24$, is dual to a conical defect with opening angle $2\pi\sqrt{1-24h/c_{\rm eff}}$~\cite{Benjamin:2020mfz}. For $h=h_{\sigma_q}$ this gives $2\pi/q$, i.e. the deficit $\delta_q=2\pi\big(1-\frac1q\big)$ of Section~\ref{twopicture}; since $h_{\sigma_q}/c_{\rm eff}=h_q/c$, counting per copy gives the same defect. A global conformal transformation of the $\hat L_j$ algebra acts in the bulk as an isometry of the background, which moves the defect.
%The two descriptions of Section~\ref{twopicture} are therefore two ways of writing the same state. 
The global and quotient descriptions are therefore two ways of writing the same state.
Using the decomposition of the Hamiltonian above, its holomorphic part gives
$$e^{-iHt}|0\rangle\;=\;e^{-it\left[\alpha L_0+\beta(L_q+L_{-q})\right]}|0\rangle\;=\;e^{iq\alpha h_{\sigma_q}t}\;e^{-iqt\left[\alpha\hat L_0+\beta(\hat L_1+\hat L_{-1})\right]}|0\rangle ,$$
together with its antiholomorphic counterpart. In the first form, in terms of $L_n$, i.e. of $F$, the exponent contains $L_{\pm q}$, which for $q>1$ lie outside the global subalgebra. The vacuum is acted on by a non-global conformal transformation, namely \eqref{mobiusmapAppendix} read as a map of $\zeta$. This is the global description: the bulk is smooth, and the Bañados metrics of the $P$ and $S$ patches depend on time. In the second form, in terms of $\hat L_j$, i.e. of $\tilde F$, the exponent contains only $\hat L_{0,\pm1}$. The vacuum, a primary of weight $h_{\sigma_q}$, is acted on by a global transformation, namely \eqref{mobiusmapAppendix} read as a Möbius map of $\zeta^q$. This is the quotient description: the bulk is a static AdS–Rindler background, and the time dependence is carried by the moving defect. These identities are exact. The shockwave description of Section~\ref{Sec:EffectiveShockwave} is an effective description of the moving defect at late times, and is compared with the exact result in Section~\ref{comparison}.

   % where, $\theta^c$ denotes the end points very near to the unstable fixed points (with a cut-off). Therefore, late time EE becomes time independent.
   % \item 
%\end{itemize}
%Bulk RT surface(?)

%\vskip 0.5cm
%\includegraphics[scale=0.2]{1.jpeg}
%\vskip 0.5cm
%\includegraphics[scale=0.2]{2.jpeg}
%\vskip 0.5cm
%\includegraphics[scale=0.2]{3.jpeg}
%\includegraphics[scale=0.2]{2.jpeg}

\section{Killing symmetry and causal horizons in the heating phase}\label{heating horizon derivation}
In this appendix, we first review the general procedure for finding horizons in a generic Banados metric \eqref{banadoslt} and then derive the causal horizon equations associated with P and S regions of the heating phase. We begin by studying the relation between the Killing symmetries of the Bañados geometry and the solution of the Hill equation, following 
\cite{Sheikh_Jabbari_2016, CompereEtAl2016}. We then use the Roberts transformation \cite{Roberts:2012aq} to determine the null hypersurfaces associated with the boundary causal diamonds considered in the main text. 

Since the B\~anados geometry is locally $\mathrm{AdS}_3$, it possesses six local Killing vectors. However, the Killing vector relevant to a particular observer must be selected through additional physical input, which in our case is the boundary interval.

% We will start with the discussion of the null hypersurfaces generated by Killing vectors for the geometry, but obviously that is not enough. We need to study the asymptotic behavior of these Killing vectors to find which ones of them correspond to an event horizon. For Banados metrics, we would see that the study of the Killing vectors reduces to the study of the solutions of a simpler equation, known as the Hill equation. It is a second-order linear differential equation. Generally, the 'potential' term of the equation has a periodicity, and studying the monodromies of the solutions helps us classify the global Hill solutions and determine whether globally defined stabilizers exist. 
%However, as we will see, the equation for finding the local null Killing vectors are intact but we have to study the asymptotic behavior  of the \textit{potential} term to find the event horizon.  The components of the generic Banados metric (\ref{banadoslt}) is
Setting the AdS radius to unity, the nonvanishing components of the metric (\ref{banadoslt}) are \cite{Banados:1998gg}
\begin{align*}
 g_{rr}&=\frac{1}{r^2},\\
 g_{++}&=\mathcal{L}_+,\qquad g_{--}=\mathcal{L}_-,\\
 g_{+-}&=-\frac12\left(r^2+\frac{\mathcal{L}_+\mathcal{L}_-}{r^2}\right).
\end{align*}
 Consider a general vector field 
 % \cite{CompereEtAl2016, Sheikh_Jabbari_2016}
 
\begin{equation*}
\zeta=R(r,x^+,x^-)\partial_r+X(r,x^+,x^-)\partial_++Y(r,x^+,x^-)\partial_-.
\end{equation*}
Demanding that this vector generate an isometry gives
\begin{equation}
(\Lie_\zeta g)_{\mu\nu}
=\zeta^\rho\partial_\rho g_{\mu\nu}
+g_{\rho\nu}\partial_\mu\zeta^\rho
+g_{\mu\rho}\partial_\nu\zeta^\rho=0.
\end{equation}
The solution has the components \cite{Sheikh_Jabbari_2016}
\begin{align}
R=-\frac r2(K_+'+K_-'),\label{zetaR}\\
X=K_++\frac{r^2K_-''+L_-K_+''}{2\Delta_{\mathrm{FG}}},\label{zetaP}\\
Y=K_-+\frac{r^2K_+''+L_+K_-''}{2\Delta_{\mathrm{FG}}}.\label{zetaM}
\end{align}
where $\Delta_{\mathrm{FG}}=(r^4-\mathcal{L}_{+}\mathcal{L}_-)$, and the functions $K_\pm $ obeys the third order stabilizer equations \cite{Sheikh_Jabbari_2016, CompereEtAl2016}
\begin{equation}
K_\pm'''-4\mathcal{L}_\pm K_\pm'-2\mathcal{L}_\pm'K_\pm=0
\label{Stabilizer}
\end{equation}
However, these equations may be solved in terms of independent solutions of \footnote{This is known as the Hill equation. It is strictly a Hill equation only if $\mathcal{L}(x)$ is periodic in $x$; however, the same equations remain applicable to our nonperiodic functions.} 
\begin{equation*}
\psi''-\mathcal{L}_+(x)\psi=0.
\label{AuxEq}
\end{equation*}
If $\psi_1,\ \psi_2$ are independent solutions,
all of three combinations $\psi_1^2,\ \psi_1\psi_2,\ \psi_2^2$ satisfies the chiral stabilizer equation \eqref{Stabilizer}. There are three more solutions for the anti-chiral mode solution. Therefore, there are six local Killing vectors. To identify which combination is relevant, we note that in the periodic case, $\psi_1 \psi_2$ is a suitable Floquet basis defined globally. Denoting the Killing vectors by $\zeta_+\equiv \zeta[\psi_1 \psi_2,0]$ and $\zeta_-\equiv \zeta[0,\varphi_1 \varphi_2]$, one considers $\zeta_{\mathcal{H}_\pm}=\zeta_+\pm \zeta_-$. The norms of these vectors are given by (after using the Banados metric)
\begin{equation*}
\zeta_{\mathcal{H}_\pm}^2=\mp \frac{\psi_1 \psi_2\varphi_1\varphi_2}{r^2}(r^2-r^2_1{_{\pm}})(r^2-r^2_2{_{\pm}})
\end{equation*} with \cite{Sheikh_Jabbari_2016}.
\begin{align}\label{horizons}
r_{1+}^2&=\frac{\psi_1'}{\psi_1}\frac{\varphi_1'}{\varphi_1},&
 r_{2+}^2&=\frac{\psi_2'}{\psi_2}\frac{\varphi_2'}{\varphi_2},\\
r_{1-}^2&=\frac{\psi_1'}{\psi_1}\frac{\varphi_2'}{\varphi_2},&
 r_{2-}^2&=\frac{\psi_2'}{\psi_2}\frac{\varphi_1'}{\varphi_1}..
\end{align}
These are algebraic candidate horizons; selecting the relevant surfaces requires specifying the Killing generator.
\subsection{Killing vectors for P and S intervals}
In our heating-phase construction, \(x^\pm\) are noncompact Hamiltonian-adapted coordinates and \(\mathcal L_\pm\) are not periodic. We select the appropriate stabilizer using the endpoints of the P or S interval.

Consider an interval with endpoints \((T,X)=(0,\pm a)\) in Poincar\'e coordinates. The corresponding AdS-Rindler Killing vector is \cite{Faulkner_2014}
\begin{equation}
 \chi_a=\frac{1}{2a}\left[(a^2-T^2-X^2-u^2)\partial_T
 -2TX\partial_X-2Tu\partial_u\right].
 \label{eq:app-interval-killing}
\end{equation}
For \(X^\pm=T\pm X\), its boundary components are \(\chi_a^\pm=[a^2-(X^\pm)^2]/(2a)\). Thus, under \(X^\pm=f_\pm(x^\pm)\), the required stabilizers are
\begin{equation}
 K_\pm(x^\pm)=\frac{a^2-f_\pm(x^\pm)^2}{2a f_\pm'(x^\pm)}.
 \label{eq:app-interval-stabilizer}
\end{equation}
These obey (\ref{Stabilizer}) provided \(\mathcal L_\pm=-\tfrac12\{f_\pm,x^\pm\}\).

Writing
\begin{equation}
 f(x)=\frac{\psi_1(x)}{\psi_2(x)},\qquad
 \psi_1'\psi_2-\psi_1\psi_2'=1,
 \label{eq:app-projective-f}
\end{equation}
we have \(f'=1/\psi_2^2\) and \(\{f,x\}=-2\mathcal L(x)\). Consequently, the stabilizer is
\begin{equation}
 K(x)=\frac{a^2\psi_2(x)^2-\psi_1(x)^2}{2a}.
 \label{eq:app-K-hill}
\end{equation}
Defining the adapted Hill basis
\begin{equation}
 \widetilde\psi_1=\frac{\psi_1+a\psi_2}{\sqrt{2a}},\qquad
 \widetilde\psi_2=\frac{a\psi_2-\psi_1}{\sqrt{2a}},
 \label{eq:app-hill-adapted}
\end{equation}
we obtain \(K=\widetilde\psi_1\widetilde\psi_2\). Thus, the product of two Hill solutions remains the relevant stabilizer, provided the basis is fixed by the endpoints of the interval. 

The associated projective coordinate is
\begin{equation}
 w(x)=\frac{\widetilde\psi_1}{\widetilde\psi_2}
 =\frac{a+f(x)}{a-f(x)}\equiv e^{k h(x)}.
 \label{eq:app-projective-w}
\end{equation}
In the unit-Wronskian normalization,
\begin{equation}
 \widetilde\psi_{1,2}(x)=\frac{e^{\pm k h_a(x)/2}}{\sqrt{k h_a'(x)}},
 \qquad K(x)=\frac{1}{k h_a'(x)},
 \label{eq:app-hill-normalized}
\end{equation}
so
% \begin{equation}
%  \frac{\widetilde\psi_{1,2}'}{\widetilde\psi_{1,2}}
%  =\frac12\left(\pm k h_a'-\frac{h_a''}{h_a'}\right).
%  \label{eq:app-hill-log}
% \end{equation}
The roots of the interval-adapted Killing vector \(\zeta[K(x^+),K(x^-)]\) are
\begin{align}
 r_{\mathcal H,1}^2&=\frac14\prod_{\sigma=\pm}
 \left(kh'(x^\sigma)-\frac{h''(x^\sigma)}{h'(x^\sigma)}\right),\nonumber\\
 r_{\mathcal H,2}^2&=\frac14\prod_{\sigma=\pm}
 \left(kh'(x^\sigma)+\frac{h''(x^\sigma)}{h'(x^\sigma)}\right).
 \label{eq:app-heating-hill-horizons}
\end{align}
These are the specialization of the general Hill construction \cite{Sheikh_Jabbari_2016} to our interval-selected basis. Their causal interpretation will be established below. For \(q>1\), \(K\) is generally position dependent, so this Killing generator need not equal the Hamiltonian-adapted coordinate flow \(\partial_t\). 

\subsection{Roberts transformation and causal horizon}
We now establish the causal interpretation. The interval with endpoints \((T, X)=(0,\pm a)\) has future and past bulk null boundaries given by the appropriate portions of 
\begin{equation}
 (T-a)^2=X^2+u^2,\qquad T\leq a,
 \label{eq:app-future-cone}
\end{equation}
\begin{equation}
 (T+a)^2=X^2+u^2,\qquad T\geq-a.
 \label{eq:app-past-cone}
\end{equation}
They are Killing horizons of (\ref{eq:app-interval-killing}). Their intersection is the bifurcation geodesic
\begin{equation}
 T=0,\qquad X^2+u^2=a^2,
 \label{eq:app-bifurcation}
\end{equation}
which is also the RT geodesic of the interval in the smooth AdS vacuum \cite{Ryu_2006, Casini_2011}.

We express these surfaces in the Hamiltonian-adapted B\~anados chart using the finite diffeomorphism of Roberts \cite{Roberts:2012aq}. With \(y=1/r\), the B\~anados metric is
\begin{equation}
 ds^2=\frac{dy^2}{y^2}+\mathcal L_+(dx^+)^2+\mathcal L_-(dx^-)^2
 -\left(\frac{1}{y^2}+y^2\mathcal L_+\mathcal L_-\right)dx^+dx^-.
\end{equation}
Let \(f_\pm=f_\pm(x^\pm)\), \(A_\pm=f_\pm''/f_\pm'\), and \(J=4r^2-A_+A_-\). On a regular branch, the Roberts map takes the form
\begin{align}
 X^+&=f_++\frac{2f_+'A_-}{J},\nonumber\\
 X^-&=f_-+\frac{2f_-'A_+}{J},\nonumber\\
 u&=\frac{4r\sqrt{f_+'f_-'}}{J}.
 \label{eq:app-roberts}
\end{align}
 A direct substitution into either null-cone equation yields
\begin{equation}
 (X^+-c)(X^--c)-u^2=
 \frac{4(f_+-c)(f_--c)}{J}
 \left[r^2-Q_c(x^+)Q_c(x^-)\right],
 \label{eq:app-roberts-cone}
\end{equation}
where
\begin{equation}
 Q_c(x)=\frac{f'(x)}{f(x)-c}-\frac12\frac{f''(x)}{f'(x)}.
 \label{eq:app-Qc}
\end{equation}
% Therefore, the two algebraic null-sheet roots are
% \begin{equation}
%  r^2=Q_{+a}(x^+)Q_{+a}(x^-),\qquad
%  r^2=Q_{-a}(x^+)Q_{-a}(x^-).
%  \label{eq:app-general-horizon-roots}
% \end{equation}
% We emphasize that (\ref{eq:app-roberts-cone}) and (\ref{eq:app-Qc}) are direct consequences of the Roberts map, rather than formulas quoted from Ref.~\cite{Roberts2012}. The inequalities in (\ref{eq:app-future-cone})--(\ref{eq:app-past-cone}) and the regular-coordinate domain select their physical causal portions.
Therefore, choosing $c=+a$ and $c=-a$, respectively,
the algebraic equations for the future and past
horizon sheets are
\begin{equation}
\begin{aligned}
\mathcal H^+:\quad
&r^2=Q_{+a}(x^+)Q_{+a}(x^-),\\
\mathcal H^-:\quad
&r^2=Q_{-a}(x^+)Q_{-a}(x^-).
\end{aligned}
\label{eq:app-general-horizon-roots}
\end{equation}
The Hill and Roberts results agree identically: from \(f=\psi_1/\psi_2\) one finds
\begin{equation}
 Q_c(x)=\frac{\psi_1'(x)-c\psi_2'(x)}
 {\psi_1(x)-c\psi_2(x)}.
 \label{eq:app-Qc-hill}
\end{equation}
In the basis (\ref{eq:app-hill-adapted}), this gives
\begin{equation}
 Q_{-a}=\frac{\widetilde\psi_1'}{\widetilde\psi_1},\qquad
 Q_{+a}=\frac{\widetilde\psi_2'}{\widetilde\psi_2}.
 \label{eq:app-Qc-adapted}
\end{equation}
Substituting these expressions into (\ref{eq:app-general-horizon-roots}) reproduces (\ref{eq:app-heating-hill-horizons}). The Hill construction therefore identifies the Killing generator, while the Roberts map demonstrates that its horizons are precisely the null boundaries of the selected interval's causal wedge.

\paragraph{Explicit P and S horizon equations.}

The heating-phase conformal map obtained in the main text is naturally expressed in terms of the angular variable on the unit circle, \(F_\pm(x^\pm)=e^{i\Theta^\pm(x^\pm)}\). To apply the Roberts transformation in real Poincaré coordinates, we 
use the Cayley coordinates \eqref{eq:cayley}.
% first compose this map with the Cayley transformation, which is a Möbius transformation mapping the unit circle to the real line,

% \[
% \chi^\pm
% =-i\,\frac{F-1}{F+1}
% =\tan\left(\frac{\Theta}{2}\right),
% \]

% where, for the P region,

% \[
% \Theta^\pm(x^\pm)=\frac{2}{q}\tan^{-1}\left[
% \tan\left(\frac{\phi}{2}\right)
% \tanh\left(\frac{kx^\pm}{2}\right)
% \right].
% \]

% Since the Schwarzian derivative is invariant under Möbius transformations, this additional transformation does not change the corresponding Bañados functions \(\mathcal L_\pm\).
We therefore use \(\chi_\pm=\widehat F_\pm\) in the Roberts transformation. In particular, the endpoints of the P interval at \(x^\pm\to\pm\infty\) are mapped to

\[
f_\pm(\pm\infty)=\pm a_P,
\qquad
a_P=\tan\left(\frac{\phi}{2q}\right).
\]
Thus, the two values of $c$ entering
Eq.~(\ref{eq:app-Qc}) are $c=\pm a_P$.

Using the Cayley-transformed heating-phase map
% \begin{equation}
%  \Theta'(x)=
%  \frac{\sqrt{\mathcal D}\sin\varphi}
%  {\cosh(kx)+\cos\varphi},
%  \qquad
%  \frac{\Theta''(x)}{\Theta'(x)}
%  =-\frac{k\sinh(kx)}
%  {\cosh(kx)+\cos\varphi},
% \end{equation}
we obtain
\begin{align}
 Q_{+a_P}(x)
 &=\frac{\sqrt{\mathcal D}}
 {2[\cosh(kx)+\cos\phi]}
 \left[
 q\sinh(kx)
 -\sin\phi\cot\left(
 \frac{\phi/q-\Theta(x)}{2}
 \right)
 \right],
 \nonumber\\
 Q_{-a_P}(x)
 &=\frac{\sqrt{\mathcal D}}
 {2[\cosh(kx)+\cos\phi]}
 \left[
 q\sinh(kx)
 +\sin\phi\cot\left(
 \frac{\phi/q+\Theta(x)}{2}
 \right)
 \right].
 \label{eq:app-P-Q}
\end{align}

Consequently, the algebraic equations for the
future and past causal horizons of the P region are
\[
\begin{aligned}
 \mathcal H_P^+:\quad
 &r^2=Q_{+a_P}(x^+)Q_{+a_P}(x^-),\\
 \mathcal H_P^-:\quad
 &r^2=Q_{-a_P}(x^+)Q_{-a_P}(x^-),
\end{aligned}\]

At \(t=0\), their common bifurcation curve is
\begin{equation}
 r_{\mathrm{bif},a_P}^2(x)=Q_{a_P,+}(x)Q_{a_P,+}(-x)
 =Q_{a_P,-}(x)Q_{a_P,-}(-x).
 \label{eq:app-bifurcation-r}
\end{equation}
In the smooth AdS$_3$ covering, this curve is the RT geodesic of the selected interval, as follows from the AdS-Rindler construction \cite{Ryu_2006, Casini_2011}. This extends the $q=1$ heating-phase interval/horizon correspondence discussed in \cite{Das:2024vqe}.

As a consistency check, for $q=1$ the
Cayley-transformed map reduces to
\begin{equation}
 f_a(x)=a_a\tanh\frac{kx}{2}.
\end{equation}
Substituting into Eq.~(\ref{eq:app-Qc}) gives
\begin{equation}
 Q_{+a_a}(x)=-\frac{k}{2},
 \qquad
 Q_{-a_a}(x)=\frac{k}{2}.
\end{equation}
Consequently, both horizon equations reduce to
\begin{equation}
 r_{\mathcal H}=\frac{k}{2}
 =\frac{\sqrt{\mathcal D}}{2}.
\end{equation}
Furthermore, the interval-adapted projective
coordinate satisfies $w_a=e^{kx}$, implying
$h_a(x)=x$ and $K_a=1/k$. Thus, the corresponding
Killing generator is proportional to
$\partial_t$. For $q>1$, the stabilizer $K_a$
is generally nonconstant.

% \paragraph{Distinction from the FG wall and the vanishing of $g_{tt}$.}

Finally, the Killing/causal horizons should be distinguished from two coordinate-dependent loci. The component along Hamiltonian time is
\begin{equation}
g_{tt}=-\frac{(r^2-\mathcal L_+)(r^2-\mathcal L_-)}{r^2},
 \label{eq:app-gtt}
\end{equation}
so \(g_{tt}=0\) at \(r^2=\mathcal L_+\) or \(r^2=\mathcal L_-\) when these roots are real. For \(q>1\), \(\partial_t\) is not a Killing vector, so such roots need not be Killing horizons.

Independently, the radial Fefferman--Graham coordinate degenerates at
\begin{equation}
 r_{\mathrm{FG}}^4=\mathcal L_+(x^+)\mathcal L_-(x^-).
 \label{eq:app-FG-wall}
\end{equation}
This is the vanishing of the metric determinant in B\~anados gauge \cite{Sheikh_Jabbari_2016}; it is a coordinate artifact, not a curvature singularity. These loci coincide with the horizon radius in the equal, constant-\(\mathcal L\) AdS-Rindler case, but need not coincide for nonconstant profiles.

In the heating phase, some horizon roots lie beyond the boundary-connected outer FG patch. These portions must not be treated as regularly covered by that patch; the relevant continuation and causal selection are instead obtained in regular Roberts, Poincar\'e, or global AdS coordinates \cite{Roberts:2012aq, Sheikh_Jabbari_2016}.

Thus the Hill-equation and Roberts-map constructions are complementary: the former determines the interval-selected Killing vector and its candidate null loci, while the latter verifies that the selected sheets are the causal horizons associated with the P/S boundary diamonds.

\section{Analysis of KSW Criteria}\label{kswa}
The main point of the KSW criteria \cite{Kontsevich:2021dmb, Witten:2021nzp} is the requirement that the real part of the kinetic terms of any fields on a complex background should be positive so that the corresponding path integral is convergent. For a generic p-form field, the kinetic term looks like 
\begin{equation*}
    S_p\sim \int d^dx\ \sqrt{\det g}\ g^{\mu_1\nu_1}\dots g^{\mu_p\nu_p}\omega_{\mu_1 \dots \mu_p}\bar{\omega}_{\nu_1 \dots \nu_p}
\end{equation*}
So, the KSW criteria demand
\begin{equation*}
    \Re S_p>0.
\end{equation*}
For $p=0\ \text{and}\ p=1$ form fields, the requirement reduces consecutively to
\begin{align}\label{ksw}
    \Re \sqrt{\det g}&>0 \\
    \Re\left(\sqrt{\det g}\ g^{-1}\right)&>0. \nonumber
\end{align}
We will first analyze it for the general Banados metric and then will focus on the specific metrics we are considering here. After $y\to\frac{1}{r}$, the metric in (\ref{banadoslt}) can be written in matrix form as :
\[
\begin{pmatrix}
r^{-2} & 0 & 0\\
0 & \mathcal{L}_+ & -\frac{P}{2}\\
0 &  -\frac{P}{2} & \mathcal{L}_- 
\end{pmatrix}
\]
where $P=r^2+\frac{\mathcal{L}_+\mathcal{L}_-}{r^2}$. Since $r^{-2}$ is a real function, it is sufficient to focus on the $2\times 2$-block-diagonal part of the matrix to study the conditions in~\eqref{ksw}. Writing $\mathcal{L}_+=a+ib,\ \mathcal{L}_-=c+id$, we write $\mathcal{L}_+\mathcal{L}_-=\underbrace{(ac-bd)}_{u}+i\underbrace{(ad+bc)}_v=u+iv$. In these variables, the determinant of the $2 \times 2$ matrix($G$) \[
\begin{pmatrix}
\mathcal{L}_+ & -\frac{P}{2}\\
-\frac{P}{2} & \mathcal{L}_- 
\end{pmatrix}
\] is $\sqrt{\text{det}G}=\frac{v+i(r^4-u)}{2r^2}$. Therefore,
\begin{equation}\label{cond1}
 \Re\sqrt{\text{det}G}>0\Rightarrow (ad+bc)>0
\end{equation}
On the other hand, the matrix $(M)$ corresponding to $\Re(\sqrt{\text{det}G}\ G^{-1})$ is 
\begin{equation}\label{matrix2}
\frac{2r^2}{\Delta}
\begin{pmatrix}
vc+(r^4-u)d & r^2v\\
r^2v & va+(r^4-u)b 
\end{pmatrix}
\end{equation}
where $\Delta=(r^4-u)^2+v^2$. Sylvester's Criterion tells us that a $2\times 2$ matrix $(M)$ is positive definite iff $\text{det}M>0\ \text{and}\ \text{one diagonal element}>0$. For (\ref{matrix2}), $\text{det}M=\frac{4r^4bd}{\Delta}>0$. On the otherhand $M_{11}=M_{11}=\frac{2r^2}{\Delta}
\left[b(c^2+d^2)+r^4d\right]$. Therefore, for $r>0$ and $\Delta>0$, the
positive-definiteness of $M$ is equivalent to
\begin{equation}
b>0,\qquad d>0.
\end{equation}
Combining this with (\ref{cond1}), we obtain
\begin{equation}
b>0,\qquad d>0,\qquad ad+bc>0.
\label{eq:KSW-corrected}
\end{equation}

% corresponds to
% \begin{equation}\label{cond2}
%    b(c^2+d^2)+r^4d>0
% \end{equation}
Therefore, to analyze the KSW allowability of a generic Banados metric, we need to verify (\ref{eq:KSW-corrected}).
We will now analyze these conditions specifically for the Banados factor corresponding to the heating phase. To avoid notational clutter, we reexpress it as
\begin{equation}\label{hlpm1}
  \mathcal{L}_{\pm}=\frac{\mathcal{D}}{4}\left(q^2+\frac{K(C_\pm^2-S_\pm^2+2iC_\pm S_\pm)}{(C_\pm^2+S_\pm^2)^2}\right)  
\end{equation}
with the following identifications:
\begin{equation*}
  \begin{split}  A=q\sqrt{\mathcal{D}}\sigma^0-\phi,\ B_\pm=q\sqrt{\mathcal{D}}x^\pm,\ K=(q^2-1)\sin{\phi}^2\ge 0,\ C_\pm=\cos{A} \cosh{B_\pm}+\cos{\phi},\\ S_\pm=\sin{A}\sinh{B_\pm},\ Q_\pm=C_\pm^2+S_\pm^2
\end{split}
\end{equation*}
Therefore, from (\ref{hlpm1}) we get that
\begin{eqnarray}
    a,c\equiv \Re(\mathcal{L}_\pm)&=&\frac{\mathcal{D}}{4}\left(q^2+K\frac{C_\pm^2-S_\pm^2}{Q_\pm^2}\right)\\
  b,d\equiv \Im(\mathcal{L}_\pm)&=& \frac{\mathcal{D}K}{2}\frac{C_\pm S_\pm}{Q_\pm^2} 
\end{eqnarray}
Using the facts that $K\ge0$ and $Q_\pm\ge0$, (\ref{eq:KSW-corrected}) turns out to be
\begin{align}
  C_+S_+>0\\[1ex]
  C_-S_->0\\[1ex]
  \frac{\mathcal{D}^2K}{8}\left[q^2(C_-S_-Q_+^2+C_+S_+Q_-^2)+K(C_-S_++C_+S_-)(C_+C_--S_+S_-)\right]>0
\end{align}
\begin{figure}[H]
    \centering
    \includegraphics[width=0.95\textwidth]{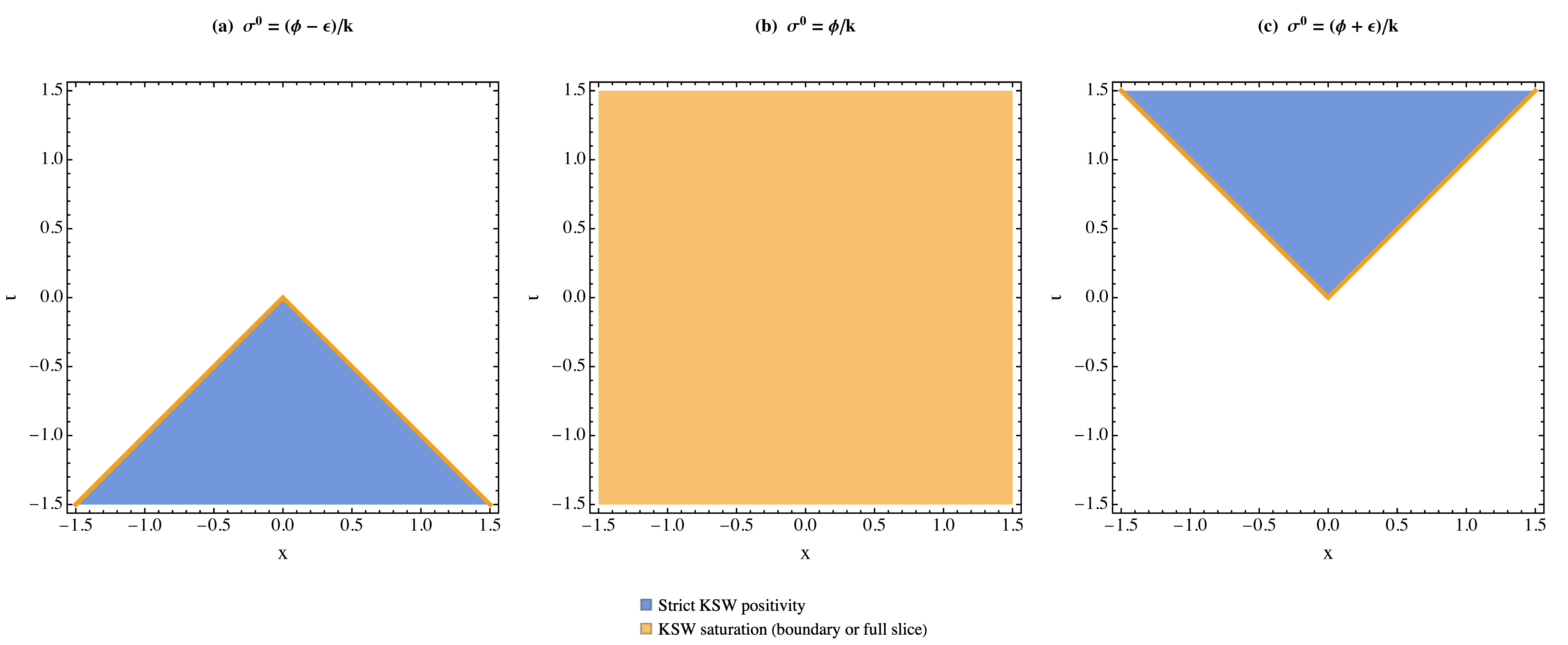}
    \caption{This figure shows the coordinate patches where the KSW conditions are either strictly followed or saturated for three different $\sigma^0$ values. }
    \label{fig:ksw}
\end{figure}
Let's check these criteria for a few cases. 
\begin{itemize}
    \item For $q=1$, $K=0$: At this value, as we see from (\ref{hlpm1}), $\mathcal{L}_\pm$ are real, so the metric is purely Lorentzian. As a result, both conditions above are exactly saturated for any $\sigma^0$.
     \item For generic $q$ at $\phi=\frac{\pi}{2}$, i.e. for $\alpha=0$, (\ref{hlpm1}) is real and both criteria are again saturated at $\sigma^0=0$. 
     %However, we can ask if there is a range of $\sigma^0$ for which the metric would be complex but would satisfy both the KSW criteria in the whole range of $x_\pm$ or if for any other fixed value of $\sigma^0$ there is a range of $x_\pm$ for which the KSW criteria are satisfied.
   \item For generic $q$ and generic $\phi$, we can check that (\ref{hlpm1}) is real at $\sigma^0=\frac{\phi}{q \sqrt{\mathcal{D}}}$ or $\sigma^0=\frac{\phi}{q\sqrt{\mathcal{D}}}+\frac{\pi}{q \sqrt{\mathcal{D}}}$ and the two conditions are also exactly saturated at these points. However, for other choices of $\sigma^0$, the conditions are either strictly satisfied or even saturated only within a particular range of the coordinates. See Fig.[\ref{fig:ksw}]. However, the vanishing of the imaginary parts of $\mathcal{L}_\pm$ at the specific values of $\sigma^0$ alone does not establish the admissibility of contours analytically continued around these points. To establish global allowability, it is necessary to specify a complex integration contour and verify the KSW positivity at every point along the contour which is beyond the scope of this paper.
\end{itemize}

\bibliographystyle{JHEP}
\bibliography{references}

\end{document}